%% file: paperPRA.tex
\documentclass[preprint,aps,pra,10pt,twocolumn,showpacs,amsmath,amssymb,superscriptaddress]{revtex4-1}

\usepackage{color}
\usepackage[english]{babel}
\usepackage{marginnote}
\usepackage{bm}
\usepackage{graphicx}
\usepackage{dcolumn}
\usepackage{epstopdf}
\usepackage[textsize=tiny]{todonotes}
\usepackage[load-configurations=abbreviations]{siunitx}
\usepackage{amsmath, amssymb, braket, mathtools}
\usepackage[version = 4]{mhchem}
\usepackage{url}
\usepackage{tikz}
\usepackage{verbatim}
\usepackage{dcolumn}

\usetikzlibrary{calc,arrows,decorations.pathmorphing,decorations.pathreplacing,backgrounds,positioning,fit,petri,shadings}

\tikzset{%
  >=latex, 
  inner sep=0pt,%
  outer sep=2pt,%
  mark coordinate/.style={inner sep=0pt,outer sep=0pt,minimum size=3pt,
    fill=black,circle}%
}

\newcommand{\vect}[1]{\mathbf{\bm{#1}}}

\newcommand{\dket}[1]{\mathinner{| #1 \rangle\rangle}}

\newcommand{\dKet}[1]{\mathinner{\left|\left. #1 \right\rangle\right\rangle}}

\newcommand{\dBraket}[1]{\Braket{\left\langle{#1}\right\rangle}}

\newcommand{\figref}[1]{figure~\ref{fig:#1}}
\newcommand{\Figref}[1]{Figure~\ref{fig:#1}}

\newcommand{\secref}[1]{section~\ref{sec:#1}}

\newcommand{\eref}[1]{Equation~\eqref{eq:#1}}

\renewcommand{\figref}[1]{Fig.~\ref{fig:#1}}
\renewcommand{\Figref}[1]{Fig.~\ref{fig:#1}}
\renewcommand{\eref}[1]{Eq.~\eqref{eq:#1}}

\renewcommand{\Im}{\operatorname{Im}}

\newcommand{\diff}[2]{\frac{d #1}{d #2}}

\newcommand{\hc}{\mathrm{H.c.}} 
\newcommand{\mO}{\textup{\AE}} 

\definecolor{darkviolet}{HTML}{8A2BE2}
\definecolor{tan}{HTML}{FFD700}

\reversemarginpar

\DeclareSIUnit\roothertz{\ensuremath{\sqrt{\text{\hertz}}}}
\graphicspath{{images/}}

\begin{document}
\title{Ultrafast coherent control of spinor Bose-Einstein condensates using stimulated Raman adiabatic passage}
\author{Andreas M. D. Thomasen}
\affiliation{New York University, 1555 Century Avenue, Pudong, Shanghai 200122, China}
\affiliation{National Institute of Informatics, 2-1-2 Hitotsubashi, Chiyoda-ku, Tokyo 101-8430, Japan}
\author{Tetsuya Mukai}
\affiliation{NTT Basic Laboratories, 3-1 Morinosato Wakamiya, Atsugi-shi, Kanagawa 243-0198, Japan}
\author{Tim Byrnes}
\affiliation{New York University, 1555 Century Avenue, Pudong, Shanghai 200122, China}
\affiliation{NYU-ECNU Institute of Physics at NYU Shanghai, 3663 Zhongshan Road North, Shanghai 200062, China}
\affiliation{National Institute of Informatics, 2-1-2
Hitotsubashi, Chiyoda-ku, Tokyo 101-8430, Japan}
\affiliation{Department of Physics, New York University, New York, NY 10003, USA}
\date{\today}

\begin{abstract}
We propose the use of stimulated Raman adiabatic passage (STIRAP) to offer a fast high fidelity method of performing SU(2) rotations on spinor Bose Einstein condensates (BEC). Past demonstrations of BEC optical control suffer from difficulties arising from 
collective enhancement of spontaneous emission and inefficient two-photon transitions
originating from selection rules.  We present here a novel scheme which allows for arbitrary coherent rotations of two-component BECs while overcoming these issues. Numerical tests of the method show that for BECs of \ce{^{87}Rb} with up to $ 10^4 $ atoms and gate times of $ \SI{1}{\micro\second} $, decoherence due to spontaneous emission can be suppressed to negligible values.
\end{abstract}
\maketitle%

\section{Introduction}
\label{sec:I}
Coherent control of Bose-Einstein Condensates (BECs) is an important task required for a variety of applications, from quantum metrology \cite{ockeloen2013quantum, riedel2010atom, byrnes2015quantum} to quantum simulation \cite{bloch2012quantum, fischer2004quantum, simon2011quantum} and quantum information \cite{zoller2003quantum, brennen1999quantum}. Experimental methods to create and coherently control them are becoming increasingly sophisticated. For instance, the precision obtainable in certain quantum magnetometry experiments is limited only by projection noise \cite{wasilewski2010quantum}. Mach-Zender type interference of BECs in microgravity has also been realized in a particle drop experiment \cite{muntinga2013interferometry}, where the BECs spatial splitting and recombination with resulting matter-wave interference has been observed. Visions for BECs investigate them as switches for nanomechanical devices \cite{treutlein2007bose} or as candidates for quantum computing \cite{timbec2, timbec3, timbec4, pyrkoventangle}. Laser cooled atoms exhibit long coherence times \cite{mukailifetime}, and as such they fulfill a key condition required of physical systems in quantum information related tasks \cite{zoller2005quantum, simon2010quantum}. In the case of applications such as atomic clocks, magnetometry, and quantum information, the desired degrees of freedom that require control are the internal hyperfine states of the atoms. Here a fundamental task is to perform this coherent control with high frequency and fidelity to improve the overall performance.

Control of the hyperfine states of atomic ensembles and BECs can be done in broadly two methods: using microwave or optical transitions.  Traditional atomic clocks use microwave resonances between clock states, which is determined by the natural hyperfine splitting between the ground states\cite{campbell2006imaging}.  Optical clocks on the other hand rely on the use of optical transition, and are now established as the most stable timekeeping devices available \cite{ludlow2015optical}. Such optical transitions are only being adapted now due to the better availability of narrow linewidth lasers, and the development of optical combs to provide measurement methods to count the laser cycles. For quantum metrology and information purposes, magnetically trapped states are typically used which to first order can be tuned so that their energy difference is insensitive to magnetic field fluctuations.  In this case, a combination of microwave and radio frequency pulses are used to control the internal states \cite{bohi2009coherent}. Optical manipulations are desirable from the point of view of faster control, and better spatial resolution via tightly focused lasers.  For atom chips \cite{mukaichip, bohi2009coherent, hofferberth2007coherent, riedel2010atom, treutlein2006microwave} this would allow for the control of multiple BECs on the same chips \cite{treutlein2004coherence,schumm2005matter}, a task that may be desirable for spatially resolved magnetometry and quantum computing.

While in principle optical methods offer a fast and spatially resolved way to perform coherent control in BECs, this approach is problematic for two reasons. First and foremost, spontaneous emission in a BEC is enhanced by a factor of $N$, where $N$ is the number of atoms in the condensate \cite{spontg}. Secondly, for magnetically trapped BECs, one usually chooses as a storage basis a pair of hyperfine sublevels that are Zeeman insensitive to fluctuations in the trapping field. For the case of \ce{^{87}Rb}, these are the $\ket{F=1, m_F=-1 }$ and $\ket{F=2, m_F=1}$ hyperfine sublevels. These have a difference of $2$ in the $m_F$ quantum number, and passage between them implies a nuclear spin-flip of the atom. As the nuclear magnetic moment is not optically accessible, only the natural hyperfine interaction can change the nuclear spin \cite{abdelrahman2014coherent}, which makes the transfer inefficient and therefore slow. This also aggravates the aforementioned problem of spontaneous emission as the internal excited states of the BEC are populated for a longer period of time. It is for this reason that high fidelity optical control of BECs has not yet been achieved.  The nuclear spin flip problem is also common to other alkali atoms where the storage states are separated by a difference in magnetic number $ \Delta m_F \ge 2 $.  

In this paper, we show that stimulated Raman adiabatic passage (STIRAP) may be used to bypass these difficulties. STIRAP is a widely used technique for population transfer in atoms and molecules \cite{kurkal2001sequential, sugny2007laser, nesterenko2009stirap, stirapb, stiraps, stirapbt}. The most elementary STIRAP configuration is typically formulated for a three-level $\Lambda$ system, where the population of some initial state is transferred completely to a target state. However, it has been shown that using additional intermediate levels, one may implement arbitrary unitary control of a two-state quantum system \cite{qubitrotationk, qubitrotationi}. Optimization of STIRAP processes is now well-understood in the most elementary configurations \cite{kis2002optimal, vasilev2009optimum, sola1999optimal}, and proposed applications include Rydberg blockade in ensembles of interacting atoms \cite{idlas2016entanglement,qubitrotationb, petrosyan2013stimulated}, as well as universal quantum computing \cite{moller2007geometric, moller2008quantum}. The attractiveness of STIRAP comes principally from its ability to transfer populations between ground states without populating excited states. We note in particular that the method is also robust against fluctuations in pulse shape and area \cite{unanyan1998robust}. This was demonstrated by the recent experimental study of Ref. \cite{xu2015coherent}, where artificial superconducting atoms were used to store quantum information in a tristate configuration. Thus precise control may be obtained with little sensitivity to experimental noise. Eliminating the excited state probability density mitigates the particle number dependence of spontaneous emission rates that BECs suffer and thus provides a natural solution to addressability issues.

This paper is organized as follows. In \secref{II} we define spinor BECs and discuss their properties. We also introduce STIRAP as well as the spin-flip problem that affects $^{87}$Rb BECs.
\secref{III} Introduces a four level STIRAP scheme that effectively implements unitary rotation of a two-component BEC by using an intermediate ground state for information storage.
In \secref{IV} we introduce a scheme for arbitrary unitary rotations that is designed to work for $^{87}$Rb BECs and involves six internal atomic states.
\secref{V} Discusses the numerical simulation of this scheme, and results are presented in \secref{VI}. We write our conclusions and outlook in \secref{VII}. Also included in the paper are two appendices that describe our simulation methods in greater detail.

\section{Spinor Bose-Einstein condensates}
\label{sec:II}

In this section we review various aspects of spinor BEC control by laser light which will serve to introduce our notation and several basic concepts. We first introduce the formalism used to describe BEC spinor quantum states. This is followed by a description of the standard STIRAP procedure as applied to BECs. Finally we explain in detail what causes the nuclear spin flip problem.  We shall see that it also affects STIRAP by limiting the number of optically accessible dark states that diagonalize a STIRAP Hamiltonian.

\subsection{Spin coherent states}
\label{sec:IIa}

Here we describe the formalism of two-component spin coherent state BECs. By this we mean a degenerate BEC that can be described completely by the configuration between two internal states of the atoms. The spatial wavefunction is assumed to be common to all atoms, and static throughout the dynamics.  In alkali metals two hyperfine ground state levels are typically used as a basis for storing quantum states. For example, \ce{^{87}Rb} BEC would typically use two of the magnetic quantum numbers $m_F $ of the $F = 1, 2$ internal state manifolds. The operators $\hat{a}^{\dagger}$, $\hat{b}^{\dagger}$ are the bosonic creation operators of atoms in such internal states. Given a coherent condensate composed of $N$ atoms, the quantum states occupy an $N+1$ dimensional Hilbert space spanned by orthonormal Fock states of the form
\begin{equation}
\ket{k} = \frac{1}{\sqrt{k!(N-k)!}}(\hat{a}^\dagger)^k (\hat{b}^\dagger)^{(N-k)} \ket{0}.
\end{equation}
We define the spin coherent states of a two-component BEC as
\begin{equation}
\dket{\alpha,\beta} \equiv \frac{1}{\sqrt{N!}}\bigl(\alpha \hat{a}^\dagger + \beta \hat{b}^\dagger\bigr)^{N}\ket{0},
\label{eq:becstates}
\end{equation}
where $ \alpha, \beta $ are normalized parameters $ |\alpha|^2 + |\beta|^2 = 1 $. 

Limiting our analysis to states of the form of \eref{becstates}, the expectation values of the Schwinger boson operators completely characterize a BEC. These operators are defined as
\begin{equation}
\begin{split}
\hat{S}_x &= \hat{a}^\dagger\hat{b} + \hat{b}^\dagger\hat{a}\\
\hat{S}_y &= -i\hat{a}^\dagger\hat{b} + i\hat{b}^\dagger\hat{a}\\
\hat{S}_z &= \hat{a}^\dagger\hat{a} - \hat{b}^\dagger\hat{b}.
\end{split}
\label{eq:SchwingerOps}
\end{equation}
By evaluating the expectation values of \eref{SchwingerOps}, one obtains
\begin{equation}
\begin{split}
\braket{\hat{S}_x} &= N (\alpha^*\beta + \alpha\beta^*) \\
\braket{\hat{S}_y} &= N (-i\alpha^*\beta + i\alpha\beta^*) \\
\braket{\hat{S}_z} &= N (|\alpha|^2 - |\beta|^2).
\end{split}
\end{equation}
The states of \eref{becstates} thus define a Bloch sphere of radius $N$ in $\bigl(\braket{\hat{S}_x},\braket{\hat{S}_y},\braket{\hat{S}_z}\bigr)$-space. We may equivalently characterize a coherent spinor in terms of a pair of angles $\phi$ and $\theta$, such that $\alpha = \cos \theta/2$  and $\beta = e^{i\phi}\sin \theta/2$. Then $\theta$ and $\phi$ are the azimuthal and polar angles respectively on the Bloch sphere.

We define a rotation about the $\braket{\hat{S}_j}$-axis of the Bloch sphere by the angle $\delta$ as
\begin{equation}
\hat{\mathcal{R}}_j(\delta) \equiv e^{-i\delta \hat{S}_j/2}.
\end{equation}
For instance,
\begin{equation}
\hat{\mathcal{R}}_z(\delta)\dket{\alpha,\beta} = \dKet{\alpha e^{-i\delta/2},\beta e^{i\delta/2}}
\end{equation}
up to an irrelevant global phase factor. 
Similarly we may also define the arbitrary rotation about a unit vector on the Bloch sphere $\vect{n}$ by an angle $\delta$ as
\begin{equation}
\hat{\mathcal{R}}_\vect{n}(\delta) \equiv e^{-i\delta \vect{n}\cdot\vect{S}/2}, \quad \vect{n} = [\cos\phi\sin\theta, \sin\phi\sin\theta, \cos\theta]^T.
\label{eq:nrotation}
\end{equation}
Here $\vect{S} = [\hat{S}_x,\hat{S}_y,\hat{S}_z]^T$. This is completely analogous to Bloch sphere rotations for qubits, where one uses Pauli spin operators in place of Schwinger boson operators. Given free choice of the angles $\phi$, $\theta$, and $\delta$, any Bloch sphere rotation is possible. Alternatively, one may write an arbitrary rotation by a sequence of rotations around the $ y $ and $ z $ axes:
\begin{equation}
\hat{U} = e^{i\alpha} \hat{\mathcal{R}}_z(\beta)\hat{\mathcal{R}}_y(\gamma)\hat{\mathcal{R}}_z(\delta) .
\label{eq:unitary}
\end{equation}
As the algebra for rotations of spin coherent states is completely equivalent to that for qubits we may also use their matrix representations. Using the vector $[\alpha,\beta]^T$ to refer to \eref{becstates}, the above rotation can be equivalently written
\begin{equation}
U = e^{i\alpha}
\begin{bmatrix}
e^{i(-\beta/2 - \delta/2)}\cos{\frac{\gamma}{2}} &
-e^{i(-\beta/2 + \delta/2)}\sin{\frac{\gamma}{2}} \\
e^{i(+\beta/2 - \delta/2)}\sin{\frac{\gamma}{2}} &
e^{i(+\beta/2 + \delta/2)}\cos{\frac{\gamma}{2}}
\end{bmatrix}.
\label{eq:arbunitary}
\end{equation}

\subsection{Three-level STIRAP for BECs}
\label{sec:IIb}

Here we describe the standard STIRAP applied to BECs in spin coherent states. Consider a BEC prepared in state $\dket{1,0}$ and let the target state be $\dket{0,1}$. The goal of a standard STIRAP is to transfer completely the quantum system from its initial state to its target state via an adiabatic procedure which only populates the ground states of the atoms. Given a hyperfine manifold with some ground states $\ket{a}$ and $\ket{b}$ as well as an excited state $\ket{e}$ (see Fig. \ref{fig:threelevelham}(a)) we apply laser transitions linking levels $\ket{a} \leftrightarrow \ket{e}$ and $\ket{b} \leftrightarrow \ket{e}$. In the rotating wave approximation (RWA), this is described by the Hamiltonian
\begin{equation}
	\hat{H}/\hbar = \bigl[\Omega_a(t)\hat{e}^\dagger \hat{a} + \Omega_b(t)\hat{e}^\dagger \hat{b} + \hc\bigr] + \Delta\hat{e}^\dagger\hat{e}.
\label{eq:threelevelham}
\end{equation}
The $\hat{e}^\dagger$ are bosonic operators which create the excited state of a three-level $\Lambda$ system, with $\hat{a}^\dagger$ and $\hat{b}^\dagger$ creating the two ground states respectively. One approach to achieve the state transfer is simply to apply (\ref{eq:threelevelham}) with constant laser amplitudes for a time corresponding to a period of the Raman oscillation with frequency $\Omega_a\Omega_b/2\Delta$.  This however creates a fractional population of the excited state of magnitude $\sim \Omega_a\Omega_b/\Delta^2$. As the excited state is susceptible to spontaneous emission, large values of $\Delta$ compared to $\Omega_{a,b}$ are used to suppress it. This results in reducing the Raman oscillation frequency. 

STIRAP differs fundamentally from this approach by using dark states of the Hamiltonian which have exactly zero population of the excited states. The dark states of an atomic Hamiltonian are often defined as the eigenstates that do not emit any light. For the systems we consider, this definition reduces to eigenstates, that have no contributions from excited states.
\begin{align}
	\ket{D_0} = \frac{1}{\sqrt{N!}}({\hat{d}_0^\dagger})^{N}\ket{0},
	\label{d0state}
\end{align}
where we have defined
\begin{align}
\hat{d}_0^\dagger \equiv \frac{\Omega_b \hat{a}^\dagger - \Omega_a \hat{b}^\dagger}{\sqrt{|\Omega_a|^2 + |\Omega_b|^2}},
\end{align}
which creates one particle in a dark state. The state (\ref{d0state}) is an eigenstate of (\ref{eq:threelevelham}) with eigenvalue zero. We now define the pulses
\begin{equation}
\begin{split}
\Omega_a &= \Omega_0f(t - T_-/2)\\
\Omega_b &= \Omega_0f(t + T_-/2),
\end{split}
\end{equation}
where $f(t)$ is a unimodal, positive function, centered on $t = 0$, that vanishes when $t$ approaches either $-\infty$ or $\infty$. It is typically chosen to be Gaussian. $T_-$ is a constant that has been added to separate the pulses in time by a length suitable for STIRAP.

To see how STIRAP transfers population between the two ground states $\dket{1,0}$ and $\dket{0,1}$, consider the following. In the limit where $t \rightarrow -\infty$, i.e. before any laser pulses strike the BEC, we have $\Omega_a/\Omega_b \rightarrow 0$. Therefore $\hat{d}^\dagger_0 \cong \hat{a}^\dagger$. When $ \Omega_a = \Omega_b $, $\hat{d}^\dagger$ becomes an even mix of $\hat{a}^\dagger$ and $\hat{b}^\dagger$. Finally, after this point as $t\rightarrow\infty$ and $\Omega_b/\Omega_a \rightarrow 0$ we have $\hat{d}^\dagger_0 \cong \hat{b}^\dagger$. Usually $f(t)$ is chosen so that the process goes from start to finish in a few $T_-$.
The above procedure transfers the quantum state from one known state entirely into another, as long as the adiabatic theorem is satisfied. To achieve this the function $f(t) $ should be chosen such that the time variation of the dark state is slow enough so that the state follows the eigenstate (\ref{d0state}) at all times.

We note that the above STIRAP procedure does not produce an arbitrary rotation such as that described in the remainder of this paper.  In this procedure the aim is to transfer the state from a known initial state to a fixed final state.  In a rotation such as \eref{unitary}, the initial or final state does not need to be known.  In this sense, the standard three-level STIRAP procedure does not achieve a SU(2) rotation. If the initial state is not in the specified initial state, the procedure fails and the state does not in general follow an adiabatic trajectory. 
However, as will be discussed further down, it is possible to modify the STIRAP procedure to perform an arbitrary rotation during the STIRAP procedure.  This was first developed in Refs. \cite{qubitrotationk, qubitrotationi}, and we will adapt this procedure to BECs in the following sections.

\subsection{The nuclear spin problem}
\label{sec:IIc}
Transitions between atomic levels due to electromagnetic fields are in general subject to  selection rules. For transitions between hyperfine states induced by laser fields, the only possible direct transitions in the $m_F$ quantum numbers are $|\Delta m_F| \leq 1$, as the photon carries an angular momentum of 1. For magnetically trapped BECs, only states with the correct sign with respect to the Zeeman shift may be used  for the storage states, as the remaining states are either anti-trapped or not trapped at all by the magnetic field. Of the trapped states, preference is given to pairs of hyperfine states whose differential Zeeman shift can be minimized. For suitable magnetic field strengths, the magnetic responses of the two storage levels can be made the same. This acts to increase the coherence time in the storage states, as to first order, fluctuations in the magnetic field do not incur additional relative phase shifts between such levels. For example, in the case of \ce{^{87}Rb}, the ground states $\ket{F = 1, m_F = -1}$ and $\ket{F = 2, m_F = 1}$ are conventionally chosen to store coherent superpositions. In this case, one may utilize transitions from \ce{5^2S_{1/2}} to the \ce{5^2P_{1/2}} (D1) or \ce{5^2P_{3/2}} (D2) excited manifold of states. For either choice, selection rules dictate the transitions pass through two excited states $ F' = 1,2 $.  Naively, the presence of two available excited states may appear to be beneficial to the Raman or STIRAP transition between the storage states, as the total transition rate is the sum of individual transitions.  However, as described in Ref. \cite{waxman2007coherent}, the two possible transitions always interfere destructively with each other, making the process inefficient.

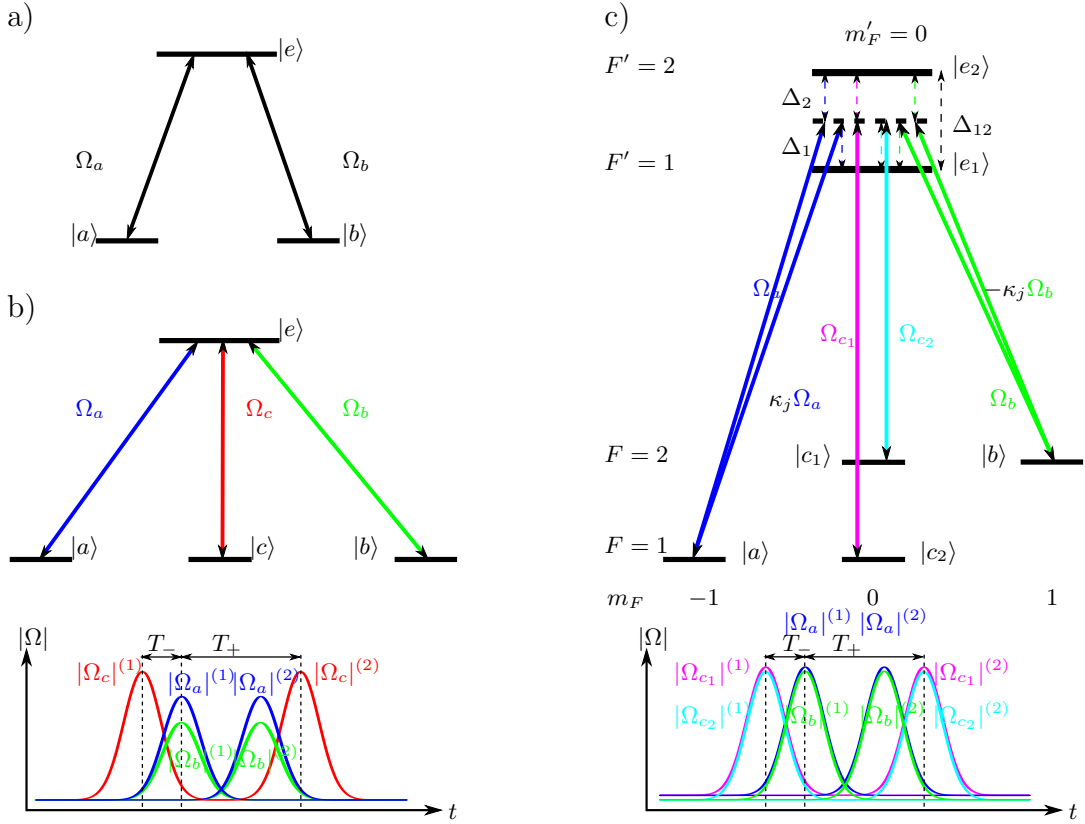
\begin{figure*}
\centering
\def\svgwidth{0.8\textwidth}
\input{LevelSchemeTotal.tex}
\caption{(Color online). a) 3-level $\Lambda$ system with two ground states and one excited state. b) STIRAP configuration for performing arbitrary SU(2) rotation. The states $\ket{a}$ and $\ket{b}$ are used to coherently store quantum information, while $\ket{c}$ is an auxiliary level. $\ket{e}$ is an excited state that is coupled to all of the ground levels. Below the level diagram the sequence of two pairs of STIRAP pulses is depicted that constitute an arbitrary rotation of a coherent spinor BEC. Relative magnitudes of the pulses and their timing with respect to each other are indicated. c) STIRAP configuration for arbitrary unitary operations for alkali atoms taking the the case of \ce{^{87}Rb}. Four laser transitions are used to couple the excited states $F'=1,2$ simultaneously. The transition amplitudes are denoted in the figure together with coefficients $\kappa_j$, which are determined by Clebsch-Gordan coefficients as found in Ref. \cite{steck2001rubidium}. Below the diagram the sequence of pulses needed for arbitrary rotation is again depicted.}
\label{fig:threelevelham}
\end{figure*}

To see how the destructive interference arises, consider \figref{threelevelham}(c) which shows the hyperfine transitions between excited and ground states, taking the example of \ce{^{87}Rb}. Similar arguments can be made for other atom species. Taking a laser transition frequency such that it is between the $ F'=1,2 $ levels, the excitations produce transitions of the form $ \ket{a} \leftrightarrow \kappa_j\ket{e_1} + \ket{e_2} $ and $ \ket{b} \leftrightarrow \ket{e_1} - \kappa_j\ket{e_2} $ (states are unnormalized).   Here $\kappa_j$ is a constant that is determined by the relative magnitudes of the dipole transition matrix elements and are entirely depend on Clebsch-Gordan coefficients. For \ce{^{87}Rb} these are $\kappa_1 = 1$ and $\kappa_2 = \sqrt{5}$ for the D1 and D2 transitions respectively. In either case the coherent superposition excited by $\ket{a}$ is orthogonal to the one that $\ket{b}$ transitions into. This limits the speed of potential Rabi oscillations between levels $\ket{a}$ and $\ket{b}$. A highly detuned system, in which the excited state populations may be adiabatically eliminated has a generalized Rabi frequency \cite{waxman2007coherent}:
\begin{align}
\Omega_{a\leftrightarrow b} &\approx \frac{1}{2}\kappa_j\Omega_a\Omega_b\Bigl(\frac{1}{\Delta_1} - \frac{1}{\Delta_2}\Bigr)\nonumber\\
	&= \kappa_j\frac{\Omega_a\Omega_b \Delta_{12}}{ 2 \Delta_1 \Delta_2},
\end{align}
where $ \Delta_{1,2} $ are the detunings of the transitions to each of the levels. 
A second order Raman transition will therefore be limited by timescales that depend on the energy difference $\Delta_{12}$ between $\ket{e_1}$ and $\ket{e_2}$. For \ce{^{87}Rb} these are given by $ \SI{816.656(30)}{\mega\hertz}$ for the D1 line and $\SI{156.947(7)}{\mega\hertz}$ for the D2 line\cite{steck2001rubidium}. But more importantly, $\Omega_{a\rightarrow b}$ vanishes proportionally to $\Delta_1\Delta_2$, i.e. the product of the two detunings. As detunings are typically much larger than $\Omega_{a,b}$ in order to enable adiabatic elimination of the excited states, the coupling between the ground states is reduced making the transition inefficient. 

We refer to this as the ``nuclear spin problem'' (also known as the ``$\Delta m_F = 2$ problem'' in Refs. \cite{waxman2007coherent,abdelrahman2014coherent}). The problem originates from the fact that the coherent superpositions excited by an application of $\Omega_{a,b}$ are mutually orthogonal. To see how this also limits a STIRAP based scheme, consider the system of \figref{threelevelham}(c). If only the laser fields $\Omega_a$ and $\Omega_b$ are turned on, the resulting Hamiltonian in the RWA is 
\begin{multline}
\hat{H}_j/\hbar = \bigl[\Omega_a(t)\bigl(\kappa_j\hat{e}_1^\dagger + \hat{e}_2^\dagger \bigr)\hat{a} + \Omega_b(t)\bigl(\hat{e}_1^\dagger - \kappa_j\hat{e}_2^\dagger \bigr)\hat{b} + \hc\bigr]\\
+ \Delta_1 \hat{e}_1^\dagger\hat{e}_1 + \Delta_2 \hat{e}_2^\dagger\hat{e}_2\bigr).
\end{multline}
None of the eigenstates of this Hamiltonian are dark states. If the two excited state superpositions created by applying $_hat{H}_j$ to $\ket{a}$ and $\ket{b}$ respectively were linearly dependent, we could choose a proper sign for $\Omega_a$ and $\Omega_b$ to make them cancel out. Thus a dark state would be obtainable by superposing $\ket{a}$ and $\ket{b}$. But because the excited states are orthogonal, this is impossible. The Hamiltonian thus cannot be used for a STIRAP. The destructive interference described earlier is seen to manifest itself here by eliminating a possible dark state, that would otherwise be useful for STIRAP.

In the following we show how this may be solved for STIRAP by introducing two additional ground states $\ket{c_1}$ and $\ket{c_2}$ as seen in \figref{threelevelham}(c).

\section{Arbitrary rotations of BECs using STIRAP}
\label{sec:III}

In this section we show how arbitrary SU(2) rotations of two-component spinor BECs may be performed, by reformulating the four-level scheme proposed in Ref. \cite{qubitrotationk}. For the purposes of this section, we do not take into account the nuclear spin problem that was introduced earlier. A full solution of this is presented in a six-level STIRAP scheme further below.

\subsection{Four-level STIRAP scheme}

Consider the operators $\hat{a}^\dagger$, $\hat{b}^\dagger$, $\hat{c}^\dagger$ and $\hat{e}^\dagger$ creating the states found in the four-level system of \figref{threelevelham}(b). The Hamiltonian is of the form
\begin{equation}
	\hat{H}/\hbar = \Omega_{a} (t)\hat{e}^\dagger \hat{a} + \Omega_{b} (t)\hat{e}^\dagger \hat{b} + \Omega_{c} (t)\hat{e}^\dagger \hat{c} + \hc 
\label{eq:fourlevelham}
\end{equation}
The dark states of the Hamiltonian are.
\begin{equation}
\begin{split}
\hat{d}^\dagger_\textup{C} &= \frac{\Omega_c\bigl(\Omega_a^* \hat{a}^\dagger + \Omega_b^* \hat{b}^\dagger \bigr) - \bigl(|\Omega_a|^2 + |\Omega_b|^2\bigr)\hat{c}^\dagger}{\sqrt{\bigl(|\Omega_a|^2 + |\Omega_b|^2\bigr)|\Omega_c|^2 + |\Omega_a^2 + \Omega_b^2|^2}}\\
\hat{d}^\dagger_\textup{NC} &= \frac{\Omega_b \hat{a}^\dagger - \Omega_a \hat{b}^\dagger}{\sqrt{|\Omega_a|^2 + |\Omega_b|^2}} . 
\end{split}
\label{eq:kislimits}
\end{equation}
These create the \emph{coupled} (C) and \emph{non-coupled} (NC) dark states respectively. In this scheme two sets of STIRAP pulses will be applied to the BEC, as shown in \figref{threelevelham}(b). We have defined the coupled dark state as the one that undergoes adiabatic evolution during this process, whilst the non-coupled dark state doesn't change. The first set of laser pulses, which define the limits of \eref{kislimits}, are
\begin{equation}
\begin{split}
\Omega_{a}^{(1)} &\equiv \Omega_0 \cos{\chi}f(t - T_-/2 + T_+/2)\\
\Omega_{b}^{(1)} &\equiv \Omega_0 \sin{\chi}e^{i\phi}f(t - T_-/2 + T_+/2)\\
\Omega_{c}^{(1)} &\equiv \Omega_0 f(t + T_-/2 + T_+/2),
\end{split}
\end{equation}
where we have added a superscript (1) to indicate that this is the first STIRAP pair. We separate the $a$ and $b$ pulses from the $c$ pulse by a length of time $T_-$. We also introduce $T_+$ as a delay between this pair of pulses and the one that we apply later in the procedure (see \figref{threelevelham}).
The coupled state has a time evolution with limiting behavior
\begin{equation}
\begin{split}
\hat{d}_\textup{C}^{(1)}{}^\dagger 	&\rightarrow
	\begin{cases} \cos\chi \hat{a}^\dagger + \sin\chi e^{-i\phi}\hat{b}^\dagger & \textup{for } t\rightarrow -\infty\\
	-\hat{c}^\dagger & \textup{for } t\rightarrow \infty\end{cases}
\end{split} ,
\label{eq:dclimits}
\end{equation}
where again we have added a superscript (1) to indicate that this is the first STIRAP pair. 
The coupled state is so called because it evolves into $\hat{c}^\dagger$. The non-coupled state is 
\begin{equation}
\hat{d}^\dagger_\textup{NC} = \sin\chi e^{i\phi} a^\dagger - \cos\chi b^\dagger,
\label{eq:ncdef}
\end{equation}
which is time-independent during the STIRAP process. 

A coherent spinor is initially in the state given by \eref{becstates}. As the two eigenstate creation operators obey $[\hat{d}_\textup{C},\hat{d}_\textup{NC}^\dagger] = 0$, they create states that form an orthonormal set. Because of that, we may write the wave function in its initial state as
\begin{align}
&  \dket{\alpha_0,\beta_0} = \nonumber \\
& \frac{1}{\sqrt{N!}}\bigl( \braket{d_\textup{C}^{(1)} (t \rightarrow -\infty)| \psi_0}
\hat{d}_\textup{C}^{(1)} {}^\dagger 
+ \braket{d_\textup{NC}| \psi_0}\hat{d}_\textup{NC}^\dagger\bigr)^{N}\ket{0},
\end{align}
where for convenience of notation, we also define the single particle quantum states
\begin{equation}
\begin{split}
\ket{\psi_0} &\equiv \alpha_0\ket{a} + \beta_0\ket{b}\\
\ket{d_{\textup{C}} (t)} &\equiv \hat{d}_{\textup{C}}^\dagger\ket{0}\\
\ket{d_\textup{NC}} &\equiv \hat{d}_\textup{NC}^\dagger\ket{0}.
\end{split}
\end{equation}
Here $\ket{\psi_0}$ is the initial state of a single atom in the BEC, and the others describe the coupled and non-coupled states respectively at any time.  
The above inner products may be evaluated to be
\begin{align}
\braket{d_\textup{C}^{(1)} (t \rightarrow -\infty)| \psi_0} &= \alpha \cos{\chi} + \beta e^{i\phi} \sin{\chi}\\
\braket{d_\textup{NC}| \psi_0} &= \alpha e^{-i\phi}\sin{\chi} - \beta \cos{\chi}.
\end{align}
After the initial set of pulses, i.e. for $t \rightarrow\infty$, the resulting wavefunction is
\begin{multline}
\dket{\alpha_1,\beta_1} = \frac{1}{\sqrt{N!}}\bigl( -\braket{d_\textup{C}^{(1)} (t \rightarrow -\infty)| \psi_0} \hat{c}^\dagger \\
+ \braket{d_\textup{NC}| \psi_0}\hat{d}_\textup{NC}^\dagger\bigr)^{N}\ket{0}.
\label{eq:coupling}
\end{multline}

After letting a suitable time pass, we may now perform a STIRAP process once more to move the coherent spinor back into a $\hat{a}$ and $\hat{b}$ superposition state. We do this using the pulses
\begin{equation}
\begin{split}
\Omega_{a}^{(2)} &\equiv \Omega_0 \cos{\chi}f(t + T_-/2 - T_+/2)\\
\Omega_{b}^{(2)} &\equiv \Omega_0 \sin{\chi}e^{i\phi}f(t + T_-/2 - T_+/2)\\
\Omega_{c}^{(2)} &\equiv \Omega_0 e^{i\delta}f(t - T_-/2 - T_+/2).
\end{split}
\end{equation}
For this STIRAP, $\Omega_{c}$ has been phase-shifted compared to the other pulses, but these are otherwise the same as the first set of STIRAP pulses with the order reversed.  This means that the limiting cases (\ref{eq:dclimits}) are also reversed, 
\begin{equation}
\begin{split}
\hat{d}_\textup{C}^{(2)} {}^\dagger 	&\rightarrow
	\begin{cases} 
	-\hat{c}^\dagger & \textup{for } t\rightarrow -\infty \\
		(\cos{\chi} \hat{a}^\dagger + \sin{\chi}e^{-i\phi} \hat{b}^\dagger)e^{i\delta} & \textup{for } t\rightarrow \infty
	\end{cases}
\end{split} ,
\end{equation}
while $ \hat{d}^\dagger_\textup{NC} $ is the same as (\ref{eq:ncdef}). Letting the condensate evolve adiabatically once again, the state is during the evolution
\begin{multline}
\dket{\alpha_2,\beta_2} = \frac{1}{\sqrt{N!}}\bigl( \braket{d_\textup{C}^{(1)}(t = -\infty)|\psi_0}\hat{d}_\textup{C}^{(2)} {} ^\dagger \\
+ \braket{d_\textup{NC}| \psi_0}\hat{d}_\textup{NC}^\dagger\bigr)^{N}\ket{0}.
\end{multline}
The resulting wavefunction is now a superposition of the interfering dark states, where the interference has been modulated by the phase $\delta$. We may calculate the final coefficients
\begin{align}
\alpha_2 =& \alpha_0\bigl( \cos^2 \chi e^{i\delta} + \sin^2\chi \bigr) \nonumber\\
	&{}+ \beta_0 e^{-i\phi}\bigl( \cos\chi\sin\chi e^{i\delta} - \cos\chi\sin\chi \bigr) \\
\beta_2 =& \alpha_0 e^{i\phi}\bigl( \cos\chi\sin\chi e^{i\delta} - \cos\chi\sin\chi \bigr) \nonumber\\
	&{}+ \beta_0\bigl( \cos^2 \chi + \sin^2\chi e^{i\delta}\bigr).
\end{align}
The entire process may be written as a linear transform between vectors $[\alpha_0, \beta_0]^T$ and $[\alpha_2, \beta_2]^T$. If one does so, the unitary matrix that connects the two has an immediate geometric interpretation as a rotation on the Bloch sphere. We may write 
\begin{equation}
\dket{\alpha_2, \beta_2} = e^{iN\delta/2}\mathcal{R}_\vect{n}(\delta)\dket{\alpha_0, \beta_0}.
\end{equation}
Here $\mathcal{R}_\vect{n}(\delta)$ is an arbitrary rotation about the unit vector $\vect{n} = [\sin{2\chi}\cos{\phi}, \sin{2\chi}\sin{\phi}, \cos{2\chi}]^T$ on the Bloch sphere. Apart from the added global phase, this is the same as \eref{nrotation}. As we have noted earlier, this is equivalent to \eref{arbunitary}.

\section{Arbitrary rotations of BECs with the nuclear spin problem}
\label{sec:IV}

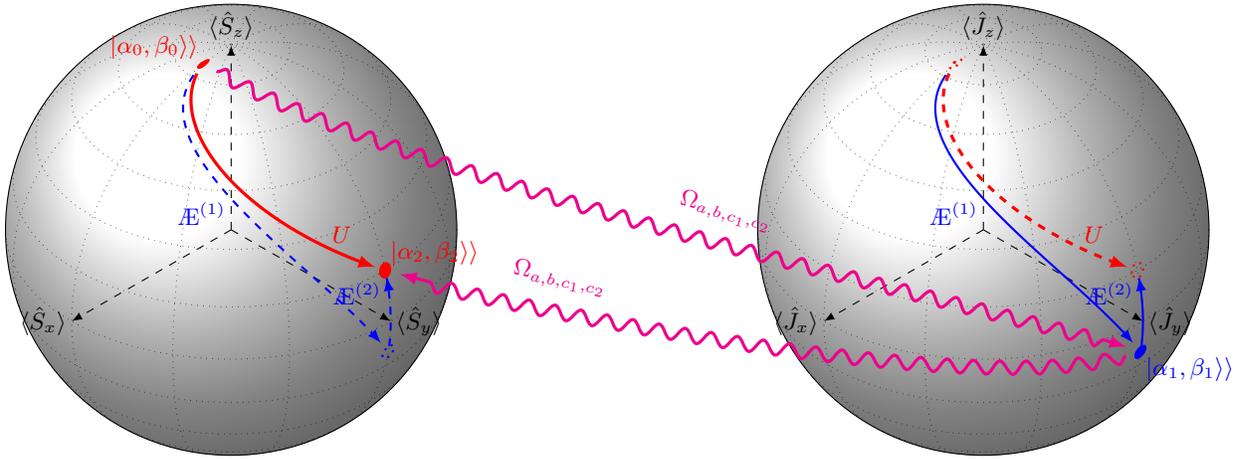
\begin{figure*}
\input{paperPRAbspheres}
\caption{(Color online). Visualization of the six level STIRAP scheme for arbitrary rotations of a spinor BEC.  The Bloch sphere on the left is defined for the coherent spinor states created by operators $\hat{a}^\dagger, \hat{b}^\dagger$ and  $\hat{c}_1^\dagger, \hat{c}_2^\dagger$ on the right. The first STIRAP pulse  transfers the state from the left Bloch sphere to the right, and vice versa for the second STIRAP pulse.   
After the first STIRAP pulse, the position of the coherent spinor on the right Bloch sphere may be described by the rotation $\mO^{(1)}$ applied to the coordinates it had on the former Bloch sphere. After the second STIRAP pulse sequence, another rotation $\mO^{(2)}$ is applied, returning the BEC to the original Hilbert space. The net effect of these two operations gives the total rotation $U = \mO^{(2)} \mO^{(1)}$.}
\label{fig:bspheres}
\end{figure*}

The scheme considered in the previous section is designed for a generic four level Hamiltonians, where the optical transition selection rules of internal atomic state manifolds are not taken into account. If it was possible to implement this scheme directly, then an efficient all-optical control of atomic BECs with negligible spontaneous emission should be possible. In practice we are limited by the nuclear spin problem, which eliminates dark states of the form presented in the last section. Therefore a new scheme must be developed to make coherent control possible, which we introduce here. Similarly to the procedure using four levels, we make use of intermediate quantum states. Here we show how this may be used to make arbitrary rotations of spin coherent states of the form \eref{arbunitary}.

\subsection{Six-level STIRAP scheme}

The transitions that we consider for this scheme involve four ground states and two excited states, as shown in \figref{threelevelham}(c). The Hamiltonian for this in the RWA is
\begin{align}
\hat{H}/\hbar =& \bigl[\Omega_a(t)\bigl(\kappa_j\hat{e}_1^\dagger + \hat{e}_2^\dagger \bigr)\hat{a} + \Omega_b(t)\bigl(\hat{e}_1^\dagger - \kappa_j\hat{e}_2^\dagger \bigr)\hat{b} \nonumber\\
& \phantom{\bigl[}	 +\Omega_{c_1}e_1^\dagger\hat{c}_1+\Omega_{c_2}e_2^\dagger\hat{c}_2 + \hc\bigr]  \nonumber\\
&{} +\Delta_1 \hat{e}_1^\dagger\hat{e}_1 + \Delta_2\hat{e}_2^\dagger\hat{e}_2.
\end{align}
Here the $\kappa_j$ terms originate from the relative dipole matrix elements of the D1 and D2 line transitions and are determined by Clebsch Gordan coefficients as indicated by \figref{threelevelham}(c). Note that the $\ket{F=j,m_F=0} \rightarrow \ket{F'=j,m_F'=0}$ transitions do not conserve total angular momentum, and are therefore not allowed.

We have furthermore used a RWA where terms that rotate as $e^{i(\omega_{c_1}-\omega_{c_2} )t}$ are dropped. This is because the characteristic frequency difference between the ground states is $\omega_{c_1} - \omega_{c_2} = \SI{6.83}{\giga\hertz}$. The period of this oscillation is about 200 times shorter than the shortest timescales we use in our simulations.

In this Hamiltonian we do not take into account AC-stark shifts whose net effect is to shift the energy of all the sublevels. As the second order Stark shift is only non-vanishing for $J=3/2$ excited states, all of the ground state sublevels are effected in the same way. The net effect of the Stark shift is therefore to shift the diagonal elements of the Hamiltonian by the same amount. The only exception to this is the case where the D2 line is used and some second order effect must be taken into account. However, this is only applicable to the excited state manifold, and thus it does not directly effect the dark states of our scheme. Neither does it induce non-adiabatic transitions as the terms affected are diagonal.

The principal difference between our scheme and the one we presented in \secref{III} is our usage of two coupled dark states compared to the one coupled and one non-coupled dark states used by their arbitrary unitary gate. They are
{\allowdisplaybreaks\begin{equation}
\begin{split}
\hat{d}_1^{(1)} {}^\dagger &= \frac{\Omega_{c_1}\Omega_{c_2}\hat{a}^\dagger - \kappa_j\Omega_{a}\Omega_{c_2}\hat{c}_1^\dagger - \Omega_{a}\Omega_{c_1}\hat{c}_2^\dagger}{\sqrt{|\Omega_{c_1}|^2|\Omega_{c_2}|^2 +|\Omega_{a}|^2\bigl(|\Omega_{c_1}|^2 +\kappa_j^2 |\Omega_{c_2}|^2\bigr)}}\\
\hat{d}_2^{(1)} {}^\dagger &= \frac{\Omega_{c_1}\Omega_{c_2}\hat{b}^\dagger - \Omega_{b}\Omega_{c_2}\hat{c}_1^\dagger + \kappa_j\Omega_{b}\Omega_{c_1}\hat{c}_2^\dagger}{\sqrt{\kappa_j^2|\Omega_{c_1}|^2|\Omega_{c_2}|^2 +|\Omega_{b}|^2\bigl(|\Omega_{c_1}|^2 + |\Omega_{c_2}|^2\bigr)}} .
\end{split}
\label{eq:eigen1}
\end{equation}}
As before there are two pairs of STIRAP pulses required to perform the rotation. The separation of the Stokes and pump pulses is $T_-$, and the two pairs are separated by a length of time $T_+$ (see \figref{threelevelham}). The set of laser pulses are assumed to take a form
\begin{equation}
\begin{split}
\Omega_a^{(1)} (t) &= \Omega_0 e^{i\theta^{(1)}_a}f(t - T_-/2 + T_+/2)\\
\Omega_b^{(1)}(t) &= \Omega_0 e^{i\theta^{(1)}_b}f(t - T_-/2 + T_+/2)\\
\Omega_{c_1}^{(1)}(t) &= \Omega_0 e^{i\theta^{(1)}_{c_1}}f(t + T_-/2 + T_+/2)\\
\Omega_{c_2}^{(1)}(t) &= \Omega_0 e^{i\theta^{(1)}_{c_1}}f(t + T_-/2 + T_+/2).
\end{split}
\end{equation}
The pulses have a fixed positive amplitude $\Omega_0$ and a unimodal temporal distribution given by $f(t)$. We also associate each pulse with a phase-angle $\theta^{(1)}$. $T_-$ is the separation between the maxima of the Stokes and pump pulses. We define the $t\rightarrow-\infty$ limit as the time where the amplitudes of $\Omega_{c_{1,2}}$ are significantly greater than $\Omega_{a,b}$. The limiting behavior of the dark states for the first STIRAP pair is then
{\allowdisplaybreaks\begin{equation}
\begin{split}
\hat{d}_1^{(1)} {}^\dagger 	&\rightarrow
	\begin{cases} 
	e^{i(\theta^{(1)}_{c_1} + \theta^{(1)}_{c_2})}\hat{a}^\dagger & \textup{for } t\rightarrow -\infty\\
	\frac{-\kappa_je^{i(\theta^{(1)}_a + \theta^{(1)}_{c_2})}\hat{c}^\dagger_1 -e^{i(\theta^{(1)}_a + \theta^{(1)}_{c_1})}\hat{c}^\dagger_2}{\sqrt{1 + \kappa_j^2}} & \textup{for } t\rightarrow \infty
	\end{cases}\\
\hat{d}_2^{(1)} {}^\dagger &\rightarrow
	\begin{cases} e^{i(\theta^{(1)}_{c_1} + \theta^{(1)}_{c_2})}\hat{b}^\dagger & \textup{for } t\rightarrow -\infty\\
	\frac{-e^{i(\theta^{(1)}_b + \theta^{(1)}_{c_2})}\hat{c}^\dagger_1 +\kappa_j e^{i(\theta^{(1)}_b + \theta^{(1)}_{c_1})}\hat{c}^\dagger_2}{\sqrt{1 + \kappa_j^2}} & \textup{for } t\rightarrow \infty.
	\end{cases}
\end{split}
\end{equation}}

A coherent spinor that starts out in the state of \eref{becstates} in the $t\rightarrow-\infty$ limit is written as
\begin{align}
\dket{\alpha_0, \beta_0} =&{}\frac{1}{\sqrt{N!}}\bigl( \alpha_0 e^{-i(\theta^{(1)}_{c_1} + \theta^{(1)}_{c_2})}\hat{d}_1^{(1)} {}^\dagger\nonumber \\
	&{}\phantom{\frac{1}{\sqrt{N!}}\bigl(}+ \beta_0 e^{-i(\theta^{(1)}_{c_1} + \theta^{(1)}_{c_2})} \hat{d}_2^{(1)} {}^\dagger \bigr)^{N}\ket{0}.
\end{align}
In the $t\rightarrow\infty$ limit this then becomes
\begin{align}
\dket{\alpha_1, \beta_1}
 =	&\frac{1}{\sqrt{N!(1 + \kappa_j^2)^{N}}}\nonumber  \\
& \times\Bigl(\bigl[ -\kappa_j\alpha_0e^{i(\theta^{(1)}_a-\theta^{(1)}_{c_1})}-\beta_0e^{i(\theta^{(1)}_b-\theta^{(1)}_{c_1})}\bigr]\hat{c}_1^\dagger  \nonumber  \\
& +\bigl[ -\alpha_0e^{i(\theta^{(1)}_a-\theta^{(1)}_{c_2})}+\kappa_j\beta_0e^{i(\theta^{(1)}_b-\theta^{(1)}_{c_2})}\bigr]\hat{c}_2^\dagger\Bigr)^{N}\ket{0} \nonumber \\
\equiv &	\frac{1}{\sqrt{N!}}(\alpha_1\hat{c}_1^\dagger + \beta_1\hat{c}_2^\dagger)^{N}\ket{0}.
\end{align}
We note that at this intermediate stage the spin coherent state is with respect to the levels $ c_1 $ and $ c_2 $, and not the storage states $ a $ and $ b $.  

A suitable time later, the second STIRAP pair is applied, the second set of lasers is assumed to take a form
\begin{equation}
\begin{split}
\Omega_a^{(2)} &= \Omega_0 e^{i\theta^{(2)}_a}f(t + T_-/2 - T_+/2)\\
\Omega_b^{(2)} &= \Omega_0 e^{i\theta^{(2)}_b}f(t + T_-/2 - T_+/2)\\
\Omega_{c_1}^{(2)} &= \Omega_0 e^{i\theta^{(2)}_{c_1}}f(t - T_-/2 - T_+/2)\\
\Omega_{c_2}^{(2)} &= \Omega_0 e^{i\theta^{(2)}_{c_2}}f(t -T_-/2 - T_+/2).
\end{split}
\end{equation}
We now define the dark states using the operators
\begin{equation}
\begin{split}
\hat{d}^{(2)}_1 {}^\dagger &= \frac{-\kappa_j\Omega_{c_1}\Omega_{b}\hat{a}^\dagger - \Omega_{c_1}\Omega_{a}\hat{b}^\dagger + (1 + \kappa_j^2)\Omega_{a}\Omega_{b}\hat{c}_1^\dagger}{\sqrt{(1 + \kappa_j^2)^2|\Omega_{a}|^2|\Omega_{b}|^2 +|\Omega_{c_1}|^2\bigl(|\Omega_{a}|^2 + \kappa_j^2|\Omega_{b}|^2\bigr)}}\\
\hat{d}^{(2)}_2 {}^\dagger &= \frac{-\Omega_{c_2}\Omega_{b}\hat{a}^\dagger + \kappa_j\Omega_{c_2}\Omega_{a}\hat{b}^\dagger + (1 + \kappa_j^2)\Omega_{a}\Omega_{b}\hat{c}_2^\dagger}{\sqrt{(1 + \kappa_j^2)^2|\Omega_{a}|^2|\Omega_{b}|^2 +|\Omega_{c_2}|^2\bigl(\kappa_j^2|\Omega_{a}|^2 + |\Omega_{b}|^2\bigr)}}\\
\end{split}
\label{eq:eigen2}
\end{equation}
We note that \eref{eigen2} is not the same form as that defined in \eref{eigen1}.  As linear combinations of dark states are also dark states, there is a freedom in the way the operators are defined.  This choice makes the analysis clearer hence we use this form for the second STIRAP pair. 
The limiting behavior is given by 
\begin{equation}
\begin{split}
\hat{d}^{(2)}_1 {}^\dagger &\rightarrow \begin{cases}
	e^{i(\theta^{(2)}_a + \theta^{(2)}_b)}\hat{c}_1^\dagger & \text{for } t\rightarrow -\infty \\
	\frac{-\kappa_j e^{i(\theta^{(2)}_{c_1} + \theta^{(2)}_{b})}\hat{a}^\dagger -e^{i(\theta^{(2)}_{c_1} + \theta^{(2)}_a)}\hat{b}^\dagger}{\sqrt{1 + \kappa_j^2}} & \text{for } t\rightarrow\infty
	\end{cases}\\
\hat{d}^{(2)}_2 {}^\dagger &\rightarrow \begin{cases}
	e^{i(\theta^{(2)}_a + \theta^{(2)}_b)}\hat{c}_2^\dagger & \text{for } t\rightarrow -\infty \\
	\frac{-e^{i(\theta^{(2)}_{c_2} + \theta^{(2)}_{b})}\hat{a}^\dagger +\kappa_j e^{i(\theta^{(2)}_{c_2} + \theta^{(2)}_a)}\hat{b}^\dagger}{\sqrt{(1 + \kappa_j^2)}} & \text{for } t\rightarrow\infty.
	\end{cases}
\end{split}
\end{equation}

Again we express $\dket{\alpha_1,\beta_1}$ in terms of the operators of \eref{eigen2} in the $t\rightarrow-\infty$ limit. This becomes
\begin{align}
\dket{\alpha_1,\beta_1} =&{}\frac{1}{\sqrt{N!}}\bigl( \alpha_1 e^{-i(\theta^{(1)}_{a} + \theta^{(1)}_{b})}\hat{d}^{(2)}_1 {}^\dagger \nonumber \\
	&{}\phantom{\frac{1}{\sqrt{N!}}\bigl(}+ \beta_1 e^{-i(\theta^{(1)}_{a} + \theta^{(1)}_{b})} \hat{d}^{(2)}_2 {}^\dagger  \bigr)^{N}\ket{0}.
\end{align}
Letting the eigenstates evolve into the $t\rightarrow\infty$ limit as before we follow the same procedure. One then obtains the final state
\begin{multline}
\dket{\alpha_2,\beta_2} = \frac{1}{\sqrt{N!(1 + \kappa_j^2)^{N}}}\\
\times \Bigl(\bigl[ -\kappa_j\alpha_1e^{i(\theta^{(2)}_{c_1}-\theta^{(2)}_{a})}-\beta_1e^{i(\theta^{(2)}_{c_2}-\theta^{(2)}_{a})}\bigr]\hat{a}^\dagger  \\
+\bigl[ -\alpha_1e^{i(\theta^{(2)}_{c_1}-\theta^{(2)}_{b})}+\kappa_j\beta_1e^{i(\theta^{(2)}_{c_2}-\theta^{(2)}_{b})}\bigr]\hat{b}^\dagger\Bigr)^{N}\ket{0}.
\label{eq:result}
\end{multline}

We define the transformations $\mO^{(1)}_j:[\alpha_0,\beta_0]^T\rightarrow[\alpha_1,\beta_1]^T$ and $\mO^{(2)}_j:[\alpha_1,\beta_1]^T\rightarrow[\alpha_2,\beta_2]^T$. Evidently these are linear transforms. They may be written as
\begin{equation}
\begin{split}
\mO^{(1)}_j &=\frac{1}{\sqrt{1 + \kappa_j^2}}
\begin{bmatrix}
-\kappa_j e^{i(\theta^{(1)}_a - \theta^{(1)}_{c_1})} & -e^{i(\theta^{(1)}_b - \theta^{(1)}_{c_1})}\\
-e^{i(\theta^{(1)}_a - \theta^{(1)}_{c_2})} & \kappa_j e^{i(\theta^{(1)}_b - \theta^{(1)}_{c_2})}
\end{bmatrix}\\
\mO^{(2)}_j &=\frac{1}{\sqrt{1 + \kappa_j^2}}
\begin{bmatrix}
-\kappa_j e^{i(\theta^{(2)}_{c_1} - \theta^{(2)}_a)} & -e^{i(\theta^{(2)}_{c_2} - \theta^{(2)}_a)}\\
-e^{i(\theta^{(2)}_{c_1} - \theta^{(2)}_b)} & \kappa_j e^{i(\theta^{(2)}_{c_2} - \theta^{(2)}_b)}
\end{bmatrix}.
\end{split}
\end{equation}
We may write the resulting unitary action of the entire transformation as the product of these two matrices. The total rotation of the whole STIRAP sequence is then
\begin{align}
U_j = \mO^{(2)}_j \mO^{(1)}_j .
\end{align}

\subsection{Examples}

Let us look at a few examples.  First, examine the case for the D1 line ($j=1$). Defining the phase angles $\theta^{(1)}_{c_1} = \theta^{(1)}_{c_2} = -\alpha/2$, $\theta^{(2)}_{c_1} = \alpha/2 + \gamma/2$ and $\theta^{(2)}_{c_2} = \alpha/2 - \gamma/2$, we obtain
\begin{equation}
U_1 =   e^{i\alpha}
\begin{bmatrix}
e^{i(\theta^{(1)}_a -  \theta^{(2)}_a)} \cos\frac{\gamma}{2} &
ie^{i(\theta^{(1)}_b -  \theta^{(2)}_a)} \sin\frac{\gamma}{2} \\
ie^{i(\theta^{(1)}_a -  \theta^{(2)}_b)} \sin\frac{\gamma}{2} &
e^{i(\theta^{(1)}_b -  \theta^{(2)}_b)} \cos\frac{\gamma}{2}
\end{bmatrix}.
\end{equation}
It is easily seen that by choosing phase angles $\theta^{(1)}_a = -\delta/2$, $\theta^{(1)}_b = \delta/2 + \pi/2$, $\theta^{(2)}_a = -\beta/2$, $\theta^{(2)}_b = \beta/2 + \pi/2$ gives an arbitrary unitary of the form of \eref{arbunitary}, i.e.
\begin{align}
U_1 &= e^{i\alpha}
\begin{bmatrix}
e^{i(-\beta/2 - \delta/2)}\cos{\frac{\gamma}{2}} &
-e^{i(-\beta/2 + \delta/2)}\sin{\frac{\gamma}{2}} \\
e^{i(\beta/2 - \delta/2)}\sin{\frac{\gamma}{2}} &
e^{i(\beta/2 + \delta/2)}\cos{\frac{\gamma}{2}}
\end{bmatrix}\nonumber \\
&= e^{i\alpha}\mathcal{R}_z(\gamma)\mathcal{R}_y(\beta)\mathcal{R}_z(\delta).
\end{align}
This is equivalent to the $\mathcal{R}_{\vect{n}}(\delta)$ rotation implemented by Kis and Renzoni in Ref. \cite{qubitrotationk}, which we described earlier in this paper for spin coherent states.
On \figref{bspheres} we show a simplified decomposition of $U_1$. Here we view the $\mO^{(j)}_1$ operators as taking the BEC into and out of the space of internal states $\ket{c_1}$ and $\ket{c_2}$. We may in this context define the corresponding Schwinger boson operators
\begin{equation}
\begin{split}
\hat{J}_x &= \hat{c}^\dagger_1\hat{c}_2 + \hat{c}_2^\dagger\hat{c}_1\\
\hat{J}_y &= -i\hat{c}^\dagger_1\hat{c}_2 + i\hat{c}_2^\dagger\hat{c}_1\\
\hat{J}_z &= \hat{c}^\dagger_1\hat{c}_2 - \hat{c}_2^\dagger\hat{c}_1.
\end{split}
\label{eq:SchwingerOps2}
\end{equation}
We may now define a Bloch sphere as we did earlier, but this time in terms of the expectation value of the above. The operation of $U_1$ may then be viewed as either a piecewise rotation performed by a combination of $\mO^{(1)}_1$ and $\mO^{(2)}_1$ or as an application simply of $U_1$.

For the D2 line, performing the matrix multiplication explicitly shows that the diagonal entries of $U_2$ cannot equal zero. Thus arbitrary rotations are not possible. However, choosing the same phase angles as for $U_1$, and setting $\gamma = \delta = 0$, one may produce
\begin{align}
U_2 &= e^{i\alpha}
\begin{bmatrix}
e^{-i\beta/2} &
0 \\
0 &
e^{i\beta/2}
\end{bmatrix}\nonumber\\
&= e^{i\alpha}\mathcal{R}_z(\beta).
\end{align}
The difference between the D1 and D2 lines arises due to the natural selection rules between transitions. This has an effect on the type of unitary operations that are possible, hence for this scheme the D1 line is preferable in terms of generality.

\section{Performance evaluation}
\label{sec:V}

The most important feature of STIRAP is the ability to perform coherent control without populating the excited states. In systems such as BECs, where the fidelity of operation is limited by stimulated emission enhanced by the particle number, avoiding excited states is crucial towards suppressing unwanted decoherence. However, this relies entirely on successfully keeping to the adiabatic approximation. In practice non-zero populations of the excited state occur due to deviations from adiabaticity, which limits the accuracy of the coherent control. To truly test the procedure numerical simulations are necessary to demonstrate the efficiency of the STIRAP gate introduced in the last section. In this section we describe the simulations including spontaneous emission
to estimate the performance of the procedure for realistic situations.

\subsection{Model including spontaneous emission}

\begin{figure*}[!tb]
\centering
\def\svgwidth{.45\textwidth}
\input{stirap.tex}
\centering
\hspace{5mm}
\def\svgwidth{.45\textwidth}
\input{stirapfail.tex}
\vspace{9mm}
\centering
\def\svgwidth{.45\textwidth}
\input{stirap2.tex}
\centering
\hspace{5mm}
\def\svgwidth{.45\textwidth}
\input{stirapfail2.tex}
\caption{(Color online). Typical six-level STIRAP time evolution performing a coherent rotation of a BEC. The simulations have been carried out using data recorded for the D1 transition in \cite{steck2001rubidium}, and the basic time unit is the decay time of the \ce{^{87}Rb} D1 transition, i.e. $\tau = \SI{27.70(4)}{\nano\second}$. The simulation ends after one $\tau$. All plots are for the case where the initial spin coherent state is $\dket{\cos{3\pi/8},\sin{3\pi/8}e^{i5\pi/8}}$ and the target rotation is the  operation $U = \mathcal{R}_z(\pi/4)\mathcal{R}_y(3\pi/4)\mathcal{R}_z(\pi/4) $. The target state after the rotation is indicated with the horizontal lines marked as $ \braket{\hat{S}_{x,y,z}}_{\text{targ}} $. The pulses have a Gaussian shape, with $T_- = t_\textup{end}/8$, $T_+ = 3t_\textup{end}/8$. After pulse width optimization we obtain for (a) (c) $N = 10^4$ a width of $T=0.1324t_\textup{end}$ ($\text{FWHM} = 0.1835t_\textup{end}$) and for (b) (d) $N=10^5$ we obtain $T=0.0877t_\textup{end}$ ($\text{FWHM} = 0.1216t_\textup{end}$). Subfigures (a) (b) show the the evolution of the expectation values of the Schwinger boson operators, $ \braket{\hat{S}_{x,y,z}}$. Subfigures (c) (d) depict the occupation of intermediate and excited states.}
\label{fig:stirap}
\end{figure*}
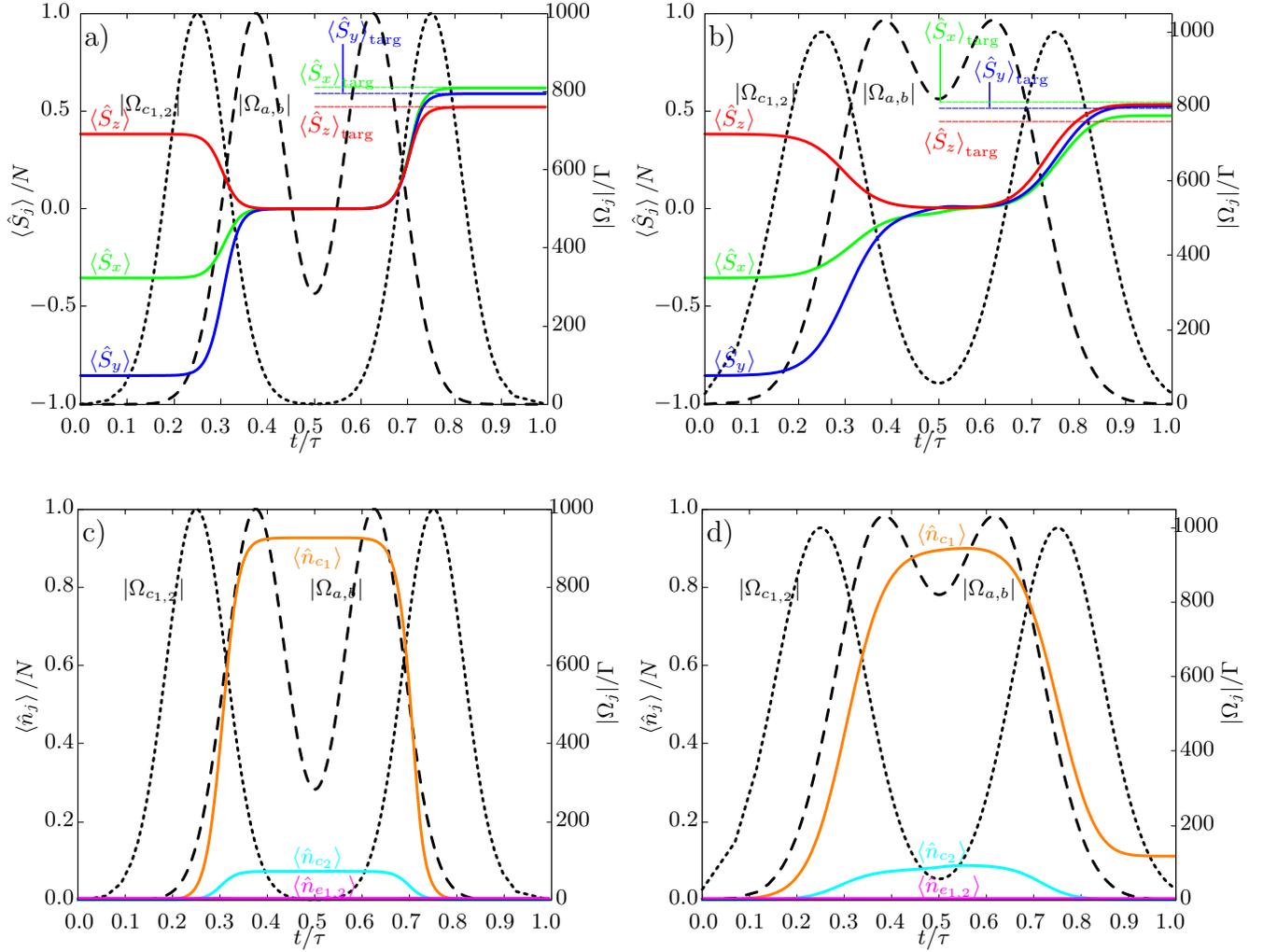

The master equation including spontaneous emission of the BEC is
\begin{multline}
\diff{\rho}{t} = \frac{i}{\hbar}[\rho,\hat{H}] - \frac{\Gamma_1}{2}\bigl( [\hat{e}_1^\dagger\hat{c}_1,\hat{c}_1^\dagger\hat{e}_1\rho] + [\rho\hat{e}_1^\dagger\hat{c}_1,\hat{c}_1^\dagger\hat{e}_1] \bigr)\\
- \frac{\Gamma_2}{2}\bigl( [\hat{e}_2^\dagger\hat{c}_2,\hat{c}_2^\dagger\hat{e}_2\rho] + [\rho\hat{e}_2^\dagger\hat{c}_2,\hat{c}_2^\dagger\hat{e}_2] \bigr),
\end{multline}
where $\Gamma_j $ is the spontaneous emission rate from the excited state $ e_j $ to the state $ c_j $. We have made a simplifying assumption that the spontaneously emitted photons only have a $\pi$-polarization.  This assumption gives us a convenient way to understand coherence in our system, as the presence of atoms in the $ c_i $ states signals an imperfect adiabatic process.  This difference should only affect the final distribution of states, and not the infidelity estimates which is insensitive to the type of spontaneous emission. 

Time evolving the complete density matrix is a numerically intensive task for large $ N $.  However, to an excellent approximation such a system can be described by a system of coupled equations in bilinear expectation values (see Appendix) \cite{timbec3}. To quantify the effect on gate performance we use
the fidelity defined as
\begin{equation}
F(\rho,\sigma) \equiv 1 - D(\rho,\sigma)
\end{equation}
where the trace distance is defined as
\begin{align}
	D(\rho,\sigma) =&{}
	\Bigl[\bigl(\braket{\hat{S}_x}_{\rho} - \braket{\hat{S}_x}_{\sigma}\bigr)^2
	+\bigl(\braket{\hat{S}_y}_{\rho} - \braket{\hat{S}_y}_{\sigma}\bigr)^2\nonumber\\
	&{}\phantom{\Bigl[}+\bigl(\braket{\hat{S}_z}_{\rho} - \braket{\hat{S}_z}_{\sigma}\bigr)^2\Bigl]/(4N^2).
\end{align}
Here $ \sigma $ is the target state and $\rho$ is the numerically evolved state.  
We use this definition of fidelity as it provides a way of comparing  fidelities for different $ N $ directly. 
The standard definition of fidelity using the inner product has an exponential dependence on the particle number
\begin{multline}
\left\lvert\dBraket{\cos\frac{\theta'}{2},e^{i\phi}\sin \frac{\theta'}{2} |\cos \frac{\theta}{2},e^{i\phi}\sin\frac{\theta}{2} }\right\rvert^2\\
= \cos^{2N}\Bigl(\frac{\theta' - \theta}{2}\Bigr) \approx e^{-\frac{N(\theta' - \theta)^2}{4}},
\end{multline}
where we have chosen the case $\phi' = \phi$ for simplicity. 
According to this definition the fidelity is exponentially sensitive to the angular difference $\theta - \theta'$.  In contrast, our fidelity expression evaluates for this case to
\begin{align}
F = \cos^2 \Bigl( \frac{\theta' - \theta}{2} \Bigr)
\end{align}
which is true for any $ N $.  For further discussions on this see Refs. \cite{timbec1,timbec4}.

The STIRAP procedure that we have presented has some free parameters which are available for optimization. 
We choose the pulse shapes $f(t)$ to have a Gaussian form and peak amplitude unity,
\begin{equation}
f(t) = e^{-\frac{t^2}{T^2}}.
\end{equation}
Here $T$ is the pulse width. The size of the overlap of successive pulses in a STIRAP process is important to the overall performance. This is because the overlap region is where the dark states evolve the most rapidly and therefore also the region where the adiabatic condition is susceptible to failure. The pulse width $T$ has been treated as a variables in an optimization problem, that can be formulated in the following way. Given some initial and target quantum states, $\rho$ and $\sigma$, minimize $D(\rho, \sigma)$ as a function of $T$.
This is a one-dimensional optimization problem, and it has been solved via golden section search under the assumption that $D$ is a unimodal function of $T$. We have not used the laser amplitudes as variables in the optimization problem. 

From our numerics we have found that it is possible to arbitrarily improve the fidelities by reducing $T$ while at the same time increasing the laser amplitudes. This may appear to violate the adiabatic condition, however, the dark states of a STIRAP do not change quickly in the neighborhood of the pulse peaks. They only change when there is some significant overlap of the pulses. If pulse amplitude is increased while $T$ is decreased at a controlled rate, the net effect is to smooth this overlap region. This ensures that the adiabatic condition is in fact followed more closely, and therefore if an optimization algorithm optimizes the fidelity as a function of both $T$ and $\Omega_0$, each iteration will just produce narrower and more intense laser pulses and the algorithm will not converge. Therefore optimization must be carried out with either fixed pulse amplitudes or fixed pulse widths, where we have chosen the latter.

\section{Numerical Results}
\label{sec:VI}

\Figref{stirap} depicts the time evolution of two coherent spinor BECs that undergo the same rotation from the same initial state $\dket{\cos 3\pi/8, \sin 3\pi/8e^{i5\pi/8}}$, but have different numbers of particles  $N=10^4,10^5$. Each are optimized separately to obtain the best target fidelities. We show two cases to illustrate one sequence that successfully produces the gate as desired with high fidelity, and another where the performance is not as good.  The four laser amplitudes are shown as two successive STIRAPs are performed. Here the $\Omega_{c_{1,2}}$ pulses form the Stokes pulse, while the $\Omega_{a,b}$ make up the pump pulse in the first STIRAP. This changes the dark state creation operators from superpositions of $\hat{a}^\dagger$ and $\hat{b}^\dagger$ to $\hat{c_1}^\dagger$ and $\hat{c_2}^\dagger$. In the second STIRAP the pulses are reversed to return to the dark state operators to a rotated superposition of $\hat{a}^\dagger$ and $\hat{b}^\dagger$. After the first STIRAP, each $\braket{\hat{S}_j}$ is reduced to zero as the coherent spinor is completely transferred. The second STIRAP then transfers the BEC to the target space. In \figref{stirap}(b) the same process is depicted for $N=10^5$. Here the pulses are broader after pulse width optimization, which gives significant overlap of the middlemost pulses. We see here that the $\braket{\hat{S}_j}$ do not go exactly to zero. Looking closely at their evolutions, we can see some irregular Rabi oscillations close to the center of the graph. The target state is not reached with a high fidelity in this case, which we attribute to the stronger effect of spontaneous emission for this case making it a more difficult case to optimize. 

\Figref{stirap}(c) shows the evolution of excited and intermediate states during the rotation of the $N=10^4$ BEC. The same laser pulse amplitudes as \figref{stirap}(a) are shown here for reference. The occupation of levels $\ket{c_1}$ and $\ket{c_2}$ increase during the first STIRAP and stay approximately constant until the second one, where they go back to $0$.  \Figref{stirap}(d) shows the evolution of the same expectation values for the $N=10^5$ BEC. Here the change is not as sharp, and the populations are not constant after the first STIRAP. They increase slightly in the central area of the graph. This is the same area where close inspection of \figref{stirap}(b) reveals Rabi oscillations of the condensate. What appears to be happening is that the first STIRAP does not transfer the condensate fully. The laser transitions $\Omega_{a,b}$ induce Rabi oscillations which populate the excited states $\ket{e_{1,2}}$. Since the spontaneous emission rate is enhanced by a factor of $10^5$, the excited state populations are effectively eliminated, and all of the excited state population is directly transferred to the $\ket{c_{1,2}}$ states. Finally, the second STIRAP leaves some population in $\ket{c_1}$, which contributes to its inefficiency.

\begin{figure}[!tb]
\centering
\def\svgwidth{\columnwidth}
\input{plotfile.tex}
\caption{(Color online). Performance of the STIRAP gate procedure as measured by the infidelity $D(\rho,\sigma)$ as a function of the time $t_\textup{end}$ it takes to finish an operation and for various combinations of $N = 10^3, 10^4, 10^5$ and $\Omega_0 = 100, 1000$ in units of $\Gamma$. The points that have been computed in this graph are spaced logarithmically from $1$ to $100$ in units of $\tau = \SI{27.70}{\nano\second}$ on the temporal axis.}
\label{fig:simulationaverages}
\end{figure}
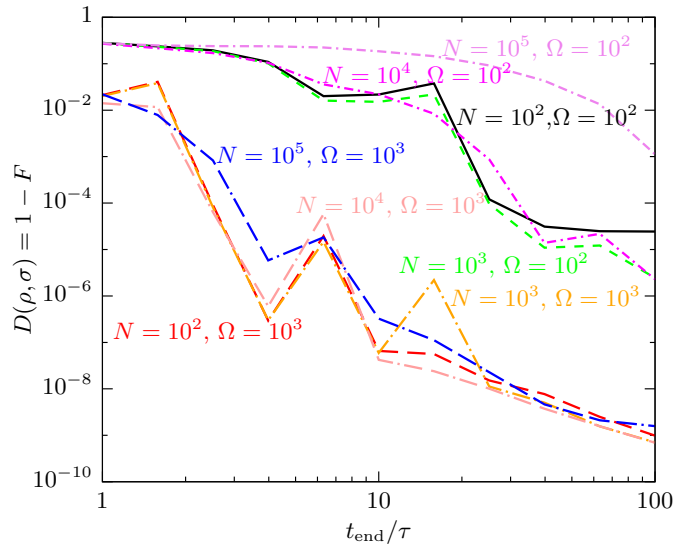 

We have performed a benchmarking of the performance of our procedure by comparing the optimized six-level STIRAP procedure for various $t_\textup{end}$, $\Omega_0$ and $N$. Here $t_\textup{end}$ defines the time the procedure takes. In the example of \figref{stirap} this is $\tau$. For each configuration of $t_\textup{end}$, $\Omega_0$ and $N$, the optimum $T$ is determined by golden section search. Then a series of \num{240} simulations with these parameters and randomized initial states $\dket{\alpha,\beta}$ and unitary rotations $U(\beta,\gamma,\delta)$ is carried out. The average fidelity is saved for that particular combination of $t_\textup{end}$, $\Omega_0$ and $N$, before the process is repeated for a different set of parameters.

We note that throughout our simulations the amplitude of any laser pulses are much lower than the D1 optical transition frequency. The detunings are constant, and $\Delta_+\equiv (\Delta_1 + \Delta_2)/2$ is chosen to be $100\Omega_0$, which is the peak amplitude of the pulses. This is to ensure that we can neglect unwanted couplings to excited states that are not accounted for in this scheme.

\Figref{simulationaverages} shows the results of the simulation for various parameters. When $\Omega_0 =10^2\Gamma$, we obtain results for $N=10^2$ -- $10^4$ that reach as low as $10^{-5}$ depending on $t_\textup{end}$. Increasing $N$ beyond this value increases the infidelity to values greater than $10^{-3}$ for the timescales we are considering. Increasing $\Omega_0$ to $10^3\Gamma$ greatly improves our results. The results show an overall trend of the infidelity to decrease with higher $t_\textup{end}$, and the results are very similar regardless of $N$. By allowing $t_\textup{end}$ to be of the order $\SI{1}{\micro\second}$, we see that the infidelity drops to below $10^{-7}$ regardless of $N$, which is effectively negligible. Note that there are several kinks in the graphs, which are likely owed to faulty or incomplete optimization of the laser pulse shapes used.

If slower gates are allowed, the fidelities can generally be improved further. We have not explored results beyond $\Omega_0 = 10^3\Gamma$, but we expect that fidelities arbitrarily close to $1$ are possible in our model. There is clearly a trade-off between fidelity and speed. We expect that if $N$ is increased to much higher values the enhanced spontaneous emission will start imposing stricter limits on the fidelity and the trade-off will likely involve $N$.

The fact that the results do not appear to depend strongly on $N$ shows that we successfully avoid excited state populations. This is helped along by choosing very large detunings that limit potential transitions to excited states. This leads to the conclusion that our fidelity in this regime is not caused primarily by enhanced spontaneous emission, but instead incomplete or faulty STIRAP procedures. This is not surprising as we have only performed a very elementary pulse optimization, where only pulse widths have been taken into account.

\section{Approximations to the model}

In real $^{87}$Rb atoms there are more ground states present as well as excited states with other quantum numbers than the ones we have considered so far. Thus some justification is needed for their absence in our simulation since two-photon transitions to omitted internal states have not been accounted for. Firstly, including all relevant $F=2$ ground states, there are a total of six two-photon transitions with $|\Delta m_F| = 2$. However, as we have already explained in \secref{IIc}, the transition that occurs between levels $a$ and $b$ is very inefficient at high detunings, and this is confirmed in our simulations, as we do not see strong evidence for non-adiabatic transitions due to this resonance. One may then consider the four possible combinations of transitions between $c_1$, $c_2$ and $\ket{F=2,m_F=-2}$, $\ket{F=2,m_F=2}$, as well as the $\ket{F=2,m_F=-1}$ to $\ket{F=1,m_F=1}$ transition. However, these five transitions are inefficient following the same arguments as in \secref{IIc}, and have the same dependence on detuning. Therefore we do not consider them in our simulations.

Secondly, linearly polarized laser light only causes transitions where the total electronic angular momentum of the atom changes. For this reason, two-photon transitions between $c_1$ and $c_2$ do not occur when the two lasers are $\pi$-polarized. Other transitions, in which both electronic and nuclear angular momentum changes, are possible, but they are inefficient for the same reason that $\Delta m_F = 2$ transitions are inefficient.

Thirdly, one may consider transitions that occur when one laser is linearly polarized and another has circular polarization. Here there is no intrinsic inefficiency as the one we have discussed for the $\Delta m_F = 2$ transitions. In our scheme, it is the $\Omega_{c_{1,2}}$ pulses that are both linearly polarized while the $\Omega_{a,b}$ pulses are $\sigma_+$, $\sigma_-$ polarized respectively. Thus for such transitions to occur there should be significant overlap between these pulses. But in reality, as shown on \figref{stirap}, the overlap of any two such pulses in our scheme is small. In our simulations, in which four such transitions are already possible when levels $a$, $b$, $c_1$ and $c_2$ are included, we believe this effect is accounted for already. In those of our simulations where the maximum pulse amplitudes were increased to $1000\Gamma$, we see that the pulse optimization step results in much narrower pulses. This means that the overlap of the pulses must be kept the same when the maximum pulse amplitudes are increased for optimum transfer. Thus our simulations confirm that we are indeed limited by the pulse overlap in terms of undesired transitions.  It may be argued that it is one of the great strengths of the counter-intuitive pulse sequence of a STIRAP that we are able to avoid such transitions by having the least possible overlap of the Stokes and pump pulses.

In summary, we believe that the fidelity is primarily limited by Raman transitions between hyperfine $^{87}$Rb sublevels when there is some overlap of stokes and pump pulses. In our simulations we show that infidelities may be driven very close to zero in the presence of four such transitions. While we have not included all possible sublevels and all possible transitions that may occur by this mechanism, we do not believe that they would change the fidelities by orders of magnitude. Although our results so far suggest that arbitrarily high fidelities may be achieved by increasing pulse amplitudes while decreasing the pulse width, it must be kept in mind that the robustness of the scheme is easily affected. This is because the overlap of the Stokes and pump pulses is very sensitive to jitter when pulse amplitudes are high. This is especially true when the pulses originate from different lasers as in our scheme. Thus we cannot realistically control this overlap as well as would be required for very intense pulses.

\section{Summary and Conclusions}
\label{sec:VII}

We have presented two methods of performing SU(2) rotations of spinor BECs using STIRAP. The first is a generic method based on three ground state levels as originally proposed in Ref. \cite{qubitrotationk} and here generalized for BECs.  The second method overcomes a known problem of alkali atoms which makes Raman transitions inefficient for states which are separated by $ \Delta m_F \ge 2 $ due to the necessity of flipping an optically inaccessible nuclear spin.  In our numerical simulations we test the method for BECs with large particle numbers in the presence of spontaneous emission, and show that very high fidelities are still achievable.  For example, for a BEC containing $ N = 10^3 $ atoms, we have shown that it is possible to perform arbitrary rotations with an effectively negligible error with gate times of $ \sim \SI{1}{\micro\second} $ while using pulse amplitudes of $\Omega_0 = 10^3\Gamma$. For these low infidelities, the source of error is no longer spontaneous emission, but imperfections intrinsic to STIRAP.  The reason for the high fidelities is that the method successfully keeps most of the population in the ground states, which suppresses spontaneous emission. Enhanced spontaneous emission is one of the primary causes of decoherence when using optical control in BECs. Our method relies on an optimization step, where the pulse width $T$ is chosen carefully. Naturally, in a real experiment other sources of decoherence, such as fluctuations of the trapping magnetic field will also contribute to decoherence.  The significance of these results is that despite the magnified effects of spontaneous emission by particle number, STIRAP methods can strongly suppress the decoherence such that very high fidelities may be achieved. 

In this work our optimizations were limited to changing just one of the parameters -- the width of the pulses -- hence there is more scope for improvement.  Already some work has been carried out on this topic in e.g. Ref. \cite{vasilev2009optimum}, optimum pulse shapes for STIRAP have been realized theoretically by reducing three level Hamiltonians to effective two-level ones. Adapting  these ideas to our scheme can potentially improve the fidelities further and thereby reduce the experimental overhead.  The ability to perform high fidelity SU(2) rotations in BECs is of fundamental interest to several potential applications such as quantum metrology and quantum information.  In magnetometry, the strength of a magnetic field may be characterized by detecting the Larmor precession. While currently  microwave and radio frequency pulses are conventionally used to obtain the desired superposition of magnetic sublevels, our approach could offer a faster and high fidelity alternative.  Our method is also naturally suited to quantum information processing applications where spin coherent states are used for quantum memories \cite{qcrevl, timbec2, timbec3, timbec4, zoller2003quantum}.  As optical laser pulses can be tightly focused, methods such as that described here would contribute to scalable atom chip configurations where there are multiple BEC on the same chip that are individually controlled. This could be combined with 
entanglement generation schemes between neighboring BECs \cite{idlas2016entanglement} to create a network of entangled BECs \cite{pyrkoventangle}.

\begin{acknowledgments}
This work is supported by the Shanghai Research Challenge Fund, New York University Global Seed Grants for Collaborative Research, National Natural Science Foundation of China grant 61571301, the Thousand Talents Program for Distinguished Young Scholars, and the  NSFC Research Fund for International Young Scientists.  
\end{acknowledgments}

\appendix

\section{Evolution of Spin Coherent States}
\label{app:a}
The spin coherent state of a BEC can be characterized in terms of its expectation values of $\hat{S}_x$, $\hat{S}_y$ and $\hat{S}_z$. There exists a bijection between the space of $(\braket{\hat{S}_x}, \braket{\hat{S}_y}, \braket{\hat{S}_z})$ and $(\braket{\hat{a}^\dagger\hat{a}}, \braket{\hat{a}^\dagger\hat{b}}, \braket{\hat{b}^\dagger\hat{b}})$. Thus for a coherent spinor, knowledge of its location in the latter space is sufficient to fully characterize it.
To obtain a closed set of equations we have here made a decoupling approximation, where expectation values may be decomposed as e.g. $\braket{\hat{e}_1^\dagger\hat{c}_1\hat{c}_1^\dagger\hat{e}_1}\approx \braket{\hat{e}_1^\dagger\hat{e}_1}(\braket{\hat{c}_1^\dagger\hat{c}_1}+1)$. We note that this decoupling of expectation values is exact for coherent spinors, but approximate for other Fock state superpositions.

The evolution of these expectation values occurs according to 
\begin{equation}
\begin{split}
\diff{\braket{\hat{A}}}{t} ={}& i\braket{[\hat{H}, \hat{A}]}\\
&- \Gamma_1\textup{Tr}\Bigl( \rho \bigl([\hat{A},\hat{e}_1^\dagger\hat{c}_1]\hat{c}_1^\dagger\hat{e}_1 + \hat{e}_1^\dagger\hat{c}_1[\hat{c}_1^\dagger\hat{e}_1, \hat{A}] \bigr) \Bigr)\\
&- \Gamma_2\textup{Tr}\Bigl( \rho \bigl([\hat{A},\hat{e}_2^\dagger\hat{c}_2]\hat{c}_2^\dagger\hat{e}_2 + \hat{e}_2^\dagger\hat{c}_2[\hat{c}_2^\dagger\hat{e}_2, \hat{A}] \bigr) \Bigr).
\label{eq:eqofmotion}
\end{split}
\end{equation}
We derive the following coupled differential equations describing the evolution of expectation values.
{\allowdisplaybreaks\begin{align}
\diff{\braket{\hat{a}^\dagger\hat{a}}}{t} ={}& 2\Im({\Omega_a(\braket{\hat{a}^\dagger\hat{e}_1} + \braket{\hat{a}^\dagger\hat{e}_2})) }\\ 
\diff{\braket{\hat{a}^\dagger\hat{b}}}{t} ={}& i\bigl( \Omega_a^*(\braket{\hat{b}^\dagger\hat{e}_1} + \braket{\hat{b}^\dagger\hat{e}_2})^*\nonumber\\
&+ \Omega_b(\braket{\hat{a}^\dagger\hat{e}_1} - \braket{\hat{a}^\dagger\hat{e}_2}) \bigr)\\ 
\diff{\braket{\hat{a}^\dagger\hat{c}_j}}{t} ={}& i\bigl((\Omega_a (\braket{\hat{c}_j^\dagger\hat{e}_1}+ \braket{\hat{c}_j^\dagger\hat{e}_2}))^*\nonumber\\
&- \Omega_{c_j} \braket{\hat{a}^\dagger\hat{e}_j}\bigr) + \Gamma \braket{\hat{a}^\dagger\hat{c}_j} \braket{\hat{e}_j^\dagger\hat{e}_j}\\ 
\diff{\braket{\hat{a}^\dagger\hat{e}_j}}{t} ={}& i(\Omega_a^*(\braket{\hat{e}_1^\dagger\hat{e}_j}^* + \braket{\hat{e}_2^\dagger\hat{e}_j} - \braket{\hat{a}^\dagger\hat{a}})\nonumber\\
& - \Omega_b^*\braket{\hat{a}^\dagger\hat{b}} - \Omega_{c_j}^*\braket{\hat{a}^\dagger\hat{c}_j} - \Delta_j\braket{\hat{a}^\dagger\hat{e}_j})\nonumber\\
&-  \Gamma\braket{\hat{a}^\dagger\hat{e}_j}(\braket{\hat{c}_j^\dagger\hat{c}_j} + 1) \\ 
%
\diff{\braket{\hat{b}^\dagger\hat{b}}}{t} ={}& 2\Im(\Omega_b(\braket{\hat{b}^\dagger\hat{e}_1} - \braket{\hat{b}^\dagger\hat{e}_2})) \\ 
\diff{\braket{\hat{b}^\dagger\hat{c}_j}}{t} ={}& i((\Omega_b(\braket{\hat{c}_j^\dagger\hat{e}_1} - \braket{\hat{c}_j^\dagger\hat{e}_2}))^* \nonumber\\
&- \Omega_{c_j}\braket{\hat{b}^\dagger\hat{e}_j}) + \Gamma\braket{\hat{b}^\dagger\hat{c}_j}\braket{\hat{e}_j^\dagger\hat{e}_j} \\ 
\diff{\braket{\hat{b}^\dagger\hat{e}_j}}{t} ={}& i(-1)^{j+1}(\Omega_b^*(\braket{\hat{e}_1^\dagger\hat{e}_j}^* - \braket{\hat{e}_2^\dagger\hat{e}_j} - \braket{\hat{b}^\dagger\hat{b}}) \nonumber\\
&- (\Omega_a\braket{\hat{a}^\dagger\hat{b}} )^* - \Omega_{c_j}^*\braket{\hat{b}^\dagger\hat{c}_j} - \Delta_j \braket{\hat{b}^\dagger\hat{e}_j}) \nonumber\\
&- \Gamma\braket{\hat{b}^\dagger\hat{e}_j}(\braket{\hat{c}_j^\dagger\hat{c}_j} + 1) \\ 
%
\diff{\braket{\hat{c}_j^\dagger\hat{c}_j}}{t} ={}& 2\Im(\Omega_{c_j}\braket{\hat{c}_j^\dagger\hat{e}_j}) \nonumber\\
&+ 2\Gamma\braket{\hat{e}_j^\dagger\hat{e}_j}(\braket{\hat{c}_j^\dagger\hat{c}_j}  + 1) \\ 
\diff{\braket{\hat{c}_1^\dagger\hat{c}_2}}{t} ={}& i((\Omega_{c_1}\braket{\hat{c}_2^\dagger\hat{e}_1})^* - \Omega_{c_2}\braket{\hat{c}_1^\dagger\hat{e}_2}) \nonumber\\
&+ \braket{\hat{c}_1^\dagger\hat{c}_2}(\Gamma_1\braket{\hat{e}_1^\dagger\hat{e}_1} + \Gamma_2\braket{\hat{e}_2^\dagger\hat{e}_2}) \\ 
\diff{\braket{\hat{c}_j^\dagger\hat{e}_j}}{t} ={}& i(\Omega_{c_j}^*(\braket{\hat{e}_j^\dagger\hat{e}_j} - \braket{\hat{c}_j^\dagger\hat{c}_j}) - \Delta_j\braket{\hat{c}_j^\dagger\hat{e}_j} \nonumber\\
&- (\Omega_a\braket{\hat{a}^\dagger\hat{c}_j})^* - (-1)^j(\Omega_b\braket{\hat{b}^\dagger\hat{c}_j})^*)\nonumber\\
&+ \Gamma\braket{\hat{c}_j^\dagger\hat{e}_j}(\braket{\hat{e}_j^\dagger\hat{e}_j} - \braket{\hat{c}_j^\dagger\hat{c}_j}) \\ 
\diff{\braket{\hat{c}_1^\dagger\hat{e}_2}}{t} ={}& i(\Omega_{c_1}^*\braket{\hat{e}_1^\dagger\hat{e}_2} - \Omega_{c_2}^*\braket{\hat{c}_1^\dagger\hat{c}_2} - (\Omega_a\braket{\hat{a}^\dagger\hat{c}_1})^* \nonumber\\
&+ (\Omega_b\braket{\hat{b}^\dagger\hat{c}_1})^* - \Delta_2\braket{\hat{c}_1^\dagger\hat{e}_2}) \nonumber\\
&+ \braket{\hat{c}_1^\dagger\hat{e}_2}(\Gamma_1\braket{\hat{e}_1^\dagger\hat{e}_1} - \Gamma_2(\braket{\hat{c}_2^\dagger\hat{c}_2} + 1)) \\ 
%
\diff{\braket{\hat{c}_2^\dagger\hat{e}_1}}{t} ={}& i((\Omega_{c_2}*\braket{\hat{e}_1^\dagger\hat{e}_2})^* - (\Omega_{c_1}\braket{\hat{c}_1^\dagger\hat{e}_1})^* \nonumber\\
&- (\Omega_a\braket{\hat{a}^\dagger\hat{c}_2})^* - (\Omega_b\braket{\hat{b}^\dagger\hat{e}_2})^* - \Delta_1\braket{\hat{c}_2^\dagger\hat{e}_1}) \nonumber\\
&+ \braket{\hat{c}_2^\dagger\hat{e}_1}(\Gamma_2\braket{\hat{e}_2^\dagger\hat{e}_2} - \Gamma_1(\braket{\hat{c}_1^\dagger\hat{c}_1} + 1)) \\ 
%
\diff{\braket{\hat{e}_j^\dagger\hat{e}_j}}{t} ={}&  -2(\Im(\Omega_a\braket{\hat{a}^\dagger\hat{e}_j}) + (-1)^j\Im(\Omega_b\braket{\hat{b}^\dagger\hat{e}_j}) \nonumber\\
&+ \Im(\Omega_{c_j}\braket{\hat{c}_j^\dagger\hat{e}_j})) \nonumber\\
&- 2\Gamma\braket{\hat{e}_j^\dagger\hat{e}_j}(\braket{\hat{c}_j^\dagger\hat{c}_j} + 1)\\ 
\diff{\braket{\hat{e}_1^\dagger\hat{e}_2}}{t} ={}& i(\Omega_a\braket{\hat{a}^\dagger\hat{e}_2} - (\Omega_a\braket{\hat{a}^\dagger\hat{e}_1})^* + (\Omega_b\braket{\hat{b}^\dagger\hat{e}_1})^* \nonumber\\
&+ \Omega_b\braket{\hat{b}^\dagger\hat{e}_2} - (\Omega_{c_2}\braket{\hat{c}_2^\dagger\hat{e}_1})^* + \Omega_{c_1}\braket{\hat{c}_1^\dagger\hat{e}_2} \nonumber\\
&+ (\Delta_1 - \Delta_2)\braket{\hat{e}_1^\dagger\hat{e}_2}) \nonumber\\
&- \braket{\hat{e}_1^\dagger\hat{e}_2}(\Gamma_1(\braket{\hat{c}_1^\dagger\hat{c}_1} + 1) \nonumber\\
&+ \Gamma_2(\braket{\hat{c}_2^\dagger\hat{c}_2} + 1)) \\ 
\diff{\braket{\hat{e}_2^\dagger\hat{e}_2}}{t} ={}& -2(\Im(\Omega_a\braket{\hat{a}^\dagger\hat{e}_2}) - \Im(\Omega_b\braket{\hat{b}^\dagger\hat{e}_2}) \nonumber\\
&+ \Im(\Omega_{c_2}\braket{\hat{c}_2^\dagger\hat{e}_2})) \nonumber\\
&- 2\Gamma_2\braket{\hat{e}_2^\dagger\hat{e}_2}(\braket{\hat{c}_2^\dagger\hat{c}_2} + 1) 
\end{align}}

\section{Validity of the mean field approximation}
\label{app:b}

The mean field approximation method used here allows us to efficiently simulate all of the relevant expectation values of our system in a very highly dimensional Fock space. We here compare the two approaches in solving in a three level bosonic system with a constant number of particles. We compare the approximate mean field model to the exact Fock state basis evolution of a condensate with $N = 8$ atoms.

\begin{figure}
\includegraphics[width = .49\linewidth]{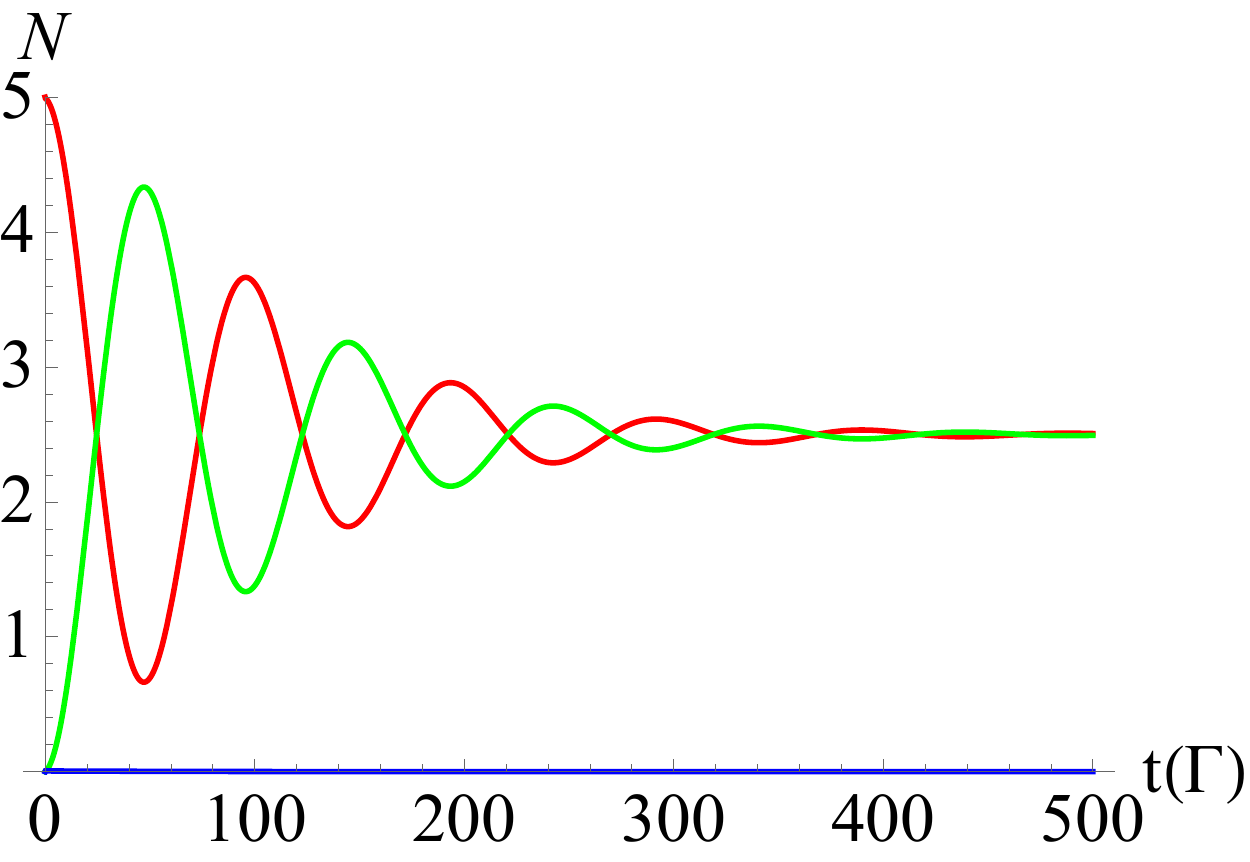}
\includegraphics[width = .49\linewidth]{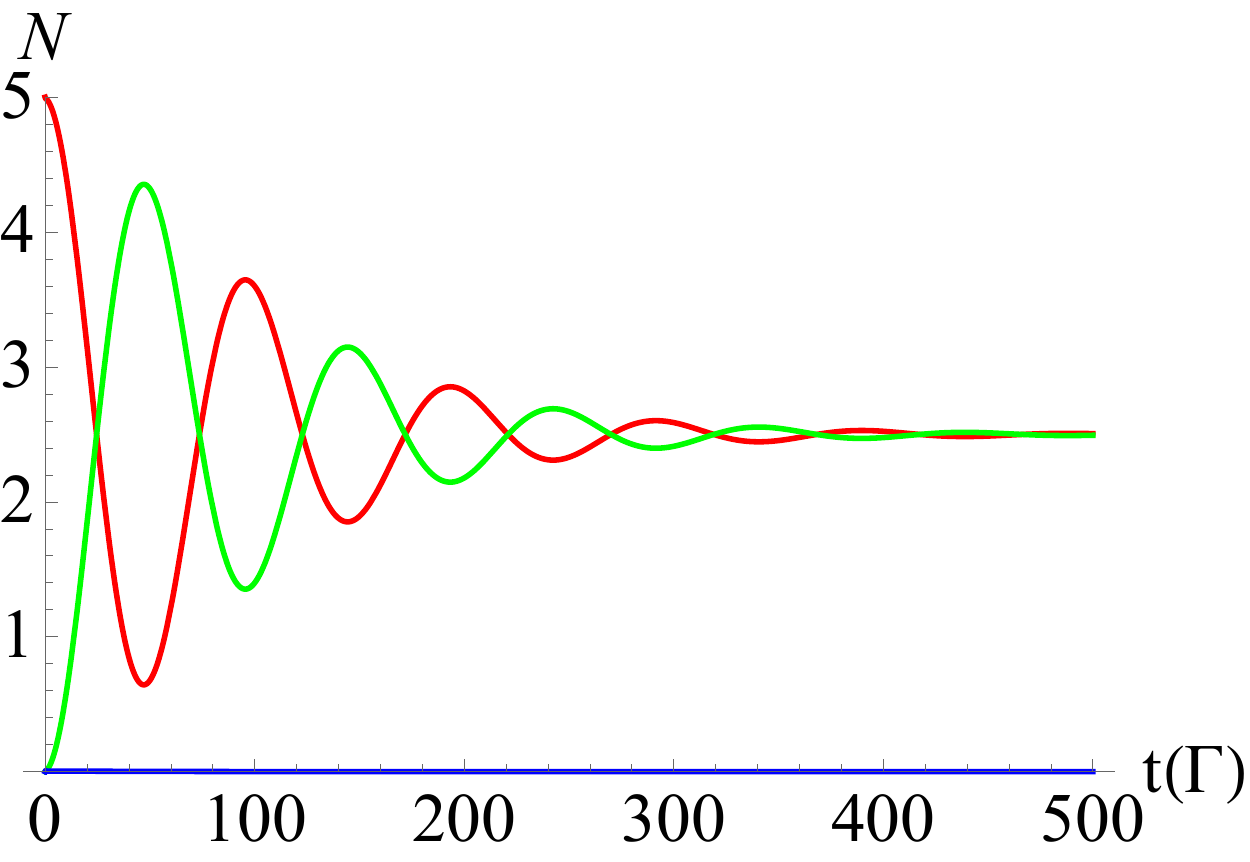}
\caption{Comparison of the mean-field solution (left) versus the exact calculation in the Fock basis (right). In both cases a coherent spinor has the initial state $\alpha=1$, $\beta=0$. The curve which initially starts at $N = 5 $ is  $\braket{\hat{a}^\dagger\hat{a}}$, while the curve which starts at zero is $\braket{\hat{b}^\dagger\hat{b}}$. The blue line, which is very close to zero throughout on both graphs is $\braket{\hat{c}^\dagger\hat{c}}$. The parameters are $
\Gamma_1 = \Gamma_2 = \Omega_a = \Omega_b = 1,\Delta_1 = \Delta_2 = 10,
N = 5,\quad t_\textup{end} = 1000$.}
\label{fig:appendix_compare}
\end{figure}

The Hamiltonian is
\begin{equation}
\begin{split}
\hat{H} ={}& \hbar \bigl[ \Omega_a\hat{c}^\dagger\hat{a} + \Omega_b\hat{c}^\dagger\hat{b} +  \Delta_1(\hat{c}^\dagger\hat{c} - \hat{a}^\dagger\hat{a})\\&\phantom{\hbar\bigl[}+ \Delta_2(\hat{c}^\dagger\hat{c} - \hat{b}^\dagger\hat{b}) \bigr].
\end{split}
\label{eq:appendix_ham}
\end{equation}
Including spontaneous emission to both ground states gives the master equation
\begin{equation}
\begin{split}
\diff{\rho}{t} ={}& -i[\hat{H},\rho] - \Gamma_1 ([\hat{c}^\dagger\hat{a},\hat{a}^\dagger\hat{c}\rho] + [\rho\hat{c}^\dagger\hat{a},\hat{a}^\dagger\hat{c}]) \\&-\Gamma_2 ([\hat{c}^\dagger\hat{b},\hat{b}^\dagger\hat{c}\rho] + [\rho\hat{c}^\dagger\hat{b},\hat{b}^\dagger\hat{c}]).
\end{split}
\label{eq:appendix_rho}
\end{equation}
A complete Fock state basis is formed by the states
\begin{equation}
\ket{m,n} = \frac{(\hat{a}^\dagger)^m(\hat{b}^\dagger)^{N-m-n}(\hat{c}^\dagger)^{N}}{\sqrt{m!n!(N-m-n)!}}\ket{0}.
\end{equation}
We have a density matrix
\begin{equation}
\rho = \sum_{m,n,m',n'} \rho_{m,n,m',n'} \ket{m,n}\bra{m',n'}.
\end{equation}
We may now obtain set of coupled differential equations in variables $  \rho_{m,n,m',n'} $ and solve them using standard numerical methods. Then the expectation values of relevant operators may be found as for instance
\begin{equation}
\begin{split}
\braket{\hat{a}^\dagger\hat{a}} =\textup{Tr}(\rho\hat{a}^\dagger\hat{a}) =\sum_{m,n}m\rho_{m,n,m,n}.
\end{split}
\end{equation}
This gives an exact solution of the problem.  

With the mean-field approach, one instead takes the trace of \eref{appendix_rho} multiplied by an operator $\hat{A}$
\begin{equation}
\begin{split}
\frac{d\braket{\hat{A}}}{dt} ={}& i\braket{[\hat{H}, \hat{A}]}\\
&- \Gamma_1\textup{Tr}\Bigl( \rho \bigl([\hat{A},\hat{c}^\dagger\hat{a}]\hat{a}^\dagger\hat{c} + \hat{c}^\dagger\hat{a}[\hat{a}^\dagger\hat{c}, \hat{A}] \bigr) \Bigr)\\
&- \Gamma_2\textup{Tr}\Bigl( \rho \bigl([\hat{A},\hat{c}^\dagger\hat{b}]\hat{b}^\dagger\hat{c} + \hat{c}^\dagger\hat{b}[\hat{b}^\dagger\hat{c}, \hat{A}] \bigr) \Bigr).
\end{split}\end{equation}
This gives a coupled set of equations in terms of the expectation values of the operators $\hat{a}^\dagger\hat{a}$, $\hat{a}^\dagger\hat{b}$, $\hat{a}^\dagger\hat{c}$, $\hat{b}^\dagger\hat{b}$, $\hat{b}^\dagger\hat{c}$, $\hat{c}^\dagger\hat{c}$. This gives fourth order products of bosonic operators for the spontaneous emisssion terms
\begin{equation}
\begin{split}
\diff{\braket{\hat{a}^\dagger\hat{a}}}{t} ={}& -2i\Omega_a\textup{Im}(\braket{\hat{a}^\dagger\hat{c}}) - 4\Gamma_1\braket{ \hat{c}^\dagger\hat{c}(\hat{a}^\dagger\hat{a}+1)}\\
\approx{}& -2i\Omega_a\textup{Im}(\braket{\hat{a}^\dagger\hat{c}})- 4\Gamma_1\braket{ \hat{c}^\dagger\hat{c}}(\braket{\hat{a}^\dagger\hat{a}}+1).
\end{split}
\end{equation}
The fourth order terms are approximated by a product of bilinear terms.  Note that the approximation is very good for spin coherent states as the variance of spin operators diminishes as $ 1/N $. This is why it is a very successful model.

A comparison of the two approaches in shown \figref{appendix_compare}. We perform a simulation where the two laser amplitudes of \eref{appendix_ham} are left constant for the duration of the simulation. We can only compare the two models for low values of $N$, as carrying out the simulation in the Fock state basis is numerically intensive when $N$ becomes large. Virtually perfect agreement is seen for the two curves.

\bibliographystyle{apsrev}
\bibliography{paperPRA}

\end{document}

%% file: LevelSchemeTotal.tex
\begingroup%
  \makeatletter%
  \providecommand\color[2][]{%
    \errmessage{(Inkscape) Color is used for the text in Inkscape, but the package 'color.sty' is not loaded}%
    \renewcommand\color[2][]{}%
  }%
  \providecommand\transparent[1]{%
    \errmessage{(Inkscape) Transparency is used (non-zero) for the text in Inkscape, but the package 'transparent.sty' is not loaded}%
    \renewcommand\transparent[1]{}%
  }%
  \providecommand\rotatebox[2]{#2}%
  \ifx\svgwidth\undefined%
    \setlength{\unitlength}{1327.46416016bp}%
    \ifx\svgscale\undefined%
      \relax%
    \else%
      \setlength{\unitlength}{\unitlength * \real{\svgscale}}%
    \fi%
  \else%
    \setlength{\unitlength}{\svgwidth}%
  \fi%
  \global\let\svgwidth\undefined%
  \global\let\svgscale\undefined%
  \makeatother%
  \begin{picture}(1,0.73584459)%
    \put(0,0){\includegraphics[width=\unitlength]{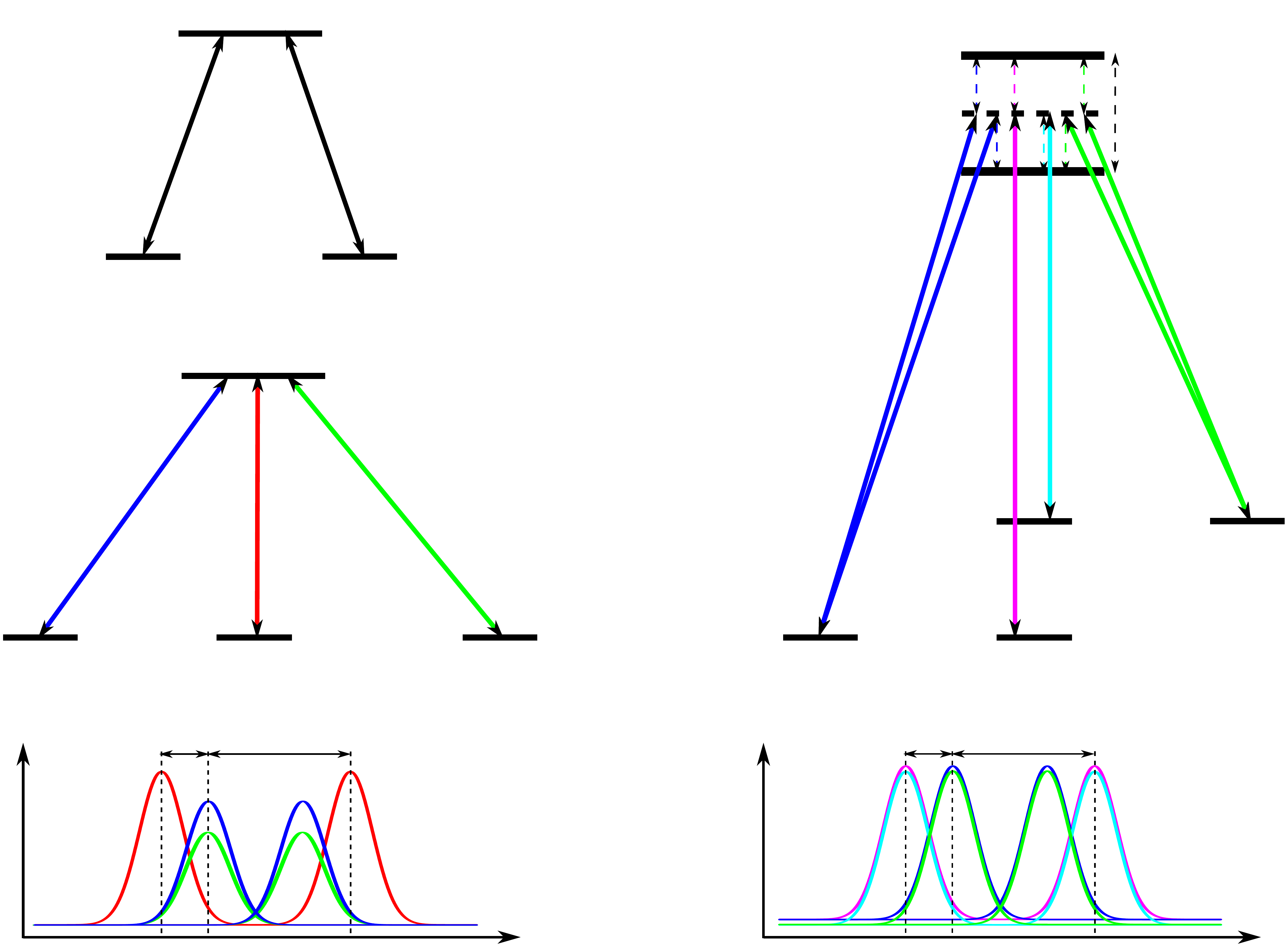}}%
    \put(0,0.73584459){\color[rgb]{0,0,0}\makebox(0,0)[lb]{\large\smash{a)}}}
    \put(0,0.46584459){\color[rgb]{0,0,0}\makebox(0,0)[lb]{\large\smash{b)}}}
    \put(0.55392612,0.73584459){\color[rgb]{0,0,0}\makebox(0,0)[lb]{\large\smash{c)}}}
    \put(0.55392612,0.60223619){\color[rgb]{0,0,0}\makebox(0,0)[lb]{\smash{$F'=1$}}}%
    \put(0.55392612,0.6926311){\color[rgb]{0,0,0}\makebox(0,0)[lb]{\smash{$F'=2$}}}%
    \put(0.55392612,0.33103941){\color[rgb]{0,0,0}\makebox(0,0)[lb]{\smash{$F=2$}}}%
    \put(0.55392612,0.247052){\color[rgb]{0,0,0}\makebox(0,0)[lb]{\smash{$F=1$}}}%
    \put(0.71777417,0.61729950){\color[rgb]{0,0,0}\makebox(0,0)[lb]{\smash{$\Delta_1$}}}%
    \put(0.71777417,0.66249846){\color[rgb]{0,0,0}\makebox(0,0)[lb]{\smash{$\Delta_2$}}}%
    \put(0.87628336,0.63749846){\color[rgb]{0,0,0}\makebox(0,0)[lb]{\smash{$\Delta_{12}$}}}%
    \put(0.90523478,0.48573709){\makebox(0,0)[lb]{\smash{$-\kappa_j{\color[rgb]{0,1,0}\Omega_b}$}}}%
    \put(0.69130708,0.48573709){\makebox(0,0)[lb]{\smash{${\color[rgb]{0,0,1}\Omega_a}$}}}%
    \put(0.70613735,0.38328611){\makebox(0,0)[lb]{\smash{$\kappa_j{\color[rgb]{0,0,1}\Omega_a}$}}}%
    \put(0.91114059,0.38328611){\makebox(0,0)[lb]{\smash{${\color[rgb]{0,1,0}\Omega_b}$}}}%
    \put(0.75392799,0.44355139){\makebox(0,0)[lb]{\smash{$\color[rgb]{1,0,1}\Omega_{c_1}$}}}%
    \put(0.82661864,0.44355139){\makebox(0,0)[lb]{\smash{$\color[rgb]{0,1,1}\Omega_{c_2}$}}}%
    \put(0.68042119,0.24174035){\color[rgb]{0,0,0}\makebox(0,0)[lb]{\smash{$\ket{a}$}}}%
    \put(0.90264519,0.33103941){\color[rgb]{0,0,0}\makebox(0,0)[lb]{\smash{$\ket{b}$}}}%
    \put(0.72987066,0.33103941){\color[rgb]{0,0,0}\makebox(0,0)[lb]{\smash{$\ket{c_1}$}}}%
    \put(0.84615072,0.24174035){\color[rgb]{0,0,0}\makebox(0,0)[lb]{\smash{$\ket{c_2}$}}}%
    \put(0.87628336,0.60223318){\color[rgb]{0,0,0}\makebox(0,0)[lb]{\smash{$\ket{e_1}$}}}%
    \put(0.87628336,0.6926311){\color[rgb]{0,0,0}\makebox(0,0)[lb]{\smash{$\ket{e_2}$}}}%
    \put(0.556292,0.1984558){\color[rgb]{0,0,0}\makebox(0,0)[lb]{\smash{$m_F$}}}%
    \put(0.63173754,0.1984558){\color[rgb]{0,0,0}\makebox(0,0)[lb]{\smash{$-1$}}}%
    \put(0.79707983,0.1984558){\color[rgb]{0,0,0}\makebox(0,0)[lb]{\smash{$0$}}}%
    \put(0.9632744,0.1984558){\color[rgb]{0,0,0}\makebox(0,0)[lb]{\smash{$1$}}}%
    \put(0.77668569,0.72242615){\color[rgb]{0,0,0}\makebox(0,0)[lb]{\smash{$m_F'=0$}}}%
    \put(0.06328684,0.60442552){\makebox(0,0)[lb]{\smash{$\Omega_a$}}}%
    \put(0.31080404,0.60442552){\makebox(0,0)[lb]{\smash{$\Omega_b$}}}%
    \put(0.0583808,0.53611212){\color[rgb]{0,0,0}\makebox(0,0)[lb]{\smash{$\ket{a}$}}}%
    \put(0.31049272,0.53611212){\color[rgb]{0,0,0}\makebox(0,0)[lb]{\smash{$\ket{b}$}}}%
    \put(0.25054827,0.70795413){\color[rgb]{0,0,0}\makebox(0,0)[lb]{\smash{$\ket{e}$}}}%
    \put(0.06328684,0.37442552){\color[rgb]{0,0,1}\makebox(0,0)[lb]{\smash{$\Omega_a$}}}%
    \put(0.31080404,0.37442552){\color[rgb]{0,1,0}\makebox(0,0)[lb]{\smash{$\Omega_b$}}}%
    \put(0.22129265,0.37442552){\color[rgb]{1,0,0}\makebox(0,0)[lb]{\smash{$\Omega_c$}}}%
    \put(0.0583808,0.24611212){\color[rgb]{0,0,0}\makebox(0,0)[lb]{\smash{$\ket{a}$}}}%
    \put(0.32049272,0.24611212){\color[rgb]{0,0,0}\makebox(0,0)[lb]{\smash{$\ket{b}$}}}%
    \put(0.22468971,0.24611212){\color[rgb]{0,0,0}\makebox(0,0)[lb]{\smash{$\ket{c}$}}}%
    \put(0.25054827,0.44795413){\color[rgb]{0,0,0}\makebox(0,0)[lb]{\smash{$\ket{e}$}}}%
    \put(0.00970876,0.16186585){\color[rgb]{0,0,0}\makebox(0,0)[lb]{\smash{$|\Omega|$}}}%
    \put(0.40934258,-0.00000001){\color[rgb]{0,0,0}\makebox(0,0)[lb]{\smash{$t$}}}%
    \put(0.06412484,0.1293851){\color[rgb]{1,0,0}\makebox(0,0)[lb]{\smash{$|\Omega_c|^{(1)}$}}}%
    \put(0.14847409,0.1193851){\color[rgb]{0,0,1}\makebox(0,0)[lb]{\smash{$|\Omega_a|^{(1)}$}}}%
    \put(0.14847409,0.0488219){\color[rgb]{0,1,0}\makebox(0,0)[lb]{\smash{$|\Omega_b|^{(1)}$}}}%
    \put(0.28412484,0.1293851){\color[rgb]{1,0,0}\makebox(0,0)[lb]{\smash{$|\Omega_c|^{(2)}$}}}%
    \put(0.20847409,0.1193851){\color[rgb]{0,0,1}\makebox(0,0)[lb]{\smash{$|\Omega_a|^{(2)}$}}}%
    \put(0.20847409,0.0488219){\color[rgb]{0,1,0}\makebox(0,0)[lb]{\smash{$|\Omega_b|^{(2)}$}}}%
    \put(0.58459408,0.16186585){\color[rgb]{0,0,0}\makebox(0,0)[lb]{\smash{$|\Omega|$}}}%
    \put(0.9842279,-0.00000001){\color[rgb]{0,0,0}\makebox(0,0)[lb]{\smash{$t$}}}%
    \put(0.6182504,0.12933247){\color[rgb]{1,0,1}\makebox(0,0)[lb]{\smash{$|\Omega_{c_1}|^{(1)}$}}}%
    \put(0.6182504,0.08933247){\color[rgb]{0,1,1}\makebox(0,0)[lb]{\smash{$|\Omega_{c_2}|^{(1)}$}}}%
    \put(0.71982008,0.17533247){\color[rgb]{0,0,1}\makebox(0,0)[lb]{\smash{$|\Omega_a|^{(1)}$}}}%
    \put(0.71982008,0.08933247){\color[rgb]{0,1,0}\makebox(0,0)[lb]{\smash{$|\Omega_b|^{(1)}$}}}%
    \put(0.78982008,0.17533247){\color[rgb]{0,0,1}\makebox(0,0)[lb]{\smash{$|\Omega_a|^{(2)}$}}}%
    \put(0.78982008,0.08933247){\color[rgb]{0,1,0}\makebox(0,0)[lb]{\smash{$|\Omega_b|^{(2)}$}}}%
    \put(0.8582504,0.12933247){\color[rgb]{1,0,1}\makebox(0,0)[lb]{\smash{$|\Omega_{c_1}|^{(2)}$}}}%
    \put(0.8582504,0.08933247){\color[rgb]{0,1,1}\makebox(0,0)[lb]{\smash{$|\Omega_{c_2}|^{(2)}$}}}%
    \put(0.719,0.156){\color[rgb]{0,0,0}\makebox(0,0)[lb]{\smash{$T_-$}}}%
    \put(0.765,0.156){\color[rgb]{0,0,0}\makebox(0,0)[lb]{\smash{$T_+$}}}%
    \put(0.13,0.156){\color[rgb]{0,0,0}\makebox(0,0)[lb]{\smash{$T_-$}}}%
    \put(0.19,0.156){\color[rgb]{0,0,0}\makebox(0,0)[lb]{\smash{$T_+$}}}%
  \end{picture}%
\endgroup%

%% file: paperPRAbspheres.tex
\newcommand\pgfmathsinandcos[3]{%
  \pgfmathsetmacro#1{sin(#3)}%
  \pgfmathsetmacro#2{cos(#3)}%
}
\newcommand\LongitudePlane[3][current plane]{%
  \pgfmathsinandcos\sinEl\cosEl{#2} 
  \pgfmathsinandcos\sint\cost{#3} 
  \tikzset{#1/.style={cm={\cost,\sint*\sinEl,0,\cosEl,(0,0)}}}
}
\newcommand\LatitudePlane[3][current plane]{%
  \pgfmathsinandcos\sinEl\cosEl{#2} 
  \pgfmathsinandcos\sint\cost{#3} 
  \pgfmathsetmacro\yshift{\cosEl*\sint}
  \tikzset{#1/.style={cm={\cost,0,0,\cost*\sinEl,(0,\yshift)}}} %
}
\newcommand\DrawLongitudeCircle[2][1]{
  \LongitudePlane{\angEl}{#2}
  \tikzset{current plane/.prefix style={scale=#1}}
  \pgfmathsetmacro\angVis{atan(sin(#2)*cos(\angEl)/sin(\angEl))} %
  \draw[current plane, dotted, very thin] (\angVis:1) arc (\angVis:\angVis+180:1);
}
\newcommand\DrawLatitudeCircle[2][1]{
  \LatitudePlane{\angEl}{#2}
  \tikzset{current plane/.prefix style={scale=#1}}
  \pgfmathsetmacro\sinVis{sin(#2)/cos(#2)*sin(\angEl)/cos(\angEl)}
  \pgfmathsetmacro\angVis{asin(min(1,max(\sinVis,-1)))}
  \draw[current plane, dotted, very thin] (\angVis:1) arc (\angVis:-\angVis-180:1);
}
\newcommand\DrawLatitudeCircleFull[2][1]{
  \LatitudePlane{\angEl}{#2}
  \tikzset{current plane/.prefix style={scale=#1}}
  \pgfmathsetmacro\sinVis{sin(#2)/cos(#2)*sin(\angEl)/cos(\angEl)}
  \pgfmathsetmacro\angVis{asin(min(1,max(\sinVis,-1)))}
  \draw[current plane] (\angVis:1) arc (\angVis:-\angVis-180:1);
}

\begin{tikzpicture} 

\def\R{3} 
\def\r{2}
\def\Lsep{10cm}
\def\angEl{35} 
\def\angAz{-45} 
\filldraw[ball color=white] (0,0) circle (\R);
\foreach \t in {-75,-60,...,75} { \DrawLatitudeCircle[\R]{\t} }
\foreach \t in {15,45,...,165} { \DrawLongitudeCircle[\R]{\t} }
\pgfmathsetmacro\H{\R*cos(\angEl)}
\pgfmathsetmacro\X{\R*cos(\angAz)}
\pgfmathsetmacro\Y{\R*sin(\angAz)*sin(\angEl)}

\coordinate (P0) at ({\R*sin(225)*cos(80)},{\R*sin(\angEl)*cos(225)*cos(80)+\R*cos(\angEl)*sin(80)});
\coordinate (P1) at ({\R*sin(135)*cos(10)},{\R*sin(\angEl)*cos(135)*cos(10)-\R*cos(\angEl)*sin(10)});
\coordinate (P2) at ({\R*sin(135)*cos(15)},{\R*sin(\angEl)*cos(135)*cos(15)+\R*cos(\angEl)*sin(15)});

\pgfmathsetmacro\xyscale{sqrt(cos(\angAz)^2+sin(\angAz)^2*sin(\angEl)^2)}

\draw[->,dashed] (0,0) -- (0,\H);
\draw[->,dashed] (0,0) -- (\X,\Y);
\draw[->,dashed] (0,0) -- (-\X,\Y);

\node[above] at (0,\H) {$\braket{\hat{S}_z}$};
\node[right] at (\X,\Y) {$\braket{\hat{S}_y}$};
\node[left] at (-\X,\Y) {$\braket{\hat{S}_x}$};

\draw [red,fill = red]
	plot [domain = 0:360, samples = 30, variable = \t]
	({\R*sin(225+2*\r*cos \t)*cos(80+\r*sin \t)},{\R*sin(\angEl)*cos(225+2*\r*cos \t)*cos(80+\r*sin \t)+\R*cos(\angEl)*sin(80+\r*sin \t)})
	-- cycle;

\draw [blue, thick, dotted]
	plot [domain = 0:360, samples = 30, variable = \t]
	({\R*sin(135+\r*cos \t)*cos(10+\r*sin \t)},{\R*sin(\angEl)*cos(135+\r*cos \t)*cos(10+\r*sin \t)-\R*cos(\angEl)*sin(10+\r*sin \t)})
	-- cycle;

\draw [red,fill = red]
	plot [domain = 0:360, samples = 30, variable = \t]
	({\R*sin(135+\r*cos \t)*cos(15+\r*sin \t)},{\R*sin(\angEl)*cos(135+\r*cos \t)*cos(15+\r*sin \t)+\R*cos(\angEl)*sin(15+\r*sin \t)})
	-- cycle;

    \draw [very thick,red,->] plot [domain = 5:87, samples = 60, variable = \k]
    			({\R*sin (225-\k)*cos (80-13*\k/18)}, {\R*sin(\angEl)*cos (225-\k)*cos (80-13*\k/18)+\R*cos(\angEl)*sin (80-13*\k/18)}) node[above left=10pt] {$U$};
    \draw [dashed,thick,blue,->] plot [domain = 5:87, samples = 60, variable = \k]
    			({\R*sin (225-\k)*cos (80-\k)}, {\R*sin(\angEl)*cos (225)*cos (80-\k)+\R*cos(\angEl)*sin (80-\k)}) node[midway, above left] {$\mO^{(1)}$};
    \draw [dashed,thick,blue,->] plot [domain = 7:83, samples = 60, variable = \k]
    			({\R*sin (135)*cos (-10+\k*5/18)}, {\R*sin(\angEl)*cos (135)*cos (-10+\k*5/18)+\R*cos(\angEl)*sin (-10+\k*5/18)}) node[below left] {$\mO^{(2)}$};

\begin{scope}[xshift = \Lsep]

\filldraw[ball color=white] (0,0) circle (\R);
\foreach \t in {-75,-60,...,75} { \DrawLatitudeCircle[\R]{\t} }
\foreach \t in {15,45,...,165} { \DrawLongitudeCircle[\R]{\t} }

\draw[->,dashed] (0,0) -- (0,\H);
\draw[->,dashed] (0,0) -- (\X,\Y);
\draw[->,dashed] (0,0) -- (-\X,\Y);

\node[above] at (0,\H) {$\braket{\hat{J}_z}$};
\node[right] at (\X,\Y) {$\braket{\hat{J}_y}$};
\node[left] at (-\X,\Y) {$\braket{\hat{J}_x}$};

\draw [red,thick,dotted]
	plot [domain = 0:360, samples = 30, variable = \t]
	({\R*sin(225+2*\r*cos \t)*cos(80+\r*sin \t)},{\R*sin(\angEl)*cos(225+2*\r*cos \t)*cos(80+\r*sin \t)+\R*cos(\angEl)*sin(80+\r*sin \t)})
	-- cycle;

\draw [blue,fill = blue]
	plot [domain = 0:360, samples = 30, variable = \t]
	({\R*sin(135+\r*cos \t)*cos(10+\r*sin \t)},{\R*sin(\angEl)*cos(135+\r*cos \t)*cos(10+\r*sin \t)-\R*cos(\angEl)*sin(10+\r*sin \t)})
	-- cycle;

\draw [red,thick,dotted]
	plot [domain = 0:360, samples = 30, variable = \t]
	({\R*sin(135+\r*cos \t)*cos(15+\r*sin \t)},{\R*sin(\angEl)*cos(135+\r*cos \t)*cos(15+\r*sin \t)+\R*cos(\angEl)*sin(15+\r*sin \t)})
	-- cycle;

    \draw [dashed,very thick,red,->] plot [domain = 5:87, samples = 60, variable = \k]
    			({\R*sin (225-\k)*cos (80-13*\k/18)}, {\R*sin(\angEl)*cos (225-\k)*cos (80-13*\k/18)+\R*cos(\angEl)*sin (80-13*\k/18)}) node[above left=10pt] {$U$};
    \draw [thick,blue,->] plot [domain = 5:87, samples = 60, variable = \k]
    			({\R*sin (225-\k)*cos (80-\k)}, {\R*sin(\angEl)*cos (225)*cos (80-\k)+\R*cos(\angEl)*sin (80-\k)}) node[midway, above left] {$\mO^{(1)}$};
    \draw [thick,blue,->] plot [domain = 7:83, samples = 60, variable = \k]
    			({\R*sin (135)*cos (-10+\k*5/18)}, {\R*sin(\angEl)*cos (135)*cos (-10+\k*5/18)+\R*cos(\angEl)*sin (-10+\k*5/18)}) node[below left] {$\mO^{(2)}$};

\end{scope}
\draw[very thick,magenta,->,decorate, decoration={snake,post length=3mm}] ($(P0) + (-30:2mm)$) .. controls +(-30:3cm) and +(160:1cm) .. ($(P1) + (\Lsep,0) + (160:2mm)$) node[midway, above=2mm, sloped] {$\Omega_{a,b,c_1,c_2}$};
\draw[very thick,magenta,->,decorate, decoration={snake,post length=3mm}] ($(P1) + (\Lsep,0) + (-165:2mm)$) .. controls +(-165:2cm) and +(-15:2cm) .. ($(P2) + (-15:2mm)$) node[near end, above=2mm, sloped] {$\Omega_{a,b,c_1,c_2}$};

\node[red,above left] (psi0) at (P0) {$\dket{\alpha_0,\beta_0}$};
\node[xshift=\Lsep,blue,below right] (psi1) at (P1) {$\dket{\alpha_1,\beta_1}$};
\node[red,above right] (psi2) at (P2) {$\dket{\alpha_2,\beta_2}$};

\end{tikzpicture}

%% file: stirap.tex
\begingroup%
  \makeatletter%
  \providecommand\color[2][]{%
    \errmessage{(Inkscape) Color is used for the text in Inkscape, but the package 'color.sty' is not loaded}%
    \renewcommand\color[2][]{}%
  }%
  \providecommand\transparent[1]{%
    \errmessage{(Inkscape) Transparency is used (non-zero) for the text in Inkscape, but the package 'transparent.sty' is not loaded}%
    \renewcommand\transparent[1]{}%
  }%
  \providecommand\rotatebox[2]{#2}%
  \ifx\svgwidth\undefined%
    \setlength{\unitlength}{611.61494141bp}%
    \ifx\svgscale\undefined%
      \relax%
    \else%
      \setlength{\unitlength}{\unitlength * \real{\svgscale}}%
    \fi%
  \else%
    \setlength{\unitlength}{\svgwidth}%
  \fi%
  \global\let\svgwidth\undefined%
  \global\let\svgscale\undefined%
  \makeatother%
  \begin{picture}(1,0.76022128)%
    \put(0.02,0){\includegraphics[width=\unitlength]{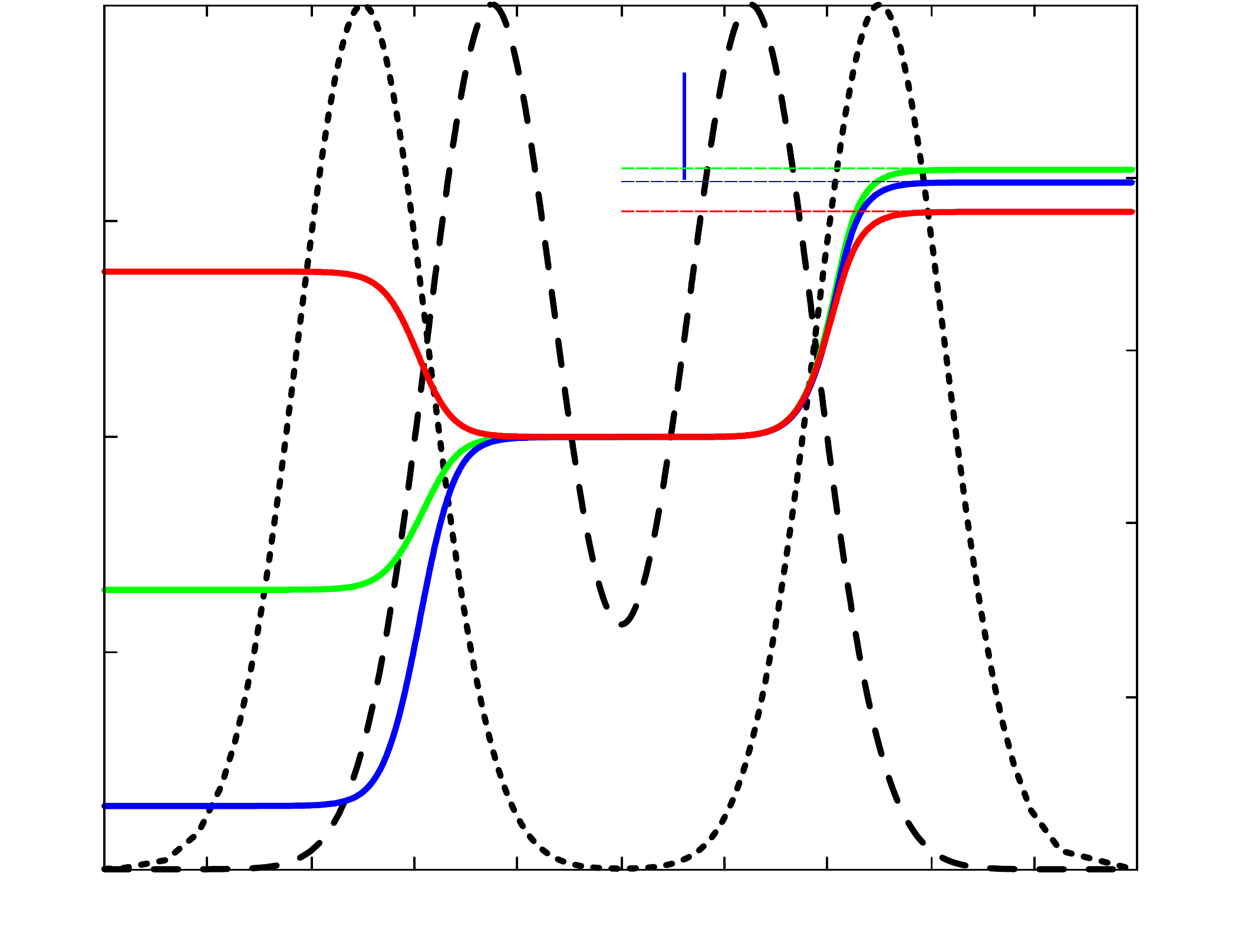}}%
\small
    \put(0.11,0.7){\color[rgb]{0,0,0}\makebox(0,0)[lb]{\large\smash{a)}}}
    \put(0.01335671,0.05877479){\color[rgb]{0,0,0}\makebox(0,0)[lb]{\smash{$-1.0$}}}%
    \put(0.01335671,0.23098747){\color[rgb]{0,0,0}\makebox(0,0)[lb]{\smash{$-0.5$}}}%
    \put(0.03335671,0.40320014){\color[rgb]{0,0,0}\makebox(0,0)[lb]{\smash{$0.0$}}}%
    \put(0.03335671,0.57705274){\color[rgb]{0,0,0}\makebox(0,0)[lb]{\smash{$0.5$}}}%
    \put(0.03335671,0.74926575){\color[rgb]{0,0,0}\makebox(0,0)[lb]{\smash{$1.0$}}}%
    \put(0.06861937,0.01581266){\color[rgb]{0,0,0}\makebox(0,0)[lb]{\smash{$0.0$}}}%
    \put(0.15062542,0.01581266){\color[rgb]{0,0,0}\makebox(0,0)[lb]{\smash{$0.1$}}}%
    \put(0.23263114,0.01581266){\color[rgb]{0,0,0}\makebox(0,0)[lb]{\smash{$0.2$}}}%
    \put(0.31627842,0.01581266){\color[rgb]{0,0,0}\makebox(0,0)[lb]{\smash{$0.3$}}}%
    \put(0.39828415,0.01581266){\color[rgb]{0,0,0}\makebox(0,0)[lb]{\smash{$0.4$}}}%
    \put(0.48028987,0.01581266){\color[rgb]{0,0,0}\makebox(0,0)[lb]{\smash{$0.5$}}}%
    \put(0.56393715,0.01581266){\color[rgb]{0,0,0}\makebox(0,0)[lb]{\smash{$0.6$}}}%
    \put(0.64594288,0.01581266){\color[rgb]{0,0,0}\makebox(0,0)[lb]{\smash{$0.7$}}}%
    \put(0.72794860,0.01581266){\color[rgb]{0,0,0}\makebox(0,0)[lb]{\smash{$0.8$}}}%
    \put(0.81159425,0.01581266){\color[rgb]{0,0,0}\makebox(0,0)[lb]{\smash{$0.9$}}}%
    \put(0.89278083,0.01581266){\color[rgb]{0,0,0}\makebox(0,0)[lb]{\smash{$1.0$}}}%
    \put(0.92230263,0.05877479){\color[rgb]{0,0,0}\makebox(0,0)[lb]{\smash{   $0$}}}%
    \put(0.92230263,0.19654428){\color[rgb]{0,0,0}\makebox(0,0)[lb]{\smash{ $200$}}}%
    \put(0.92230263,0.3343154){\color[rgb]{0,0,0}\makebox(0,0)[lb]{\smash{ $400$}}}%
    \put(0.92230263,0.47208489){\color[rgb]{0,0,0}\makebox(0,0)[lb]{\smash{ $600$}}}%
    \put(0.92230263,0.61149594){\color[rgb]{0,0,0}\makebox(0,0)[lb]{\smash{ $800$}}}%
    \put(0.92266271,0.74926575){\color[rgb]{0,0,0}\makebox(0,0)[lb]{\smash{$1000$}}}%
    \put(0.01121504,0.35809667){\color[rgb]{0,0,0}\rotatebox{90}{\makebox(0,0)[lb]{\smash{$\braket{\hat{S}_j}/N$}}}}%
    \put(1.03692954,0.37531826){\color[rgb]{0,0,0}\rotatebox{90}{\makebox(0,0)[lb]{\smash{$|\Omega_j|/\Gamma$}}}}%
    \put(0.47290942,-0.01136947){\color[rgb]{0,0,0}\makebox(0,0)[lb]{\smash{$t/\tau$}}}%
    \put(0.17092669,0.58404036){\color[rgb]{0,0,0}\makebox(0,0)[lb]{\smash{$|\Omega_{c_{1,2}}|$}}}%
    \put(0.38082162,0.58404036){\color[rgb]{0,0,0}\makebox(0,0)[lb]{\smash{$|\Omega_{a,b}|$}}}%
    \put(0.11799862,0.30448237){\color[rgb]{0,0,0}\makebox(0,0)[lb]{\textcolor{green}{\smash{$\braket{\hat{S}_x}$}}}}%
    \put(0.11799862,0.1310333){\color[rgb]{0,0,0}\makebox(0,0)[lb]{\textcolor{blue}{\smash{$\braket{\hat{S}_y}$}}}}%
    \put(0.11799862,0.55821343){\color[rgb]{0,0,0}\makebox(0,0)[lb]{\textcolor{red}{\smash{$\braket{\hat{S}_z}$}}}}%
    \put(0.48842359,0.63831599){\color[rgb]{0,1,0}\makebox(0,0)[lb]{\small\smash{$\braket{\hat{S}_x}_\textup{targ}$}}}%
    \put(0.53726597,0.71418077){\color[rgb]{0,0,1}\makebox(0,0)[lb]{\small\smash{$\braket{\hat{S}_y}_\textup{targ}$}}}%
    \put(0.48842359,0.54693212){\color[rgb]{1,0,0}\makebox(0,0)[lb]{\small\smash{$\braket{\hat{S}_z}_\textup{targ}$}}}%
  \end{picture}%
\endgroup%

%% file: stirapfail.tex
\begingroup%
  \makeatletter%
  \providecommand\color[2][]{%
    \errmessage{(Inkscape) Color is used for the text in Inkscape, but the package 'color.sty' is not loaded}%
    \renewcommand\color[2][]{}%
  }%
  \providecommand\transparent[1]{%
    \errmessage{(Inkscape) Transparency is used (non-zero) for the text in Inkscape, but the package 'transparent.sty' is not loaded}%
    \renewcommand\transparent[1]{}%
  }%
  \providecommand\rotatebox[2]{#2}%
  \ifx\svgwidth\undefined%
    \setlength{\unitlength}{611.61494141bp}%
    \ifx\svgscale\undefined%
      \relax%
    \else%
      \setlength{\unitlength}{\unitlength * \real{\svgscale}}%
    \fi%
  \else%
    \setlength{\unitlength}{\svgwidth}%
  \fi%
  \global\let\svgwidth\undefined%
  \global\let\svgscale\undefined%
  \makeatother%
  \begin{picture}(1,0.76022128)%
    \put(0.02,0){\includegraphics[width=\unitlength]{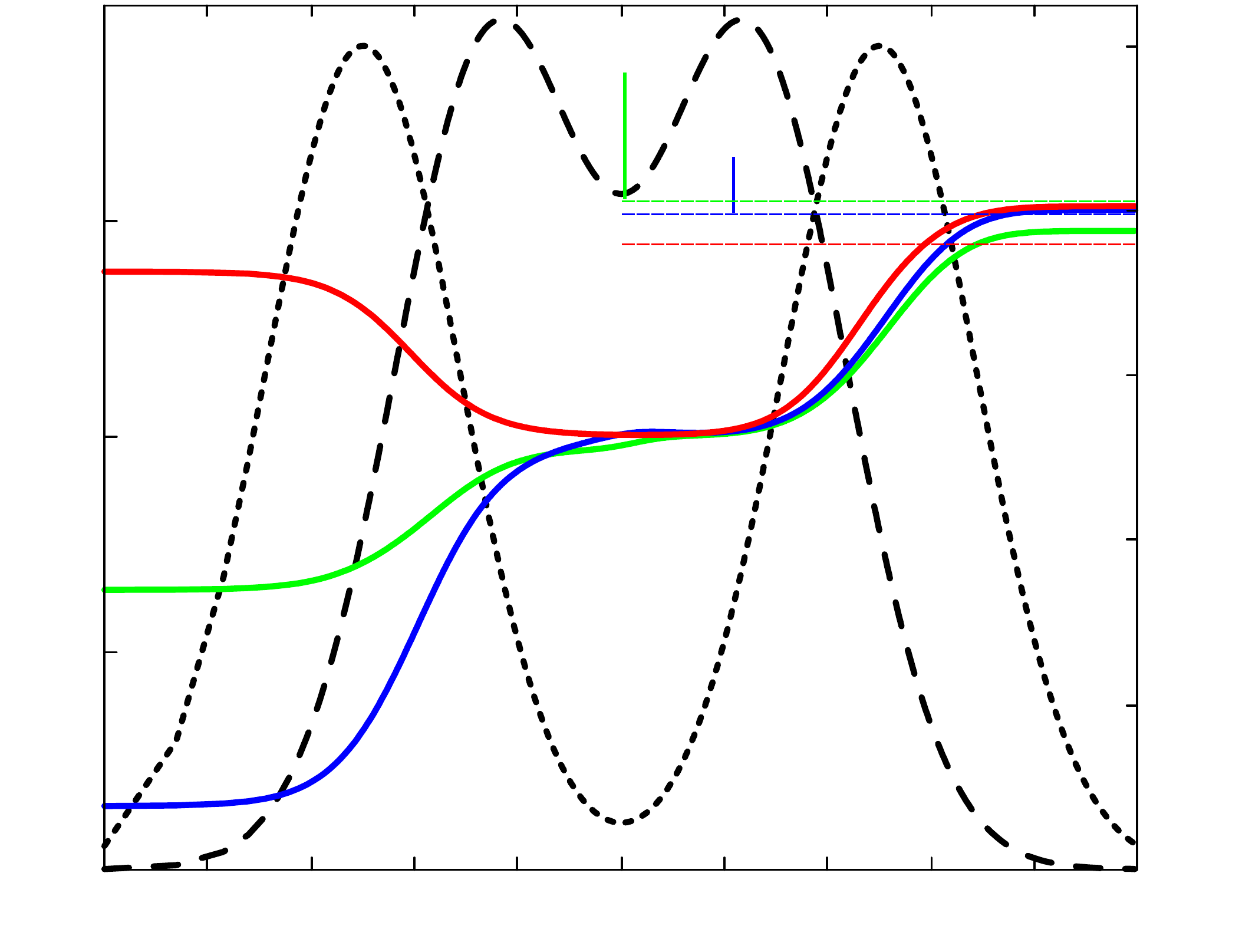}}%
\small
    \put(0.11,0.7){\color[rgb]{0,0,0}\makebox(0,0)[lb]{\large\smash{b)}}}
    \put(0.01335671,0.05877479){\color[rgb]{0,0,0}\makebox(0,0)[lb]{\smash{$-1.0$}}}%
    \put(0.01335671,0.23098747){\color[rgb]{0,0,0}\makebox(0,0)[lb]{\smash{$-0.5$}}}%
    \put(0.03335671,0.40320014){\color[rgb]{0,0,0}\makebox(0,0)[lb]{\smash{$0.0$}}}%
    \put(0.03335671,0.57705274){\color[rgb]{0,0,0}\makebox(0,0)[lb]{\smash{$0.5$}}}%
    \put(0.03335671,0.74926575){\color[rgb]{0,0,0}\makebox(0,0)[lb]{\smash{$1.0$}}}%
    \put(0.06861937,0.01581266){\color[rgb]{0,0,0}\makebox(0,0)[lb]{\smash{$0.0$}}}%
    \put(0.15062542,0.01581266){\color[rgb]{0,0,0}\makebox(0,0)[lb]{\smash{$0.1$}}}%
    \put(0.23263114,0.01581266){\color[rgb]{0,0,0}\makebox(0,0)[lb]{\smash{$0.2$}}}%
    \put(0.31627842,0.01581266){\color[rgb]{0,0,0}\makebox(0,0)[lb]{\smash{$0.3$}}}%
    \put(0.39828415,0.01581266){\color[rgb]{0,0,0}\makebox(0,0)[lb]{\smash{$0.4$}}}%
    \put(0.48028987,0.01581266){\color[rgb]{0,0,0}\makebox(0,0)[lb]{\smash{$0.5$}}}%
    \put(0.56393715,0.01581266){\color[rgb]{0,0,0}\makebox(0,0)[lb]{\smash{$0.6$}}}%
    \put(0.64594288,0.01581266){\color[rgb]{0,0,0}\makebox(0,0)[lb]{\smash{$0.7$}}}%
    \put(0.72794860,0.01581266){\color[rgb]{0,0,0}\makebox(0,0)[lb]{\smash{$0.8$}}}%
    \put(0.81159425,0.01581266){\color[rgb]{0,0,0}\makebox(0,0)[lb]{\smash{$0.9$}}}%
    \put(0.89278083,0.01581266){\color[rgb]{0,0,0}\makebox(0,0)[lb]{\smash{$1.0$}}}%
    \put(0.92230263,0.05877479){\color[rgb]{0,0,0}\makebox(0,0)[lb]{\smash{   $0$}}}%
    \put(0.92230263,0.1899846){\color[rgb]{0,0,0}\makebox(0,0)[lb]{\smash{ $200$}}}%
    \put(0.92230263,0.32119442){\color[rgb]{0,0,0}\makebox(0,0)[lb]{\smash{ $400$}}}%
    \put(0.92230263,0.45240423){\color[rgb]{0,0,0}\makebox(0,0)[lb]{\smash{ $600$}}}%
    \put(0.92230263,0.58525397){\color[rgb]{0,0,0}\makebox(0,0)[lb]{\smash{ $800$}}}%
    \put(0.92266271,0.71646330){\color[rgb]{0,0,0}\makebox(0,0)[lb]{\smash{$1000$}}}%
    \put(0.01121504,0.35809667){\color[rgb]{0,0,0}\rotatebox{90}{\makebox(0,0)[lb]{\smash{$\braket{\hat{S}_j}/N$}}}}%
    \put(1.03692954,0.37531826){\color[rgb]{0,0,0}\rotatebox{90}{\makebox(0,0)[lb]{\smash{$|\Omega_j|/\Gamma$}}}}%
    \put(0.47290942,-0.01136947){\color[rgb]{0,0,0}\makebox(0,0)[lb]{\smash{$t/\tau$}}}%
    \put(0.15592669,0.60404036){\color[rgb]{0,0,0}\makebox(0,0)[lb]{\smash{$|\Omega_{c_{1,2}}|$}}}%
    \put(0.38082162,0.60404036){\color[rgb]{0,0,0}\makebox(0,0)[lb]{\smash{$|\Omega_{a,b}|$}}}%
    \put(0.11799862,0.30448237){\color[rgb]{0,0,0}\makebox(0,0)[lb]{\textcolor{green}{\smash{$\braket{\hat{S}_x}$}}}}%
    \put(0.11799862,0.1310333){\color[rgb]{0,0,0}\makebox(0,0)[lb]{\textcolor{blue}{\smash{$\braket{\hat{S}_y}$}}}}%
    \put(0.11799862,0.55821343){\color[rgb]{0,0,0}\makebox(0,0)[lb]{\textcolor{red}{\smash{$\braket{\hat{S}_z}$}}}}%
    \put(0.48866404,0.71215423){\color[rgb]{0,1,0}\makebox(0,0)[lb]{\small\smash{$\braket{\hat{S}_x}_\textup{targ}$}}}%
    \put(0.57674689,0.64413755){\color[rgb]{0,0,1}\makebox(0,0)[lb]{\small\smash{$\braket{\hat{S}_y}_\textup{targ}$}}}%
    \put(0.48866404,0.51693212){\color[rgb]{1,0,0}\makebox(0,0)[lb]{\small\smash{$\braket{\hat{S}_z}_\textup{targ}$}}}%
\end{picture}%
\endgroup%

%% file: stirap2.tex
\begingroup%
  \makeatletter%
  \providecommand\color[2][]{%
    \errmessage{(Inkscape) Color is used for the text in Inkscape, but the package 'color.sty' is not loaded}%
    \renewcommand\color[2][]{}%
  }%
  \providecommand\transparent[1]{%
    \errmessage{(Inkscape) Transparency is used (non-zero) for the text in Inkscape, but the package 'transparent.sty' is not loaded}%
    \renewcommand\transparent[1]{}%
  }%
  \providecommand\rotatebox[2]{#2}%
  \ifx\svgwidth\undefined%
    \setlength{\unitlength}{611.61494141bp}%
    \ifx\svgscale\undefined%
      \relax%
    \else%
      \setlength{\unitlength}{\unitlength * \real{\svgscale}}%
    \fi%
  \else%
    \setlength{\unitlength}{\svgwidth}%
  \fi%
  \global\let\svgwidth\undefined%
  \global\let\svgscale\undefined%
  \makeatother%
  \begin{picture}(1,0.76022128)%
    \put(0.02,0){\includegraphics[width=\unitlength]{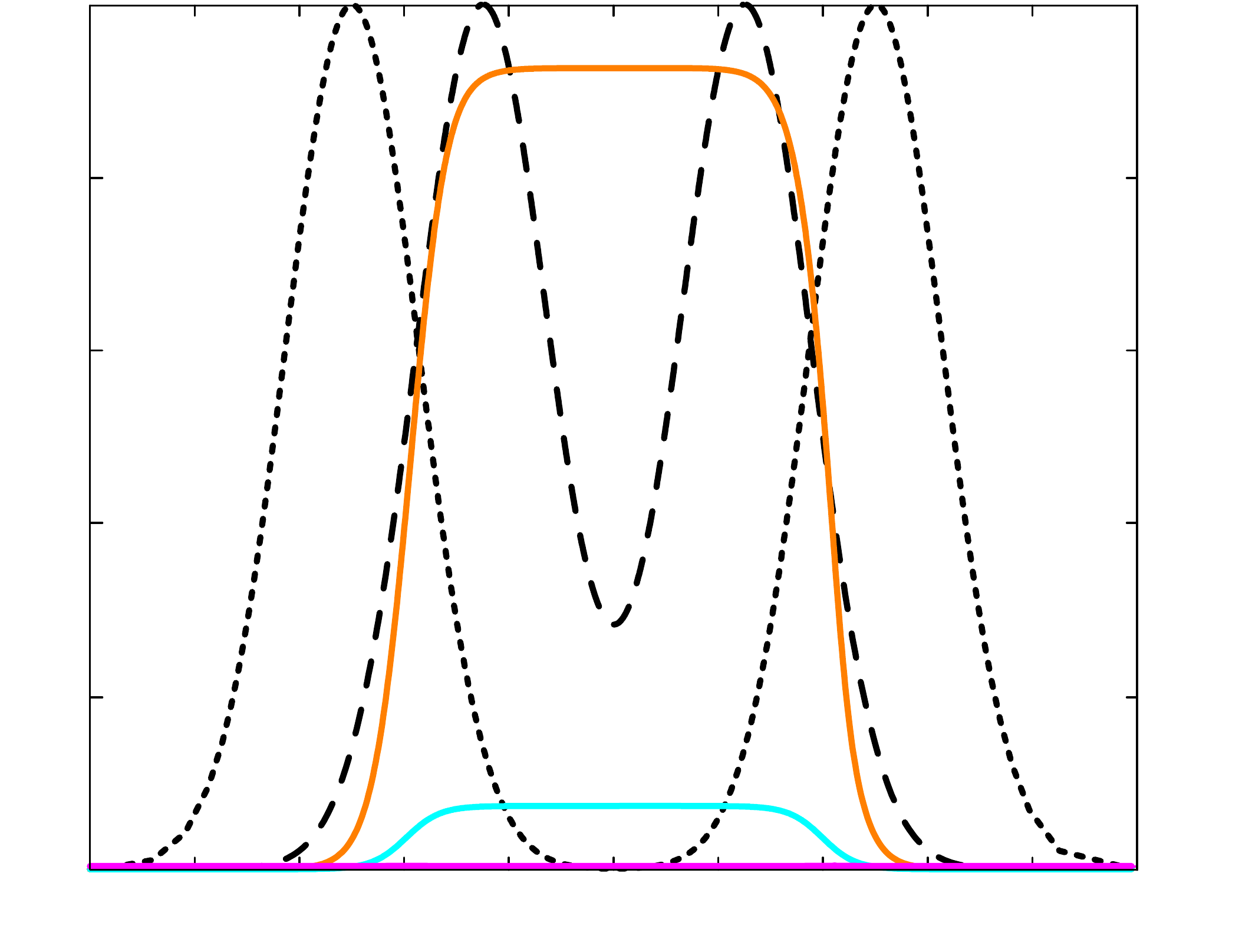}}%
\small
    \put(0.1,0.7){\color[rgb]{0,0,0}\makebox(0,0)[lb]{\large\smash{c)}}}
    \put(0.03335671,0.05877479){\color[rgb]{0,0,0}\makebox(0,0)[lb]{\smash{$0.0$}}}%
    \put(0.03335671,0.19654428){\color[rgb]{0,0,0}\makebox(0,0)[lb]{\smash{$0.2$}}}%
    \put(0.03335671,0.3343154){\color[rgb]{0,0,0}\makebox(0,0)[lb]{\smash{$0.4$}}}%
    \put(0.03335671,0.47208489){\color[rgb]{0,0,0}\makebox(0,0)[lb]{\smash{$0.6$}}}%
    \put(0.03335671,0.61149594){\color[rgb]{0,0,0}\makebox(0,0)[lb]{\smash{$0.8$}}}%
    \put(0.03335671,0.74926575){\color[rgb]{0,0,0}\makebox(0,0)[lb]{\smash{$1.0$}}}%
    \put(0.06861937,0.01581266){\color[rgb]{0,0,0}\makebox(0,0)[lb]{\smash{$0.0$}}}%
    \put(0.15062542,0.01581266){\color[rgb]{0,0,0}\makebox(0,0)[lb]{\smash{$0.1$}}}%
    \put(0.23263114,0.01581266){\color[rgb]{0,0,0}\makebox(0,0)[lb]{\smash{$0.2$}}}%
    \put(0.31627842,0.01581266){\color[rgb]{0,0,0}\makebox(0,0)[lb]{\smash{$0.3$}}}%
    \put(0.39828415,0.01581266){\color[rgb]{0,0,0}\makebox(0,0)[lb]{\smash{$0.4$}}}%
    \put(0.48028987,0.01581266){\color[rgb]{0,0,0}\makebox(0,0)[lb]{\smash{$0.5$}}}%
    \put(0.56393715,0.01581266){\color[rgb]{0,0,0}\makebox(0,0)[lb]{\smash{$0.6$}}}%
    \put(0.64594288,0.01581266){\color[rgb]{0,0,0}\makebox(0,0)[lb]{\smash{$0.7$}}}%
    \put(0.72794860,0.01581266){\color[rgb]{0,0,0}\makebox(0,0)[lb]{\smash{$0.8$}}}%
    \put(0.81159425,0.01581266){\color[rgb]{0,0,0}\makebox(0,0)[lb]{\smash{$0.9$}}}%
    \put(0.89278083,0.01581266){\color[rgb]{0,0,0}\makebox(0,0)[lb]{\smash{$1.0$}}}%
    \put(0.92230263,0.05877479){\color[rgb]{0,0,0}\makebox(0,0)[lb]{\smash{   $0$}}}%
    \put(0.92230263,0.19654428){\color[rgb]{0,0,0}\makebox(0,0)[lb]{\smash{ $200$}}}%
    \put(0.92230263,0.3343154){\color[rgb]{0,0,0}\makebox(0,0)[lb]{\smash{ $400$}}}%
    \put(0.92230263,0.47208489){\color[rgb]{0,0,0}\makebox(0,0)[lb]{\smash{ $600$}}}%
    \put(0.92230263,0.61149594){\color[rgb]{0,0,0}\makebox(0,0)[lb]{\smash{ $800$}}}%
    \put(0.92266271,0.74926575){\color[rgb]{0,0,0}\makebox(0,0)[lb]{\smash{$1000$}}}%
    \put(0.01121504,0.35809667){\color[rgb]{0,0,0}\rotatebox{90}{\makebox(0,0)[lb]{\smash{$\braket{\hat{n}_j}/N$}}}}%
    \put(1.03692954,0.37531826){\color[rgb]{0,0,0}\rotatebox{90}{\makebox(0,0)[lb]{\smash{$|\Omega_j|/\Gamma$}}}}%
    \put(0.47290942,-0.01136947){\color[rgb]{0,0,0}\makebox(0,0)[lb]{\smash{$t/\tau$}}}%
    \put(0.17092669,0.60404036){\color[rgb]{0,0,0}\makebox(0,0)[lb]{\smash{$|\Omega_{c_{1,2}}|$}}}%
    \put(0.50082162,0.60404036){\color[rgb]{0,0,0}\makebox(0,0)[lb]{\smash{$|\Omega_{a,b}|$}}}%
    \put(0.4702566,0.66108496){\color[rgb]{1,0.5,0}\makebox(0,0)[lb]{\smash{$\braket{\hat{n}_{c_1}}$}}}%
    \put(0.4702566,0.13099257){\color[rgb]{0,1,1}\makebox(0,0)[lb]{\smash{$\braket{\hat{n}_{c_2}}$}}}%
    \put(0.4702566,0.08030945){\color[rgb]{1,0,1}\makebox(0,0)[lb]{\smash{$\braket{\hat{n}_{e_{1,2}}}$}}}%
  \end{picture}%
\endgroup%

%% file: stirapfail2.tex
\begingroup%
  \makeatletter%
  \providecommand\color[2][]{%
    \errmessage{(Inkscape) Color is used for the text in Inkscape, but the package 'color.sty' is not loaded}%
    \renewcommand\color[2][]{}%
  }%
  \providecommand\transparent[1]{%
    \errmessage{(Inkscape) Transparency is used (non-zero) for the text in Inkscape, but the package 'transparent.sty' is not loaded}%
    \renewcommand\transparent[1]{}%
  }%
  \providecommand\rotatebox[2]{#2}%
  \ifx\svgwidth\undefined%
    \setlength{\unitlength}{611.61494141bp}%
    \ifx\svgscale\undefined%
      \relax%
    \else%
      \setlength{\unitlength}{\unitlength * \real{\svgscale}}%
    \fi%
  \else%
    \setlength{\unitlength}{\svgwidth}%
  \fi%
  \global\let\svgwidth\undefined%
  \global\let\svgscale\undefined%
  \makeatother%
  \begin{picture}(1,0.76022128)%
    \put(0.02,0){\includegraphics[width=\unitlength]{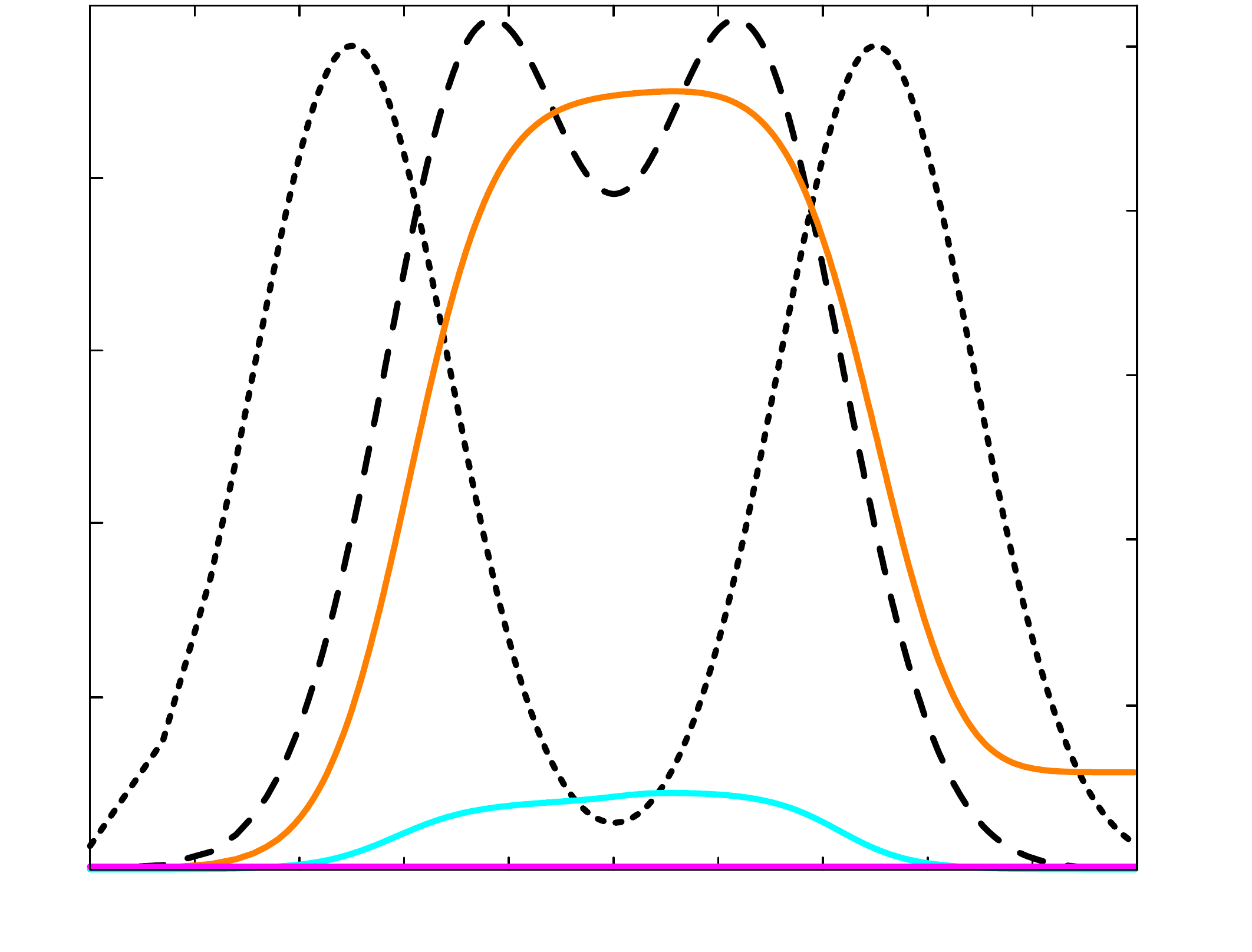}}%
\small
    \put(0.1,0.7){\color[rgb]{0,0,0}\makebox(0,0)[lb]{\large\smash{d)}}}
    \put(0.03335671,0.05877479){\color[rgb]{0,0,0}\makebox(0,0)[lb]{\smash{$0.0$}}}%
    \put(0.03335671,0.19654428){\color[rgb]{0,0,0}\makebox(0,0)[lb]{\smash{$0.2$}}}%
    \put(0.03335671,0.3343154){\color[rgb]{0,0,0}\makebox(0,0)[lb]{\smash{$0.4$}}}%
    \put(0.03335671,0.47208489){\color[rgb]{0,0,0}\makebox(0,0)[lb]{\smash{$0.6$}}}%
    \put(0.03335671,0.61149594){\color[rgb]{0,0,0}\makebox(0,0)[lb]{\smash{$0.8$}}}%
    \put(0.03335671,0.74926575){\color[rgb]{0,0,0}\makebox(0,0)[lb]{\smash{$1.0$}}}%
    \put(0.06861937,0.01581266){\color[rgb]{0,0,0}\makebox(0,0)[lb]{\smash{$0.0$}}}%
    \put(0.15062542,0.01581266){\color[rgb]{0,0,0}\makebox(0,0)[lb]{\smash{$0.1$}}}%
    \put(0.23263114,0.01581266){\color[rgb]{0,0,0}\makebox(0,0)[lb]{\smash{$0.2$}}}%
    \put(0.31627842,0.01581266){\color[rgb]{0,0,0}\makebox(0,0)[lb]{\smash{$0.3$}}}%
    \put(0.39828415,0.01581266){\color[rgb]{0,0,0}\makebox(0,0)[lb]{\smash{$0.4$}}}%
    \put(0.48028987,0.01581266){\color[rgb]{0,0,0}\makebox(0,0)[lb]{\smash{$0.5$}}}%
    \put(0.56393715,0.01581266){\color[rgb]{0,0,0}\makebox(0,0)[lb]{\smash{$0.6$}}}%
    \put(0.64594288,0.01581266){\color[rgb]{0,0,0}\makebox(0,0)[lb]{\smash{$0.7$}}}%
    \put(0.72794860,0.01581266){\color[rgb]{0,0,0}\makebox(0,0)[lb]{\smash{$0.8$}}}%
    \put(0.81159425,0.01581266){\color[rgb]{0,0,0}\makebox(0,0)[lb]{\smash{$0.9$}}}%
    \put(0.89278083,0.01581266){\color[rgb]{0,0,0}\makebox(0,0)[lb]{\smash{$1.0$}}}%
    \put(0.92230263,0.05877479){\color[rgb]{0,0,0}\makebox(0,0)[lb]{\smash{   $0$}}}%
    \put(0.92230263,0.1899846){\color[rgb]{0,0,0}\makebox(0,0)[lb]{\smash{ $200$}}}%
    \put(0.92230263,0.32119442){\color[rgb]{0,0,0}\makebox(0,0)[lb]{\smash{ $400$}}}%
    \put(0.92230263,0.45240423){\color[rgb]{0,0,0}\makebox(0,0)[lb]{\smash{ $600$}}}%
    \put(0.92230263,0.58525397){\color[rgb]{0,0,0}\makebox(0,0)[lb]{\smash{ $800$}}}%
    \put(0.92266271,0.71646330){\color[rgb]{0,0,0}\makebox(0,0)[lb]{\smash{$1000$}}}%
    \put(0.01121504,0.35809667){\color[rgb]{0,0,0}\rotatebox{90}{\makebox(0,0)[lb]{\smash{$\braket{\hat{n}_j}/N$}}}}%
    \put(1.03692954,0.37531826){\color[rgb]{0,0,0}\rotatebox{90}{\makebox(0,0)[lb]{\smash{$|\Omega_j|/\Gamma$}}}}%
    \put(0.47290942,-0.01136947){\color[rgb]{0,0,0}\makebox(0,0)[lb]{\smash{$t/\tau$}}}%
    \put(0.15592669,0.60404036){\color[rgb]{0,0,0}\makebox(0,0)[lb]{\smash{$|\Omega_{c_{1,2}}|$}}}%
    \put(0.55082162,0.60404036){\color[rgb]{0,0,0}\makebox(0,0)[lb]{\smash{$|\Omega_{a,b}|$}}}%
    \put(0.4702566,0.70108496){\color[rgb]{1,0.5,0}\makebox(0,0)[lb]{\smash{$\braket{\hat{n}_{c_1}}$}}}%
    \put(0.4702566,0.14099257){\color[rgb]{0,1,1}\makebox(0,0)[lb]{\smash{$\braket{\hat{n}_{c_2}}$}}}%
    \put(0.4702566,0.08030945){\color[rgb]{1,0,1}\makebox(0,0)[lb]{\smash{$\braket{\hat{n}_{e_{1,2}}}$}}}%
  \end{picture}%
\endgroup%

%% file: plotfile.tex
\begingroup%
  \makeatletter%
  \providecommand\color[2][]{%
    \errmessage{(Inkscape) Color is used for the text in Inkscape, but the package 'color.sty' is not loaded}%
    \renewcommand\color[2][]{}%
  }%
  \providecommand\transparent[1]{%
    \errmessage{(Inkscape) Transparency is used (non-zero) for the text in Inkscape, but the package 'transparent.sty' is not loaded}%
    \renewcommand\transparent[1]{}%
  }%
  \providecommand\rotatebox[2]{#2}%
  \ifx\svgwidth\undefined%
    \setlength{\unitlength}{454.17578125bp}%
    \ifx\svgscale\undefined%
      \relax%
    \else%
      \setlength{\unitlength}{\unitlength * \real{\svgscale}}%
    \fi%
  \else%
    \setlength{\unitlength}{\svgwidth}%
  \fi%
  \global\let\svgwidth\undefined%
  \global\let\svgscale\undefined%
  \makeatother%
  \begin{picture}(1,0.81336674)%
    \put(0,0){\includegraphics[width=\unitlength,page=1]{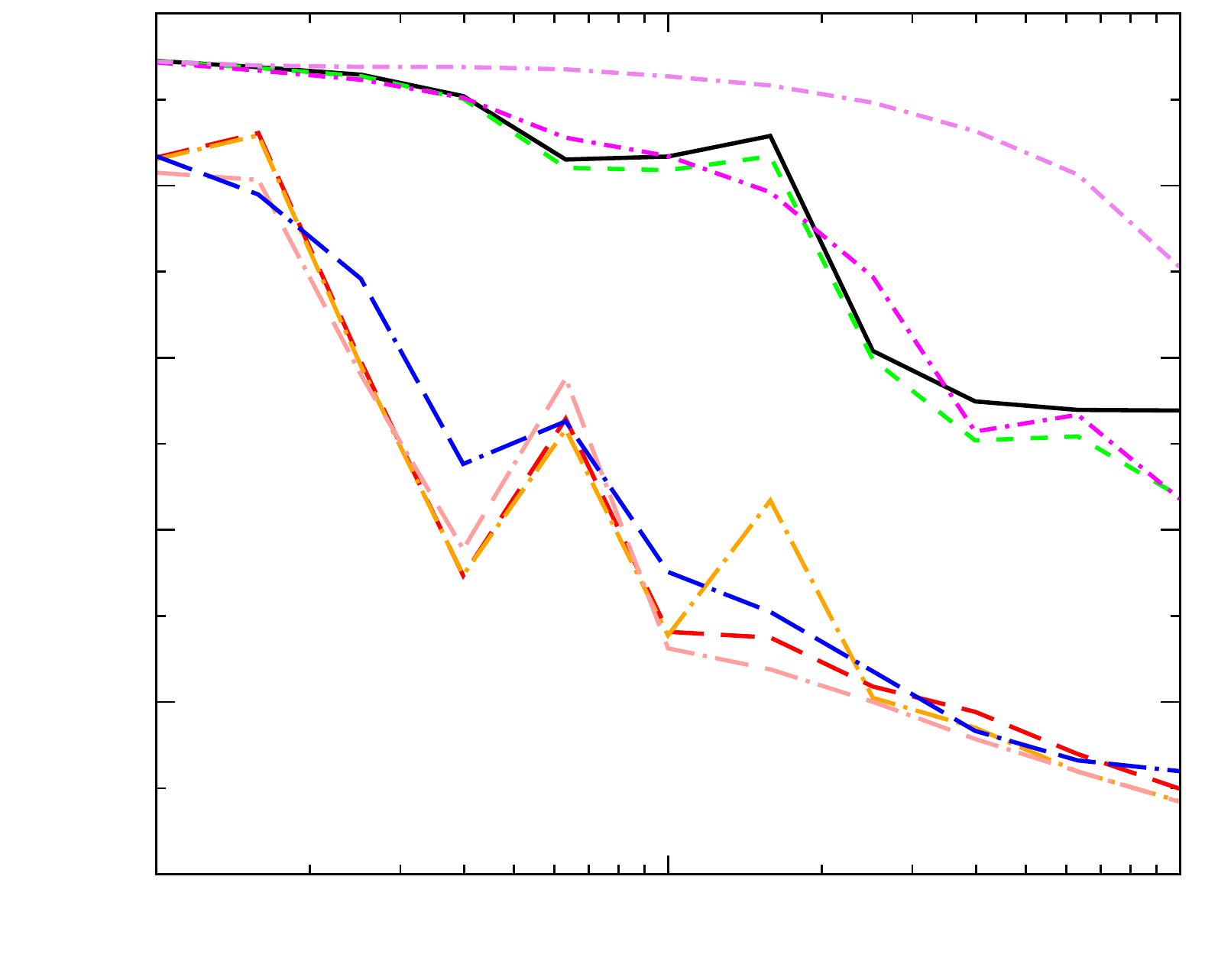}}%
    \put(0.11488135,0.07958697){\makebox(0,0)[rb]{\smash{$10^{-10}$}}}%
    \put(0.11488135,0.22261529){\makebox(0,0)[rb]{\smash{$10^{-8}$}}}%
    \put(0.11488135,0.36546747){\makebox(0,0)[rb]{\smash{$10^{-6}$}}}%
    \put(0.11488135,0.50849579){\makebox(0,0)[rb]{\smash{$10^{-4}$}}}%
    \put(0.11488135,0.65134797){\makebox(0,0)[rb]{\smash{$10^{-2}$}}}%
    \put(0.11488135,0.7943763){\makebox(0,0)[rb]{\smash{$1$}}}%
    \put(0.12950124,0.04788119){\makebox(0,0)[b]{\smash{1}}}%
    \put(0.55435877,0.04788119){\makebox(0,0)[b]{\smash{10}}}%
    \put(0.9792163,0.04788119){\makebox(0,0)[b]{\smash{100}}}%
    \put(0.01923557,0.44490808){\rotatebox{90}{\makebox(0,0)[b]{\smash{$D(\rho,\sigma) = 1 - F$}}}}%
    \put(0.55435877,0.00032251){\makebox(0,0)[b]{\smash{$t_\textup{end}/\tau$}}}%
    \put(0.67272875,0.6363604){\makebox(0,0)[lb]{\smash{$N = 10^2$,$\Omega = 10^2$}}}%
    \put(0.58460553,0.40824118){\color[rgb]{0,1,0}\makebox(0,0)[lb]{\smash{$N = 10^3$, $\Omega = 10^2$}}}%
    \put(0.46688158,0.70060262){\color[rgb]{1,0,1}\makebox(0,0)[lb]{\smash{$N = 10^4$, $\Omega = 10^2$}}}%
    \put(0.650114,0.74216264){\color[rgb]{0.93333333,0.50980392,0.93333333}\makebox(0,0)[lb]{\smash{$N = 10^5$, $\Omega = 10^2$}}}%
    \put(0.14547418,0.3070094){\color[rgb]{1,0,0}\makebox(0,0)[lb]{\smash{$N = 10^2$, $\Omega = 10^3$}}}%
    \put(0.66373688,0.35764019){\color[rgb]{1,0.64705882,0}\makebox(0,0)[lb]{\smash{$N = 10^3$, $\Omega = 10^3$}}}%
    \put(0.42630946,0.5010858){\color[rgb]{1,0.62352941,0.62352941}\makebox(0,0)[lb]{\smash{$N = 10^4$, $\Omega = 10^3$}}}%
    \put(0.3037035,0.57836309){\color[rgb]{0,0,1}\makebox(0,0)[lb]{\smash{$N = 10^5$, $\Omega = 10^3$}}}%
  \end{picture}%
\endgroup%

%% file: paperPRA.bbl
\begin{thebibliography}{50}
\expandafter\ifx\csname natexlab\endcsname\relax\def\natexlab#1{#1}\fi
\expandafter\ifx\csname bibnamefont\endcsname\relax
  \def\bibnamefont#1{#1}\fi
\expandafter\ifx\csname bibfnamefont\endcsname\relax
  \def\bibfnamefont#1{#1}\fi
\expandafter\ifx\csname citenamefont\endcsname\relax
  \def\citenamefont#1{#1}\fi
\expandafter\ifx\csname url\endcsname\relax
  \def\url#1{\texttt{#1}}\fi
\expandafter\ifx\csname urlprefix\endcsname\relax\def\urlprefix{URL }\fi
\providecommand{\bibinfo}[2]{#2}
\providecommand{\eprint}[2][]{\url{#2}}

\bibitem[{\citenamefont{Ockeloen et~al.}(2013)\citenamefont{Ockeloen, Schmied,
  Riedel, and Treutlein}}]{ockeloen2013quantum}
\bibinfo{author}{\bibfnamefont{C.~F.} \bibnamefont{Ockeloen}},
  \bibinfo{author}{\bibfnamefont{R.}~\bibnamefont{Schmied}},
  \bibinfo{author}{\bibfnamefont{M.~F.} \bibnamefont{Riedel}},
  \bibnamefont{and}
  \bibinfo{author}{\bibfnamefont{P.}~\bibnamefont{Treutlein}},
  \bibinfo{journal}{Physical review letters} \textbf{\bibinfo{volume}{111}},
  \bibinfo{pages}{143001} (\bibinfo{year}{2013}).

\bibitem[{\citenamefont{Riedel et~al.}(2010)\citenamefont{Riedel, B{\"o}hi, Li,
  H{\"a}nsch, Sinatra, and Treutlein}}]{riedel2010atom}
\bibinfo{author}{\bibfnamefont{M.~F.} \bibnamefont{Riedel}},
  \bibinfo{author}{\bibfnamefont{P.}~\bibnamefont{B{\"o}hi}},
  \bibinfo{author}{\bibfnamefont{Y.}~\bibnamefont{Li}},
  \bibinfo{author}{\bibfnamefont{T.~W.} \bibnamefont{H{\"a}nsch}},
  \bibinfo{author}{\bibfnamefont{A.}~\bibnamefont{Sinatra}}, \bibnamefont{and}
  \bibinfo{author}{\bibfnamefont{P.}~\bibnamefont{Treutlein}},
  \bibinfo{journal}{Nature} \textbf{\bibinfo{volume}{464}},
  \bibinfo{pages}{1170} (\bibinfo{year}{2010}).

\bibitem[{\citenamefont{Byrnes and Dowling}(2015)}]{byrnes2015quantum}
\bibinfo{author}{\bibfnamefont{T.}~\bibnamefont{Byrnes}} \bibnamefont{and}
  \bibinfo{author}{\bibfnamefont{J.~P.} \bibnamefont{Dowling}},
  \bibinfo{journal}{Physical Review A} \textbf{\bibinfo{volume}{92}},
  \bibinfo{pages}{023629} (\bibinfo{year}{2015}).

\bibitem[{\citenamefont{Bloch et~al.}(2012)\citenamefont{Bloch, Dalibard, and
  Nascimb{\`e}ne}}]{bloch2012quantum}
\bibinfo{author}{\bibfnamefont{I.}~\bibnamefont{Bloch}},
  \bibinfo{author}{\bibfnamefont{J.}~\bibnamefont{Dalibard}}, \bibnamefont{and}
  \bibinfo{author}{\bibfnamefont{S.}~\bibnamefont{Nascimb{\`e}ne}},
  \bibinfo{journal}{Nature Physics} \textbf{\bibinfo{volume}{8}},
  \bibinfo{pages}{267} (\bibinfo{year}{2012}).

\bibitem[{\citenamefont{Fischer and Sch{\"u}tzhold}(2004)}]{fischer2004quantum}
\bibinfo{author}{\bibfnamefont{U.~R.} \bibnamefont{Fischer}} \bibnamefont{and}
  \bibinfo{author}{\bibfnamefont{R.}~\bibnamefont{Sch{\"u}tzhold}},
  \bibinfo{journal}{Physical Review A} \textbf{\bibinfo{volume}{70}},
  \bibinfo{pages}{063615} (\bibinfo{year}{2004}).

\bibitem[{\citenamefont{Simon et~al.}(2011)\citenamefont{Simon, Bakr, Ma, Tai,
  Preiss, and Greiner}}]{simon2011quantum}
\bibinfo{author}{\bibfnamefont{J.}~\bibnamefont{Simon}},
  \bibinfo{author}{\bibfnamefont{W.~S.} \bibnamefont{Bakr}},
  \bibinfo{author}{\bibfnamefont{R.}~\bibnamefont{Ma}},
  \bibinfo{author}{\bibfnamefont{M.~E.} \bibnamefont{Tai}},
  \bibinfo{author}{\bibfnamefont{P.~M.} \bibnamefont{Preiss}},
  \bibnamefont{and} \bibinfo{author}{\bibfnamefont{M.}~\bibnamefont{Greiner}},
  \bibinfo{journal}{Nature} \textbf{\bibinfo{volume}{472}},
  \bibinfo{pages}{307} (\bibinfo{year}{2011}).

\bibitem[{\citenamefont{Micheli et~al.}(2003)\citenamefont{Micheli, Jaksch,
  Cirac, and Zoller}}]{zoller2003quantum}
\bibinfo{author}{\bibfnamefont{A.}~\bibnamefont{Micheli}},
  \bibinfo{author}{\bibfnamefont{D.}~\bibnamefont{Jaksch}},
  \bibinfo{author}{\bibfnamefont{J.~I.} \bibnamefont{Cirac}}, \bibnamefont{and}
  \bibinfo{author}{\bibfnamefont{P.}~\bibnamefont{Zoller}},
  \bibinfo{journal}{Phys. Rev. A} \textbf{\bibinfo{volume}{67}},
  \bibinfo{pages}{013607} (\bibinfo{year}{2003}).

\bibitem[{\citenamefont{Brennen et~al.}(1999)\citenamefont{Brennen, Caves,
  Jessen, and Deutsch}}]{brennen1999quantum}
\bibinfo{author}{\bibfnamefont{G.~K.} \bibnamefont{Brennen}},
  \bibinfo{author}{\bibfnamefont{C.~M.} \bibnamefont{Caves}},
  \bibinfo{author}{\bibfnamefont{P.~S.} \bibnamefont{Jessen}},
  \bibnamefont{and} \bibinfo{author}{\bibfnamefont{I.~H.}
  \bibnamefont{Deutsch}}, \bibinfo{journal}{Physical Review Letters}
  \textbf{\bibinfo{volume}{82}}, \bibinfo{pages}{1060} (\bibinfo{year}{1999}).

\bibitem[{\citenamefont{Wasilewski et~al.}(2010)\citenamefont{Wasilewski,
  Jensen, Krauter, Renema, Balabas, and Polzik}}]{wasilewski2010quantum}
\bibinfo{author}{\bibfnamefont{W.}~\bibnamefont{Wasilewski}},
  \bibinfo{author}{\bibfnamefont{K.}~\bibnamefont{Jensen}},
  \bibinfo{author}{\bibfnamefont{H.}~\bibnamefont{Krauter}},
  \bibinfo{author}{\bibfnamefont{J.~J.} \bibnamefont{Renema}},
  \bibinfo{author}{\bibfnamefont{M.}~\bibnamefont{Balabas}}, \bibnamefont{and}
  \bibinfo{author}{\bibfnamefont{E.~S.} \bibnamefont{Polzik}},
  \bibinfo{journal}{Physical review letters} \textbf{\bibinfo{volume}{104}},
  \bibinfo{pages}{133601} (\bibinfo{year}{2010}).

\bibitem[{\citenamefont{M{\"u}ntinga et~al.}(2013)\citenamefont{M{\"u}ntinga,
  Ahlers, Krutzik, Wenzlawski, Arnold, Becker, Bongs, Dittus, Duncker, Gaaloul
  et~al.}}]{muntinga2013interferometry}
\bibinfo{author}{\bibfnamefont{H.}~\bibnamefont{M{\"u}ntinga}},
  \bibinfo{author}{\bibfnamefont{H.}~\bibnamefont{Ahlers}},
  \bibinfo{author}{\bibfnamefont{M.}~\bibnamefont{Krutzik}},
  \bibinfo{author}{\bibfnamefont{A.}~\bibnamefont{Wenzlawski}},
  \bibinfo{author}{\bibfnamefont{S.}~\bibnamefont{Arnold}},
  \bibinfo{author}{\bibfnamefont{D.}~\bibnamefont{Becker}},
  \bibinfo{author}{\bibfnamefont{K.}~\bibnamefont{Bongs}},
  \bibinfo{author}{\bibfnamefont{H.}~\bibnamefont{Dittus}},
  \bibinfo{author}{\bibfnamefont{H.}~\bibnamefont{Duncker}},
  \bibinfo{author}{\bibfnamefont{N.}~\bibnamefont{Gaaloul}},
  \bibnamefont{et~al.}, \bibinfo{journal}{Physical review letters}
  \textbf{\bibinfo{volume}{110}}, \bibinfo{pages}{093602}
  (\bibinfo{year}{2013}).

\bibitem[{\citenamefont{Treutlein et~al.}(2007)\citenamefont{Treutlein, Hunger,
  Camerer, H{\"a}nsch, and Reichel}}]{treutlein2007bose}
\bibinfo{author}{\bibfnamefont{P.}~\bibnamefont{Treutlein}},
  \bibinfo{author}{\bibfnamefont{D.}~\bibnamefont{Hunger}},
  \bibinfo{author}{\bibfnamefont{S.}~\bibnamefont{Camerer}},
  \bibinfo{author}{\bibfnamefont{T.~W.} \bibnamefont{H{\"a}nsch}},
  \bibnamefont{and} \bibinfo{author}{\bibfnamefont{J.}~\bibnamefont{Reichel}},
  \bibinfo{journal}{Physical Review Letters} \textbf{\bibinfo{volume}{99}},
  \bibinfo{pages}{140403} (\bibinfo{year}{2007}).

\bibitem[{\citenamefont{Byrnes}(2012)}]{timbec2}
\bibinfo{author}{\bibfnamefont{T.}~\bibnamefont{Byrnes}},
  \bibinfo{journal}{World Acad. Sci. Eng. Technol}
  \textbf{\bibinfo{volume}{63}}, \bibinfo{pages}{542} (\bibinfo{year}{2012}).

\bibitem[{\citenamefont{Pyrkov and Byrnes}(2013{\natexlab{a}})}]{timbec3}
\bibinfo{author}{\bibfnamefont{A.}~\bibnamefont{Pyrkov}} \bibnamefont{and}
  \bibinfo{author}{\bibfnamefont{T.}~\bibnamefont{Byrnes}}, in
  \emph{\bibinfo{booktitle}{International Conference on Micro-and
  Nano-Electronics 2012}} (\bibinfo{organization}{International Society for
  Optics and Photonics}, \bibinfo{year}{2013}{\natexlab{a}}), pp.
  \bibinfo{pages}{87001E--87001E}.

\bibitem[{\citenamefont{Byrnes et~al.}(2015)\citenamefont{Byrnes, Rosseau,
  Khosla, Pyrkov, Thomasen, Mukai, Koyama, Abdelrahman, and
  Ilo-Okeke}}]{timbec4}
\bibinfo{author}{\bibfnamefont{T.}~\bibnamefont{Byrnes}},
  \bibinfo{author}{\bibfnamefont{D.}~\bibnamefont{Rosseau}},
  \bibinfo{author}{\bibfnamefont{M.}~\bibnamefont{Khosla}},
  \bibinfo{author}{\bibfnamefont{A.}~\bibnamefont{Pyrkov}},
  \bibinfo{author}{\bibfnamefont{A.}~\bibnamefont{Thomasen}},
  \bibinfo{author}{\bibfnamefont{T.}~\bibnamefont{Mukai}},
  \bibinfo{author}{\bibfnamefont{S.}~\bibnamefont{Koyama}},
  \bibinfo{author}{\bibfnamefont{A.}~\bibnamefont{Abdelrahman}},
  \bibnamefont{and}
  \bibinfo{author}{\bibfnamefont{E.}~\bibnamefont{Ilo-Okeke}},
  \bibinfo{journal}{Optics Communications} \textbf{\bibinfo{volume}{337}},
  \bibinfo{pages}{102} (\bibinfo{year}{2015}).

\bibitem[{\citenamefont{Pyrkov and
  Byrnes}(2013{\natexlab{b}})}]{pyrkoventangle}
\bibinfo{author}{\bibfnamefont{A.~N.} \bibnamefont{Pyrkov}} \bibnamefont{and}
  \bibinfo{author}{\bibfnamefont{T.}~\bibnamefont{Byrnes}},
  \bibinfo{journal}{New Journal of Physics} \textbf{\bibinfo{volume}{15}},
  \bibinfo{pages}{093019} (\bibinfo{year}{2013}{\natexlab{b}}).

\bibitem[{\citenamefont{Hufnagel et~al.}(2009)\citenamefont{Hufnagel, Mukai,
  and Shimizu}}]{mukailifetime}
\bibinfo{author}{\bibfnamefont{C.}~\bibnamefont{Hufnagel}},
  \bibinfo{author}{\bibfnamefont{T.}~\bibnamefont{Mukai}}, \bibnamefont{and}
  \bibinfo{author}{\bibfnamefont{F.}~\bibnamefont{Shimizu}},
  \bibinfo{journal}{Phys. Rev. A} \textbf{\bibinfo{volume}{79}},
  \bibinfo{pages}{053641} (\bibinfo{year}{2009}).

\bibitem[{\citenamefont{Zoller et~al.}(2005)\citenamefont{Zoller, Beth, Binosi,
  Blatt, Briegel, Bruss, Calarco, Cirac, Deutsch, Eisert
  et~al.}}]{zoller2005quantum}
\bibinfo{author}{\bibfnamefont{P.}~\bibnamefont{Zoller}},
  \bibinfo{author}{\bibfnamefont{T.}~\bibnamefont{Beth}},
  \bibinfo{author}{\bibfnamefont{D.}~\bibnamefont{Binosi}},
  \bibinfo{author}{\bibfnamefont{R.}~\bibnamefont{Blatt}},
  \bibinfo{author}{\bibfnamefont{H.}~\bibnamefont{Briegel}},
  \bibinfo{author}{\bibfnamefont{D.}~\bibnamefont{Bruss}},
  \bibinfo{author}{\bibfnamefont{T.}~\bibnamefont{Calarco}},
  \bibinfo{author}{\bibfnamefont{J.~I.} \bibnamefont{Cirac}},
  \bibinfo{author}{\bibfnamefont{D.}~\bibnamefont{Deutsch}},
  \bibinfo{author}{\bibfnamefont{J.}~\bibnamefont{Eisert}},
  \bibnamefont{et~al.}, \bibinfo{journal}{The European Physical Journal
  D-Atomic, Molecular, Optical and Plasma Physics}
  \textbf{\bibinfo{volume}{36}}, \bibinfo{pages}{203} (\bibinfo{year}{2005}).

\bibitem[{\citenamefont{Simon et~al.}(2010)\citenamefont{Simon, Afzelius,
  Appel, de~La~Giroday, Dewhurst, Gisin, Hu, Jelezko, Kr{\"o}ll, M{\"u}ller
  et~al.}}]{simon2010quantum}
\bibinfo{author}{\bibfnamefont{C.}~\bibnamefont{Simon}},
  \bibinfo{author}{\bibfnamefont{M.}~\bibnamefont{Afzelius}},
  \bibinfo{author}{\bibfnamefont{J.}~\bibnamefont{Appel}},
  \bibinfo{author}{\bibfnamefont{A.~B.} \bibnamefont{de~La~Giroday}},
  \bibinfo{author}{\bibfnamefont{S.}~\bibnamefont{Dewhurst}},
  \bibinfo{author}{\bibfnamefont{N.}~\bibnamefont{Gisin}},
  \bibinfo{author}{\bibfnamefont{C.}~\bibnamefont{Hu}},
  \bibinfo{author}{\bibfnamefont{F.}~\bibnamefont{Jelezko}},
  \bibinfo{author}{\bibfnamefont{S.}~\bibnamefont{Kr{\"o}ll}},
  \bibinfo{author}{\bibfnamefont{J.}~\bibnamefont{M{\"u}ller}},
  \bibnamefont{et~al.}, \bibinfo{journal}{The European Physical Journal D}
  \textbf{\bibinfo{volume}{58}}, \bibinfo{pages}{1} (\bibinfo{year}{2010}).

\bibitem[{\citenamefont{Campbell et~al.}(2006)\citenamefont{Campbell, Mun,
  Boyd, Medley, Leanhardt, Marcassa, Pritchard, and
  Ketterle}}]{campbell2006imaging}
\bibinfo{author}{\bibfnamefont{G.~K.} \bibnamefont{Campbell}},
  \bibinfo{author}{\bibfnamefont{J.}~\bibnamefont{Mun}},
  \bibinfo{author}{\bibfnamefont{M.}~\bibnamefont{Boyd}},
  \bibinfo{author}{\bibfnamefont{P.}~\bibnamefont{Medley}},
  \bibinfo{author}{\bibfnamefont{A.~E.} \bibnamefont{Leanhardt}},
  \bibinfo{author}{\bibfnamefont{L.~G.} \bibnamefont{Marcassa}},
  \bibinfo{author}{\bibfnamefont{D.~E.} \bibnamefont{Pritchard}},
  \bibnamefont{and} \bibinfo{author}{\bibfnamefont{W.}~\bibnamefont{Ketterle}},
  \bibinfo{journal}{Science} \textbf{\bibinfo{volume}{313}},
  \bibinfo{pages}{649} (\bibinfo{year}{2006}).

\bibitem[{\citenamefont{Ludlow et~al.}(2015)\citenamefont{Ludlow, Boyd, Ye,
  Peik, and Schmidt}}]{ludlow2015optical}
\bibinfo{author}{\bibfnamefont{A.~D.} \bibnamefont{Ludlow}},
  \bibinfo{author}{\bibfnamefont{M.~M.} \bibnamefont{Boyd}},
  \bibinfo{author}{\bibfnamefont{J.}~\bibnamefont{Ye}},
  \bibinfo{author}{\bibfnamefont{E.}~\bibnamefont{Peik}}, \bibnamefont{and}
  \bibinfo{author}{\bibfnamefont{P.~O.} \bibnamefont{Schmidt}},
  \bibinfo{journal}{Reviews of Modern Physics} \textbf{\bibinfo{volume}{87}},
  \bibinfo{pages}{637} (\bibinfo{year}{2015}).

\bibitem[{\citenamefont{B{\"o}hi et~al.}(2009)\citenamefont{B{\"o}hi, Riedel,
  Hoffrogge, Reichel, H{\"a}nsch, and Treutlein}}]{bohi2009coherent}
\bibinfo{author}{\bibfnamefont{P.}~\bibnamefont{B{\"o}hi}},
  \bibinfo{author}{\bibfnamefont{M.~F.} \bibnamefont{Riedel}},
  \bibinfo{author}{\bibfnamefont{J.}~\bibnamefont{Hoffrogge}},
  \bibinfo{author}{\bibfnamefont{J.}~\bibnamefont{Reichel}},
  \bibinfo{author}{\bibfnamefont{T.~W.} \bibnamefont{H{\"a}nsch}},
  \bibnamefont{and}
  \bibinfo{author}{\bibfnamefont{P.}~\bibnamefont{Treutlein}},
  \bibinfo{journal}{Nature Physics} \textbf{\bibinfo{volume}{5}},
  \bibinfo{pages}{592} (\bibinfo{year}{2009}).

\bibitem[{\citenamefont{Mukai et~al.}(2007)\citenamefont{Mukai, Hufnagel,
  Kasper, Meno, Tsukada, Semba, and Shimizu}}]{mukaichip}
\bibinfo{author}{\bibfnamefont{T.}~\bibnamefont{Mukai}},
  \bibinfo{author}{\bibfnamefont{C.}~\bibnamefont{Hufnagel}},
  \bibinfo{author}{\bibfnamefont{A.}~\bibnamefont{Kasper}},
  \bibinfo{author}{\bibfnamefont{T.}~\bibnamefont{Meno}},
  \bibinfo{author}{\bibfnamefont{A.}~\bibnamefont{Tsukada}},
  \bibinfo{author}{\bibfnamefont{K.}~\bibnamefont{Semba}}, \bibnamefont{and}
  \bibinfo{author}{\bibfnamefont{F.}~\bibnamefont{Shimizu}},
  \bibinfo{journal}{Phys. Rev. Lett.} \textbf{\bibinfo{volume}{98}},
  \bibinfo{pages}{260407} (\bibinfo{year}{2007}).

\bibitem[{\citenamefont{Hofferberth}(2007)}]{hofferberth2007coherent}
\bibinfo{author}{\bibfnamefont{S.}~\bibnamefont{Hofferberth}}, Ph.D. thesis,
  \bibinfo{school}{University of Heidelberg} (\bibinfo{year}{2007}).

\bibitem[{\citenamefont{Treutlein et~al.}(2006)\citenamefont{Treutlein,
  H{\"a}nsch, Reichel, Negretti, Cirone, and Calarco}}]{treutlein2006microwave}
\bibinfo{author}{\bibfnamefont{P.}~\bibnamefont{Treutlein}},
  \bibinfo{author}{\bibfnamefont{T.~W.} \bibnamefont{H{\"a}nsch}},
  \bibinfo{author}{\bibfnamefont{J.}~\bibnamefont{Reichel}},
  \bibinfo{author}{\bibfnamefont{A.}~\bibnamefont{Negretti}},
  \bibinfo{author}{\bibfnamefont{M.~A.} \bibnamefont{Cirone}},
  \bibnamefont{and} \bibinfo{author}{\bibfnamefont{T.}~\bibnamefont{Calarco}},
  \bibinfo{journal}{Physical Review A} \textbf{\bibinfo{volume}{74}},
  \bibinfo{pages}{022312} (\bibinfo{year}{2006}).

\bibitem[{\citenamefont{Treutlein et~al.}(2004)\citenamefont{Treutlein,
  Hommelhoff, Steinmetz, H{\"a}nsch, and Reichel}}]{treutlein2004coherence}
\bibinfo{author}{\bibfnamefont{P.}~\bibnamefont{Treutlein}},
  \bibinfo{author}{\bibfnamefont{P.}~\bibnamefont{Hommelhoff}},
  \bibinfo{author}{\bibfnamefont{T.}~\bibnamefont{Steinmetz}},
  \bibinfo{author}{\bibfnamefont{T.~W.} \bibnamefont{H{\"a}nsch}},
  \bibnamefont{and} \bibinfo{author}{\bibfnamefont{J.}~\bibnamefont{Reichel}},
  \bibinfo{journal}{Physical review letters} \textbf{\bibinfo{volume}{92}},
  \bibinfo{pages}{203005} (\bibinfo{year}{2004}).

\bibitem[{\citenamefont{Schumm et~al.}(2005)\citenamefont{Schumm, Hofferberth,
  Andersson, Wildermuth, Groth, Bar-Joseph, Schmiedmayer, and
  Kr{\"u}ger}}]{schumm2005matter}
\bibinfo{author}{\bibfnamefont{T.}~\bibnamefont{Schumm}},
  \bibinfo{author}{\bibfnamefont{S.}~\bibnamefont{Hofferberth}},
  \bibinfo{author}{\bibfnamefont{L.~M.} \bibnamefont{Andersson}},
  \bibinfo{author}{\bibfnamefont{S.}~\bibnamefont{Wildermuth}},
  \bibinfo{author}{\bibfnamefont{S.}~\bibnamefont{Groth}},
  \bibinfo{author}{\bibfnamefont{I.}~\bibnamefont{Bar-Joseph}},
  \bibinfo{author}{\bibfnamefont{J.}~\bibnamefont{Schmiedmayer}},
  \bibnamefont{and}
  \bibinfo{author}{\bibfnamefont{P.}~\bibnamefont{Kr{\"u}ger}},
  \bibinfo{journal}{Nature physics} \textbf{\bibinfo{volume}{1}},
  \bibinfo{pages}{57} (\bibinfo{year}{2005}).

\bibitem[{\citenamefont{G{\"o}rlitz et~al.}(2001)\citenamefont{G{\"o}rlitz,
  Chikkatur, and Ketterle}}]{spontg}
\bibinfo{author}{\bibfnamefont{A.}~\bibnamefont{G{\"o}rlitz}},
  \bibinfo{author}{\bibfnamefont{A.}~\bibnamefont{Chikkatur}},
  \bibnamefont{and} \bibinfo{author}{\bibfnamefont{W.}~\bibnamefont{Ketterle}},
  \bibinfo{journal}{Physical Review A} \textbf{\bibinfo{volume}{63}},
  \bibinfo{pages}{041601} (\bibinfo{year}{2001}).

\bibitem[{\citenamefont{Abdelrahman et~al.}(2014)\citenamefont{Abdelrahman,
  Mukai, H{\"a}ffner, and Byrnes}}]{abdelrahman2014coherent}
\bibinfo{author}{\bibfnamefont{A.}~\bibnamefont{Abdelrahman}},
  \bibinfo{author}{\bibfnamefont{T.}~\bibnamefont{Mukai}},
  \bibinfo{author}{\bibfnamefont{H.}~\bibnamefont{H{\"a}ffner}},
  \bibnamefont{and} \bibinfo{author}{\bibfnamefont{T.}~\bibnamefont{Byrnes}},
  \bibinfo{journal}{Optics express} \textbf{\bibinfo{volume}{22}},
  \bibinfo{pages}{3501} (\bibinfo{year}{2014}).

\bibitem[{\citenamefont{Kurkal and Rice}(2001)}]{kurkal2001sequential}
\bibinfo{author}{\bibfnamefont{V.}~\bibnamefont{Kurkal}} \bibnamefont{and}
  \bibinfo{author}{\bibfnamefont{S.~A.} \bibnamefont{Rice}},
  \bibinfo{journal}{Chemical physics letters} \textbf{\bibinfo{volume}{344}},
  \bibinfo{pages}{125} (\bibinfo{year}{2001}).

\bibitem[{\citenamefont{Sugny et~al.}(2007)\citenamefont{Sugny, Ndong,
  Lauvergnat, Justum, and Desouter-Lecomte}}]{sugny2007laser}
\bibinfo{author}{\bibfnamefont{D.}~\bibnamefont{Sugny}},
  \bibinfo{author}{\bibfnamefont{M.}~\bibnamefont{Ndong}},
  \bibinfo{author}{\bibfnamefont{D.}~\bibnamefont{Lauvergnat}},
  \bibinfo{author}{\bibfnamefont{Y.}~\bibnamefont{Justum}}, \bibnamefont{and}
  \bibinfo{author}{\bibfnamefont{M.}~\bibnamefont{Desouter-Lecomte}},
  \bibinfo{journal}{Journal of Photochemistry and Photobiology A: Chemistry}
  \textbf{\bibinfo{volume}{190}}, \bibinfo{pages}{359} (\bibinfo{year}{2007}).

\bibitem[{\citenamefont{Nesterenko et~al.}(2009)\citenamefont{Nesterenko,
  Novikov, de~Souza~Cruz, and Lapolli}}]{nesterenko2009stirap}
\bibinfo{author}{\bibfnamefont{V.}~\bibnamefont{Nesterenko}},
  \bibinfo{author}{\bibfnamefont{A.}~\bibnamefont{Novikov}},
  \bibinfo{author}{\bibfnamefont{F.}~\bibnamefont{de~Souza~Cruz}},
  \bibnamefont{and} \bibinfo{author}{\bibfnamefont{E.}~\bibnamefont{Lapolli}},
  \bibinfo{journal}{Laser physics} \textbf{\bibinfo{volume}{19}},
  \bibinfo{pages}{616} (\bibinfo{year}{2009}).

\bibitem[{\citenamefont{Bevilacqua et~al.}(2013)\citenamefont{Bevilacqua,
  Schaller, Brandes, and Renzoni}}]{stirapb}
\bibinfo{author}{\bibfnamefont{G.}~\bibnamefont{Bevilacqua}},
  \bibinfo{author}{\bibfnamefont{G.}~\bibnamefont{Schaller}},
  \bibinfo{author}{\bibfnamefont{T.}~\bibnamefont{Brandes}}, \bibnamefont{and}
  \bibinfo{author}{\bibfnamefont{F.}~\bibnamefont{Renzoni}},
  \bibinfo{journal}{Phys. Rev. A} \textbf{\bibinfo{volume}{88}},
  \bibinfo{pages}{013404} (\bibinfo{year}{2013}).

\bibitem[{\citenamefont{Schaller et~al.}(2006)\citenamefont{Schaller, Mostame,
  and Sch\"utzhold}}]{stiraps}
\bibinfo{author}{\bibfnamefont{G.}~\bibnamefont{Schaller}},
  \bibinfo{author}{\bibfnamefont{S.}~\bibnamefont{Mostame}}, \bibnamefont{and}
  \bibinfo{author}{\bibfnamefont{R.}~\bibnamefont{Sch\"utzhold}},
  \bibinfo{journal}{Phys. Rev. A} \textbf{\bibinfo{volume}{73}},
  \bibinfo{pages}{062307} (\bibinfo{year}{2006}).

\bibitem[{\citenamefont{Bergmann et~al.}(1998)\citenamefont{Bergmann, Theuer,
  and Shore}}]{stirapbt}
\bibinfo{author}{\bibfnamefont{K.}~\bibnamefont{Bergmann}},
  \bibinfo{author}{\bibfnamefont{H.}~\bibnamefont{Theuer}}, \bibnamefont{and}
  \bibinfo{author}{\bibfnamefont{B.~W.} \bibnamefont{Shore}},
  \bibinfo{journal}{Rev. Mod. Phys.} \textbf{\bibinfo{volume}{70}},
  \bibinfo{pages}{1003} (\bibinfo{year}{1998}).

\bibitem[{\citenamefont{Kis and Renzoni}(2002)}]{qubitrotationk}
\bibinfo{author}{\bibfnamefont{Z.}~\bibnamefont{Kis}} \bibnamefont{and}
  \bibinfo{author}{\bibfnamefont{F.}~\bibnamefont{Renzoni}},
  \bibinfo{journal}{Phys. Rev. A} \textbf{\bibinfo{volume}{65}},
  \bibinfo{pages}{032318} (\bibinfo{year}{2002}).

\bibitem[{\citenamefont{Issouffa et~al.}(2013)\citenamefont{Issouffa, Messikh,
  Wahiddin, Umarov, and Gharib}}]{qubitrotationi}
\bibinfo{author}{\bibfnamefont{Y.}~\bibnamefont{Issouffa}},
  \bibinfo{author}{\bibfnamefont{A.}~\bibnamefont{Messikh}},
  \bibinfo{author}{\bibfnamefont{M.}~\bibnamefont{Wahiddin}},
  \bibinfo{author}{\bibfnamefont{B.}~\bibnamefont{Umarov}}, \bibnamefont{and}
  \bibinfo{author}{\bibfnamefont{M.}~\bibnamefont{Gharib}}, in
  \emph{\bibinfo{booktitle}{Information and Communication Technology for the
  Muslim World (ICT4M)}} (\bibinfo{year}{2013}), pp. \bibinfo{pages}{1--6}.

\bibitem[{\citenamefont{Kis and Stenholm}(2002)}]{kis2002optimal}
\bibinfo{author}{\bibfnamefont{Z.}~\bibnamefont{Kis}} \bibnamefont{and}
  \bibinfo{author}{\bibfnamefont{S.}~\bibnamefont{Stenholm}},
  \bibinfo{journal}{Journal of Modern Optics} \textbf{\bibinfo{volume}{49}},
  \bibinfo{pages}{111} (\bibinfo{year}{2002}).

\bibitem[{\citenamefont{Vasilev et~al.}(2009)\citenamefont{Vasilev, Kuhn, and
  Vitanov}}]{vasilev2009optimum}
\bibinfo{author}{\bibfnamefont{G.}~\bibnamefont{Vasilev}},
  \bibinfo{author}{\bibfnamefont{A.}~\bibnamefont{Kuhn}}, \bibnamefont{and}
  \bibinfo{author}{\bibfnamefont{N.}~\bibnamefont{Vitanov}},
  \bibinfo{journal}{Physical Review A} \textbf{\bibinfo{volume}{80}},
  \bibinfo{pages}{013417} (\bibinfo{year}{2009}).

\bibitem[{\citenamefont{Sol{\'a} et~al.}(1999)\citenamefont{Sol{\'a},
  Malinovsky, and Tannor}}]{sola1999optimal}
\bibinfo{author}{\bibfnamefont{I.~R.} \bibnamefont{Sol{\'a}}},
  \bibinfo{author}{\bibfnamefont{V.~S.} \bibnamefont{Malinovsky}},
  \bibnamefont{and} \bibinfo{author}{\bibfnamefont{D.~J.}
  \bibnamefont{Tannor}}, \bibinfo{journal}{Physical Review A}
  \textbf{\bibinfo{volume}{60}}, \bibinfo{pages}{3081} (\bibinfo{year}{1999}).

\bibitem[{\citenamefont{Idlas et~al.}(2016)\citenamefont{Idlas, Domenzain,
  Spreeuw, and Byrnes}}]{idlas2016entanglement}
\bibinfo{author}{\bibfnamefont{S.}~\bibnamefont{Idlas}},
  \bibinfo{author}{\bibfnamefont{L.}~\bibnamefont{Domenzain}},
  \bibinfo{author}{\bibfnamefont{R.}~\bibnamefont{Spreeuw}}, \bibnamefont{and}
  \bibinfo{author}{\bibfnamefont{T.}~\bibnamefont{Byrnes}},
  \bibinfo{journal}{Physical Review A} \textbf{\bibinfo{volume}{93}},
  \bibinfo{pages}{022319} (\bibinfo{year}{2016}).

\bibitem[{\citenamefont{Beterov et~al.}(2013)\citenamefont{Beterov, Saffman,
  Yakshina, Zhukov, Tretyakov, Entin, Ryabtsev, Mansell, MacCormick, Bergamini
  et~al.}}]{qubitrotationb}
\bibinfo{author}{\bibfnamefont{I.~I.} \bibnamefont{Beterov}},
  \bibinfo{author}{\bibfnamefont{M.}~\bibnamefont{Saffman}},
  \bibinfo{author}{\bibfnamefont{E.~A.} \bibnamefont{Yakshina}},
  \bibinfo{author}{\bibfnamefont{V.~P.} \bibnamefont{Zhukov}},
  \bibinfo{author}{\bibfnamefont{D.~B.} \bibnamefont{Tretyakov}},
  \bibinfo{author}{\bibfnamefont{V.~M.} \bibnamefont{Entin}},
  \bibinfo{author}{\bibfnamefont{I.~I.} \bibnamefont{Ryabtsev}},
  \bibinfo{author}{\bibfnamefont{C.~W.} \bibnamefont{Mansell}},
  \bibinfo{author}{\bibfnamefont{C.}~\bibnamefont{MacCormick}},
  \bibinfo{author}{\bibfnamefont{S.}~\bibnamefont{Bergamini}},
  \bibnamefont{et~al.}, \bibinfo{journal}{Phys. Rev. A}
  \textbf{\bibinfo{volume}{88}}, \bibinfo{pages}{010303}
  (\bibinfo{year}{2013}).

\bibitem[{\citenamefont{Petrosyan and
  M{\o}lmer}(2013)}]{petrosyan2013stimulated}
\bibinfo{author}{\bibfnamefont{D.}~\bibnamefont{Petrosyan}} \bibnamefont{and}
  \bibinfo{author}{\bibfnamefont{K.}~\bibnamefont{M{\o}lmer}},
  \bibinfo{journal}{Physical Review A} \textbf{\bibinfo{volume}{87}},
  \bibinfo{pages}{033416} (\bibinfo{year}{2013}).

\bibitem[{\citenamefont{M{\o}ller et~al.}(2007)\citenamefont{M{\o}ller, Madsen,
  and M{\o}lmer}}]{moller2007geometric}
\bibinfo{author}{\bibfnamefont{D.}~\bibnamefont{M{\o}ller}},
  \bibinfo{author}{\bibfnamefont{L.~B.} \bibnamefont{Madsen}},
  \bibnamefont{and}
  \bibinfo{author}{\bibfnamefont{K.}~\bibnamefont{M{\o}lmer}},
  \bibinfo{journal}{Physical Review A} \textbf{\bibinfo{volume}{75}},
  \bibinfo{pages}{062302} (\bibinfo{year}{2007}).

\bibitem[{\citenamefont{M{\o}ller et~al.}(2008)\citenamefont{M{\o}ller, Madsen,
  and M{\o}lmer}}]{moller2008quantum}
\bibinfo{author}{\bibfnamefont{D.}~\bibnamefont{M{\o}ller}},
  \bibinfo{author}{\bibfnamefont{L.~B.} \bibnamefont{Madsen}},
  \bibnamefont{and}
  \bibinfo{author}{\bibfnamefont{K.}~\bibnamefont{M{\o}lmer}},
  \bibinfo{journal}{Physical review letters} \textbf{\bibinfo{volume}{100}},
  \bibinfo{pages}{170504} (\bibinfo{year}{2008}).

\bibitem[{\citenamefont{Unanyan et~al.}(1998)\citenamefont{Unanyan,
  Fleischhauer, Shore, and Bergmann}}]{unanyan1998robust}
\bibinfo{author}{\bibfnamefont{R.}~\bibnamefont{Unanyan}},
  \bibinfo{author}{\bibfnamefont{M.}~\bibnamefont{Fleischhauer}},
  \bibinfo{author}{\bibfnamefont{B.}~\bibnamefont{Shore}}, \bibnamefont{and}
  \bibinfo{author}{\bibfnamefont{K.}~\bibnamefont{Bergmann}},
  \bibinfo{journal}{Optics Communications} \textbf{\bibinfo{volume}{155}},
  \bibinfo{pages}{144} (\bibinfo{year}{1998}).

\bibitem[{\citenamefont{Xu et~al.}(2015)\citenamefont{Xu, Liu, Xue, Su, Deng,
  Tian, Zheng, Han, Zhong, Wang et~al.}}]{xu2015coherent}
\bibinfo{author}{\bibfnamefont{H.}~\bibnamefont{Xu}},
  \bibinfo{author}{\bibfnamefont{W.}~\bibnamefont{Liu}},
  \bibinfo{author}{\bibfnamefont{G.}~\bibnamefont{Xue}},
  \bibinfo{author}{\bibfnamefont{F.}~\bibnamefont{Su}},
  \bibinfo{author}{\bibfnamefont{H.}~\bibnamefont{Deng}},
  \bibinfo{author}{\bibfnamefont{Y.}~\bibnamefont{Tian}},
  \bibinfo{author}{\bibfnamefont{D.}~\bibnamefont{Zheng}},
  \bibinfo{author}{\bibfnamefont{S.}~\bibnamefont{Han}},
  \bibinfo{author}{\bibfnamefont{Y.}~\bibnamefont{Zhong}},
  \bibinfo{author}{\bibfnamefont{H.}~\bibnamefont{Wang}}, \bibnamefont{et~al.},
  \bibinfo{journal}{arXiv preprint arXiv:1508.01849}  (\bibinfo{year}{2015}).

\bibitem[{\citenamefont{Waxman and Folman}(2007)}]{waxman2007coherent}
\bibinfo{author}{\bibfnamefont{A.}~\bibnamefont{Waxman}} \bibnamefont{and}
  \bibinfo{author}{\bibfnamefont{R.}~\bibnamefont{Folman}}, Master's thesis,
  \bibinfo{school}{Ben-Guiron University of the Negev} (\bibinfo{year}{2007}).

\bibitem[{\citenamefont{Steck}(2001)}]{steck2001rubidium}
\bibinfo{author}{\bibfnamefont{D.~A.} \bibnamefont{Steck}},
  \emph{\bibinfo{title}{Rubidium 87 d line data}} (\bibinfo{year}{2001}).

\bibitem[{\citenamefont{Byrnes et~al.}(2012)\citenamefont{Byrnes, Wen, and
  Yamamoto}}]{timbec1}
\bibinfo{author}{\bibfnamefont{T.}~\bibnamefont{Byrnes}},
  \bibinfo{author}{\bibfnamefont{K.}~\bibnamefont{Wen}}, \bibnamefont{and}
  \bibinfo{author}{\bibfnamefont{Y.}~\bibnamefont{Yamamoto}},
  \bibinfo{journal}{Phys. Rev. A} \textbf{\bibinfo{volume}{85}},
  \bibinfo{pages}{040306} (\bibinfo{year}{2012}).

\bibitem[{\citenamefont{Ladd et~al.}(2010)\citenamefont{Ladd, Jelezko,
  Laflamme, Nakamura, Monroe, and O’Brien}}]{qcrevl}
\bibinfo{author}{\bibfnamefont{T.~D.} \bibnamefont{Ladd}},
  \bibinfo{author}{\bibfnamefont{F.}~\bibnamefont{Jelezko}},
  \bibinfo{author}{\bibfnamefont{R.}~\bibnamefont{Laflamme}},
  \bibinfo{author}{\bibfnamefont{Y.}~\bibnamefont{Nakamura}},
  \bibinfo{author}{\bibfnamefont{C.}~\bibnamefont{Monroe}}, \bibnamefont{and}
  \bibinfo{author}{\bibfnamefont{J.~L.} \bibnamefont{O’Brien}},
  \bibinfo{journal}{Nature} \textbf{\bibinfo{volume}{464}}, \bibinfo{pages}{45}
  (\bibinfo{year}{2010}).

\end{thebibliography}
